\pdfoutput=1
\documentclass[a4paper,11pt]{article}

\usepackage{jheppub}
\usepackage[T1]{fontenc} 
\usepackage{atlasphysics}
\usepackage{mydefs}
\usepackage{subfigure}
\usepackage{rotating}
\usepackage{morefloats} 
\usepackage{afterpage}
\usepackage{fancyvrb}
\usepackage{placeins}
\usepackage{multirow}

\newif\ifdraft

\ifdraft
\usepackage{epstopdf}
\usepackage{lineno}
\linenumbers
\else
\DeclareGraphicsExtensions{.pdf}
\fi

\ifdraft
\else
\usepackage{preprintcover}  
\PreprintCoverPaperTitle{Measurement of dijet cross-sections in $pp$ collisions at 7 TeV centre-of-mass energy using the ATLAS detector}  
\PreprintIdNumber{CERN-PH-EP-2013-192}  
\PreprintCoverAbstract{
Double-differential dijet cross-sections measured in $pp$ collisions at the LHC with a \onlyseventev centre-of-mass energy are presented as functions of dijet mass and half the rapidity separation of the two highest-\pt jets.
These measurements are obtained using data corresponding to an integrated luminosity of \lumifbshort{}, recorded by the ATLAS detector in 2011.
The data are corrected for detector effects so that cross-sections are presented at the particle level.
Cross-sections are measured up to $5\TeV$ dijet mass using jets reconstructed with the \AKT algorithm for values of the jet radius parameter of $0.4$ and $0.6$.
The cross-sections are compared with next-to-leading-order perturbative QCD calculations by \nlojet corrected to account for non-perturbative effects.
Comparisons with \powheg predictions, using a next-to-leading-order matrix element calculation interfaced to a parton-shower Monte Carlo simulation, are also shown.
Electroweak effects are accounted for in both cases.
The quantitative comparison of data and theoretical predictions obtained using various parameterizations of the parton distribution functions is performed using a frequentist method.
In general, good agreement with data is observed for the \nlojet theoretical predictions when using the CT10, NNPDF2.1 and MSTW 2008 PDF sets.
Disagreement is observed when using the ABM11 and HERAPDF1.5 PDF sets for some ranges of dijet mass and half the rapidity separation.
An example setting a lower limit on the compositeness scale for a model of contact interactions is presented, showing that the unfolded results can be used to constrain contributions to dijet production beyond that predicted by the Standard Model.
}  
\PreprintJournalName{JHEP}  
\fi

\title{\boldmath Measurement of dijet cross-sections in $pp$ collisions at 7 TeV centre-of-mass energy using the ATLAS detector}

\author{The ATLAS Collaboration}

\emailAdd{atlas.publications@cern.ch}

\abstract{
Double-differential dijet cross-sections measured in $pp$ collisions at the LHC with a \onlyseventev centre-of-mass energy are presented as functions of dijet mass and half the rapidity separation of the two highest-\pt jets.
These measurements are obtained using data corresponding to an integrated luminosity of \lumifbshort{}, recorded by the ATLAS detector in 2011.
The data are corrected for detector effects so that cross-sections are presented at the particle level.
Cross-sections are measured up to $5\TeV$ dijet mass using jets reconstructed with the \AKT algorithm for values of the jet radius parameter of $0.4$ and $0.6$.
The cross-sections are compared with next-to-leading-order perturbative QCD calculations by \nlojet corrected to account for non-perturbative effects.
Comparisons with \powheg predictions, using a next-to-leading-order matrix element calculation interfaced to a parton-shower Monte Carlo simulation, are also shown.
Electroweak effects are accounted for in both cases.
The quantitative comparison of data and theoretical predictions obtained using various parameterizations of the parton distribution functions is performed using a frequentist method.
In general, good agreement with data is observed for the \nlojet theoretical predictions when using the CT10, NNPDF2.1 and MSTW 2008 PDF sets.
Disagreement is observed when using the ABM11 and HERAPDF1.5 PDF sets for some ranges of dijet mass and half the rapidity separation.
An example setting a lower limit on the compositeness scale for a model of contact interactions is presented, showing that the unfolded results can be used to constrain contributions to dijet production beyond that predicted by the Standard Model.
}

\VerbatimFootnotes

\begin{document}

\maketitle

\flushbottom

\section{Introduction}
\label{sec:intro}

Measurements of jet production in $pp$ collisions at the LHC \cite{Evans:2008zzb} test the predictions of Quantum Chromodynamics (QCD) at the largest centre-of-mass energies explored thus far in colliders.
They are useful for understanding the strong interaction and its coupling strength \alphas, one of the fundamental parameters of the Standard 
Model.
The measurement of cross-sections as a function of dijet mass is also sensitive to resonances and new interactions, and can be employed in searches for physics beyond the Standard Model.
Where no new contribution is found, the cross-section results can be exploited to study the partonic structure of the proton.
In particular, the high dijet-mass region can be used to constrain the parton distribution function (PDF) of gluons in the proton at high momentum fraction.

Previous measurements at the Tevatron in $p\bar{p}$ collisions have shown good agreement between predictions from QCD calculations and data for lower dijet mass \cite{Aaltonen:2008dn,Abazov:2010fr}.
Recent results from the LHC using $pp$ collisions have extended the measurement of dijet production cross-sections to higher dijet-mass values \cite{Chatrchyan:2011qta,Aad:2011fc,Chatrchyan:2012bja}.
While good agreement between data and several theoretical predictions has been observed in general, the predicted cross-section values using the CT10 PDF set \cite{Lai:2010vv} at high dijet mass tend to be larger than those measured in data.

In this paper, measurements of the double-differential dijet cross-sections are presented as functions of the dijet mass and half the rapidity separation of the two highest-\pt jets.
The measurements are made at the particle level (see section \ref{sec:xsdef} for a definition), using the iterative, dynamically stabilized (IDS) unfolding method \cite{Malaescu:2011yg} to correct for detector effects.
The use of an integrated luminosity more than a factor of 100 larger than in the previous ATLAS publication \cite{Aad:2011fc} improves the statistical power in the high dijet-mass region.
In spite of the increased number of simultaneous proton--proton interactions in the same beam bunch crossing during 2011 data taking, improvements in the jet energy calibration result in an overall systematic uncertainty smaller than previously achieved \cite{ATLAS-CONF-2013-004}.
The measurements are compared to next-to-leading-order (NLO) QCD calculations \cite{Nagy:2003tz}, as well as to a NLO matrix element calculation by POWHEG \cite{Nason:2004rx}, which is interfaced to the parton-shower Monte Carlo generator \pythia.
Furthermore, a quantitative comparison of data and theory predictions is made using a frequentist method.

New particles, predicted by theories beyond the Standard Model (SM), may decay into dijets that can be observed as narrow resonances in dijet mass spectra \cite{Baur:1987ga}.
New interactions at higher energy scales, parameterized using an effective theory of contact interactions \cite{Eichten:1983hw}, may lead to modifications of dijet cross-sections at high dijet mass.
Searches for these deviations have been performed by ATLAS \cite{ATLAS:2012pu} and CMS \cite{Chatrchyan:2012bf}.
The approach followed here is to constrain physics beyond the SM using unfolded cross-sections and the full information on their uncertainties and correlations. 
This information is also provided in HepData \cite{Buckley:2010jn}.
This has the advantage of allowing new models to be confronted with data without the need for additional detector simulations.
This paper considers only one theory of contact interactions, rather than presenting a comprehensive list of exclusion ranges for various models.
The results are an illustration of what can be achieved using unfolded data, and do not seek to improve the current exclusion range.

The content of the paper is as follows.
The ATLAS detector and data-taking conditions are briefly described in sections \ref{sec:atlasdet} and \ref{sec:conditions}, followed by the cross-section definition in section \ref{sec:xsdef}.
Sections \ref{sec:mcsamp} and \ref{sec:theory} describe the Monte Carlo samples and theoretical predictions, respectively.
The trigger, jet reconstruction, and data selection are presented in section \ref{sec:datasel}, followed by studies of the stability of the results under different luminosity conditions (pileup) in section \ref{sec:pileup}.
Sections \ref{sec:unfolding} and \ref{sec:sysunc} discuss the data unfolding and systematic uncertainties on the measurement, respectively.
These are followed by the introduction of a frequentist method for the quantitative comparison of data and theory predictions in section \ref{sec:stattest}.
The cross-section results are presented in section \ref{sec:results}, along with quantitative statements of the ability of the theory prediction to describe the data.
An application of the frequentist method for setting a lower limit on the compositeness scale of a model of contact interactions is shown in section \ref{sec:setlimits}.
Finally, the conclusions are given in section \ref{sec:conclusions}.

\section{The ATLAS experiment}
\label{sec:atlasdet}

The ATLAS detector is described in detail elsewhere \cite{Aad:2008zzm}.
The main system used for this analysis is the calorimeter, divided into electromagnetic and hadronic parts.
The lead/liquid-argon (LAr) electromagnetic calorimeter is split into three regions:\footnote{
   ATLAS uses a right-handed coordinate system with its origin at
   the nominal interaction point (IP) in the centre of the detector and the $z$-axis pointing along the beam axis.
   The $x$-axis points from the IP to the centre of the LHC ring, and the $y$-axis points upward.
   Cylindrical coordinates ($r$, $\phi$) are used in the transverse plane, $\phi$ being the azimuthal angle around the beam axis, referred to the $x$-axis.
   The pseudorapidity is defined in terms of the polar angle $\theta$ with respect to the beamline as $\eta=-\ln\tan(\theta/2)$. 
   When dealing with massive jets and particles, the rapidity $y = 1/2 \ln (E+p_z)/(E-p_z)$ is used, where $E$ is the jet energy and $p_z$ is the $z$-component of the jet momentum.
   The transverse momentum \pt is defined as the component of the momentum transverse to the beam axis.
}
the barrel ($\abseta < 1.475$), the end-cap ($1.375 < \abseta < 3.2$), and the forward ($3.1< \abseta <4.9$) regions.
The hadronic calorimeter is divided into four regions:
the barrel ($\abseta < 0.8$) and the extended barrel ($0.8 < \abseta < 1.7$) made of scintillator/steel, the end-cap ($1.5 < \abseta < 3.2$) with LAr/copper modules, and the forward calorimeter covering ($3.1 < \abseta < 4.9$) composed of LAr/copper and LAr/tungsten modules.
The tracking detectors, consisting of silicon pixels, silicon microstrips, and transition radiation tracking detectors immersed in a 2\T axial magnetic field provided by a solenoid, are used to reconstruct charged-particle tracks in the pseudorapidity region $\abseta < 2.5$.
Outside the calorimeter, a large muon spectrometer measures the momenta and trajectories of muons deflected by a large air-core toroidal magnetic system.

\section{Data-taking conditions}
\label{sec:conditions}

Data-taking periods in ATLAS are divided into intervals of approximately uniform running conditions called luminosity blocks, with a typical duration of one minute.
For a given pair of colliding beam bunches, the expected number of $pp$ collisions per bunch crossing averaged over a luminosity block is referred to as \actmu.
The average of \actmu over all bunches in the collider for a given luminosity block is denoted by \avgmu.
In 2011 the peak luminosity delivered by the accelerator at the start of an LHC fill increased, causing \avgmu to change from 5 at the beginning of the data-taking period to more than 18 by the end.
A \emph{bunch train} in the accelerator is generally composed of 36 proton bunches with 50\ns \emph{bunch spacing}, followed by a larger 250\ns window before the next bunch train.

The overlay of multiple collisions, either from the same bunch crossing or due to electronic signals present from adjacent bunch crossings, is termed \emph{pileup}.
There are two separate, although related, contributions to consider.
In-time pileup refers to additional energy deposited in the detector due to simultaneous collisions in the same bunch crossing.
Out-of-time pileup is a result of the 500\ns pulse length of the LAr calorimeter readout, compared to the 50\ns spacing between bunch crossings, leaving residual electronic signals in the detector, predominantly from previous interactions.
The shape of the LAr calorimeter pulse is roughly a 100\ns peak of positive amplitude, followed by a shallower 400\ns trough of negative amplitude.
Due to this trough, out-of-time pileup results by design in a subtraction of measured energy in the LAr calorimeter, providing some compensation for in-time pileup.

\section{Cross-section definition}
\label{sec:xsdef}

Jet cross-sections are defined using jets reconstructed by the \AKT algorithm \cite{Cacciari:2008gp} implemented in the FastJet \cite{Cacciari:2011ma} package.
In this analysis, jets are clustered using two different values of the radius parameter, \rfour and \rsix.

Non-perturbative (hadronization and underlying event) and perturbative (higher-order corrections and parton showers) effects are different, depending on the choice of the jet radius parameter.
By performing the measurement for two values of the jet radius parameter, different contributions from perturbative and non-perturbative effects are probed.
The fragmentation process leads to a dispersion of the jet constituents, more of which are collected for jets with a larger radius parameter.
Particles from the underlying event make additional contributions to the jet, and have less of an effect on jets with a smaller radius parameter.
Additionally, hard emissions lead to higher jet multiplicity when using a smaller jet radius parameter.

Measured cross-sections are corrected for all experimental effects, and thus are defined at the \emph{particle-level} final state.
Here particle level refers to stable particles, defined as those with a proper lifetime longer than $10\ps$, including muons and neutrinos from decaying hadrons \cite{Buttar:2008jx}.

Events containing two or more jets are considered, with the leading (subleading) jet defined as the one within the range $\absy<3.0$ with the highest (second highest) \pt.
Dijet double-differential cross-sections are measured as functions of the dijet mass \mass and half the rapidity separation $\ystar = |y_1 - y_2|/2$ of the two leading jets.
This rapidity separation is invariant under a Lorentz boost along the $z$-direction, so that in the dijet rest frame $y_1^\prime = -y_2^\prime = \ystar$.
The leading (subleading) jet is required to have $\pt > 100 \GeV$ ($\pt > 50 \GeV$).
The requirements on \pt for the two jets are asymmetric to improve the stability of the NLO calculation~\cite{Frixione:1997ks}.
The measurement is made in six ranges of $\ystar < 3.0$, in equal steps of $0.5$.
In each range of \ystar, a lower limit on the dijet mass is chosen in order to avoid the region of phase space affected by the requirements on the \pt of the two leading jets.

\section{Monte Carlo samples}
\label{sec:mcsamp}

The default Monte Carlo (MC) generator used to simulate jet events is \pythia 6.425 \cite{Sjostrand:2006za} with the ATLAS Underlying Event Tune AUET2B \cite{ATL-PHYS-PUB-2011-009}.
It is a leading-order (LO) generator with $2 \to 2$ matrix element calculations, supplemented by leading-logarithmic parton showers ordered in \pt.
A simulation of the underlying event is also provided, including multiple parton interactions.
The Lund string model \cite{Andersson:1979ij} is used to simulate the hadronization process.
To simulate pileup, minimum bias events\footnote{
   Events passing a trigger with minimum requirements, and which correspond mostly to inelastic $pp$ collisions, are called minimum-bias events.
} are generated using \pythia 8 \cite{Sjostrand:2007gs} with the 4C tune \cite{Corke:2010yf} and MRST LO$^{**}$ proton PDF set \cite{Sherstnev:2008dm}.
The number of minimum bias events overlaid on each signal event is chosen to model the distribution of \avgmu throughout the data-taking period.

To estimate the uncertainties on the hadronization and parton-shower modelling, events are also generated by \herwigpp 2.5.2 \cite{Bahr:2008pv,Gieseke:2011na,Corcella:2000bw} using the UE-EE-3 tune \cite{Gieseke:2012ft}.
In this LO generator, the parton shower follows an angular ordering, and a clustering model \cite{Webber:1983if} is used for the hadronization. 
The effect of the underlying event is included using the eikonal multiple-scattering model \cite{Bahr:2008dy}.

Two PDF sets are considered for both MC generators.
For the nominal detector simulation, the MRST LO$^{**}$ proton PDF set is used.
Additionally, versions of the same tunes based on the CTEQ6L1 \cite{Pumplin:2002vw} proton PDF set are used to assess uncertainties on the non-perturbative corrections (see section \ref{subsec:theoryunc}).

The output four-vectors from these event generators are passed to a detector simulation \cite{Aad:2010ah} based on \geant \cite{Agostinelli:2002hh}.
Simulated events are digitized to model the detector responses, and then reconstructed using the software used to process data.

\section{Theoretical predictions and uncertainties}
\label{sec:theory}

The measured dijet cross-sections are compared to fixed-order NLO QCD predictions by \nlojet \cite{Nagy:2003tz}, corrected for non-perturbative effects in the fragmentation process and in the underlying event using calculations by \pythia 6.425.
A NLO matrix element calculation by \powheg, which is interfaced to the \pythia parton-shower MC generator, is also considered.
Both the \nlojet and \powheg predictions are corrected to account for NLO electroweak effects.
When used in this paper, the term \emph{Standard Model predictions} refers to NLO QCD calculations corrected for non-perturbative and electroweak effects.

\subsection{Theoretical predictions}
\label{subsec:theorycalc}

The fixed-order $O(\alphas^3)$ QCD calculations are performed with the \nlojet  program interfaced to APPLGRID~\cite{Carli:2010rw} for fast convolution with various PDF sets.
The following proton PDF sets are considered for the theoretical predictions:
CT10 \cite{Lai:2010vv}, HERAPDF1.5 \cite{HERAPDF15}, epATLJet13 \cite{Carli:1447073}, MSTW 2008 \cite{Martin:2009iq}, NNPDF2.1 \cite{Ball:2010de,Forte:2010ta} and NNPDF2.3 \cite{Ball:2012cx}, and ABM11 \cite{Alekhin:2013dmy}.
The epATLJet13 PDF set is the result of simultaneously using in the fit ATLAS jet data, collected at centre-of-mass energies of $2.76\TeV$ and $7\TeV$, and HERA-I $ep$ data.
In the previous ATLAS measurement \cite{Aad:2011fc}, a scale choice was introduced to ensure that the cross-sections remained positive for large values of \ystar.
Although values of \ystar greater than 3.0 are not considered here, the same choice is used for consistency.
The renormalization ($\mu_\mathrm{R}$) and factorization ($\mu_\mathrm{F}$) scales are set to
\ifdraft \begin{linenomath} \fi
\begin{equation}
  \mu = \mu_\mathrm{R} = \mu_\mathrm{F} = \pt^\mathrm{max} \mathrm{e}^{0.3 \ystar},
\end{equation}
\ifdraft \end{linenomath} \fi
where $\pt^\mathrm{max}$ is the \pt of the leading jet.
Further details can be found in ref. \cite{Ellis:1992en}.

\begin{figure}[tbp]
\centering 
\subfigure[Non-perturbative corrections]{
  \includegraphics[width=.45\textwidth]{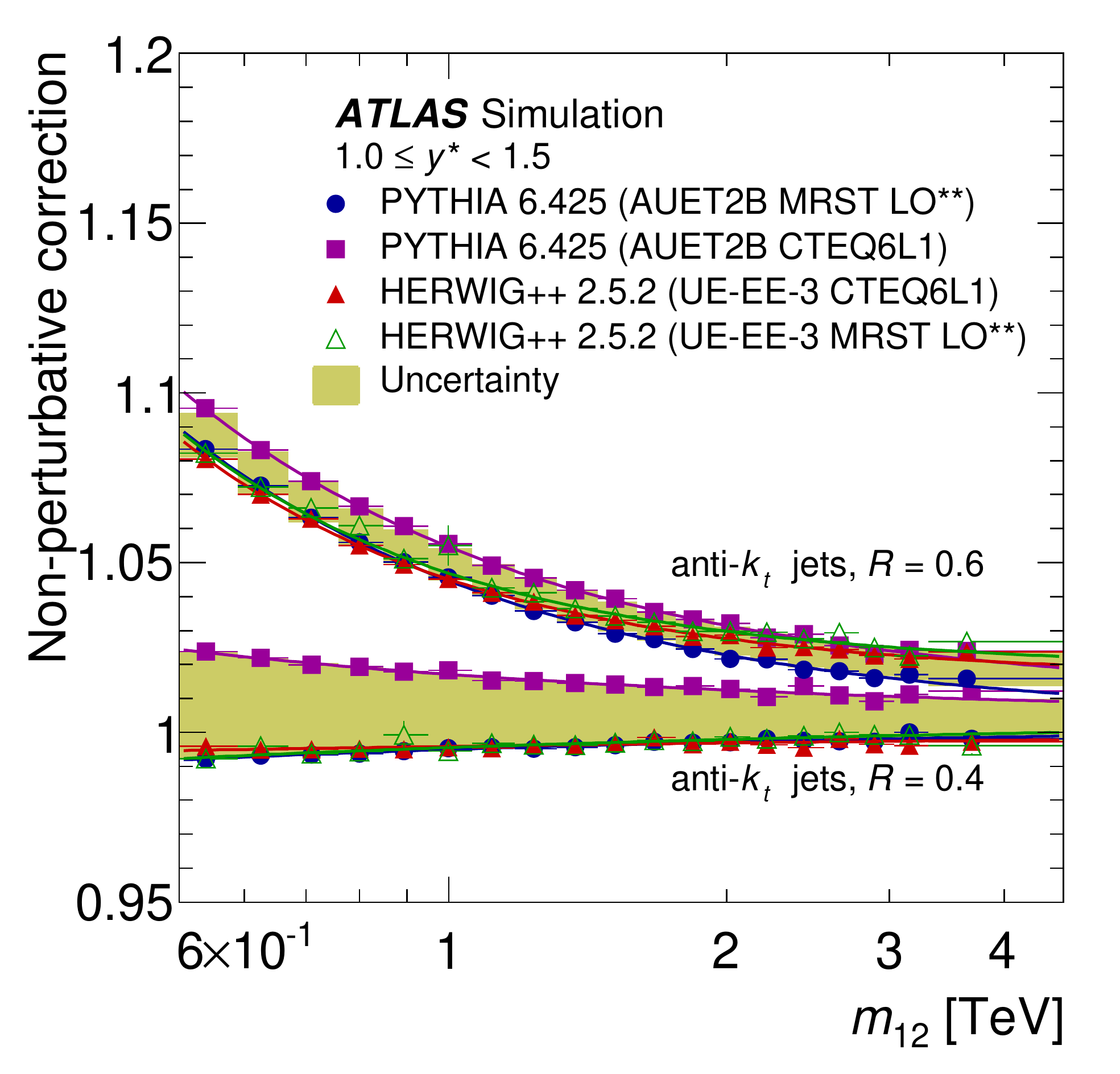}
  \label{fig:npcorr:a}
}
\subfigure[Electroweak corrections]{
  \includegraphics[width=.45\textwidth]{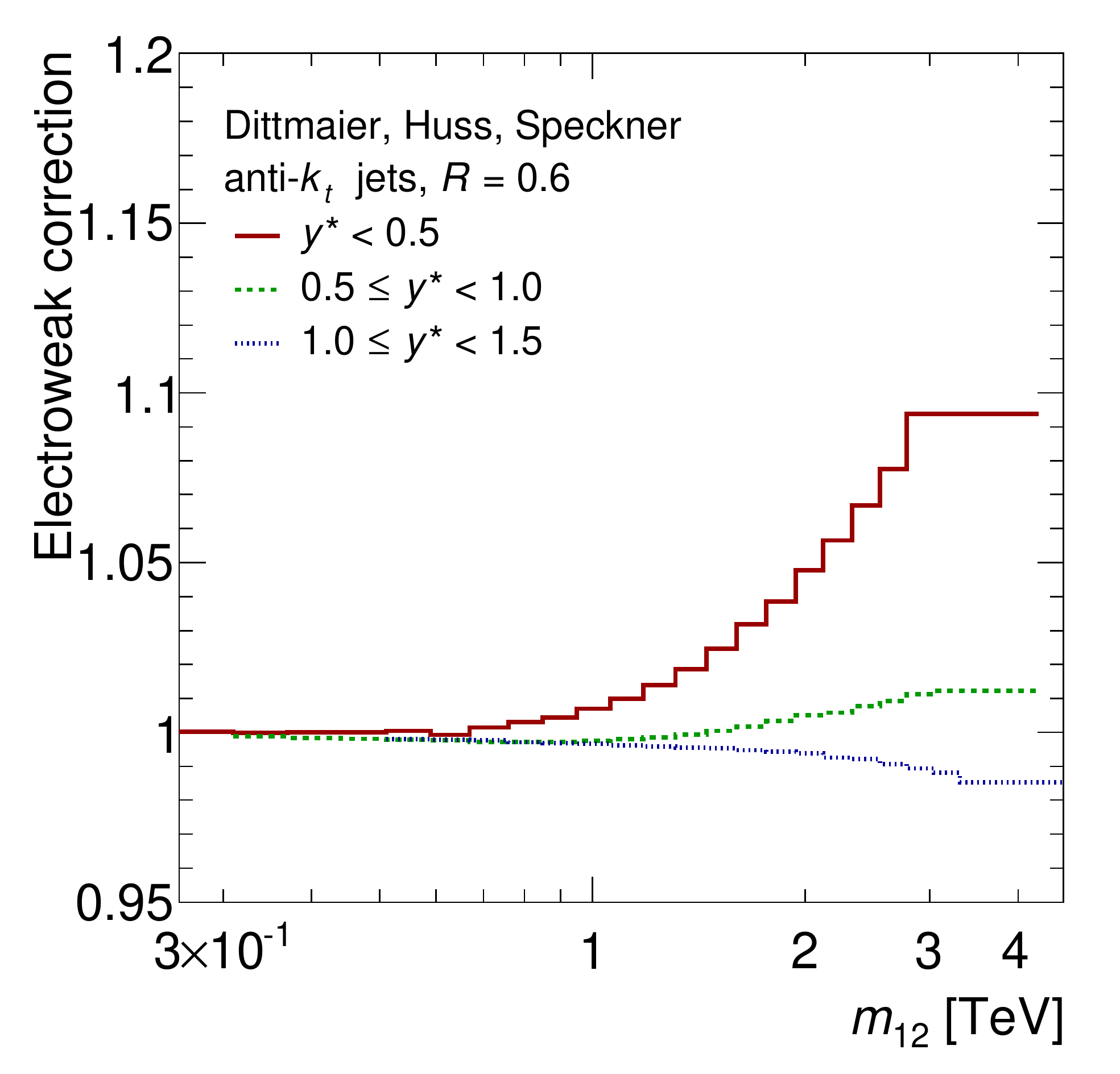}
  \label{fig:npcorr:b}
}
\caption{
Non-perturbative corrections (ratio of particle-level cross-sections to parton-level cross-sections) obtained using various MC generators and tunes are shown in (a), for the differential dijet cross-sections as a function of dijet mass in the range \ystarthree with values of jet radius parameter \rfour and \rsix. Uncertainties are taken as the envelope of the various curves.
Electroweak corrections are shown in (b) as a function of dijet mass in multiple ranges of \ystar \cite{Dittmaier:2012kx}, for jet radius parameter \rsix.
}
\label{fig:npcorr}
\end{figure}

Non-perturbative corrections are evaluated using leading-logarithmic parton-shower generators, separately for each value of the jet radius parameter.
The corrections are calculated as bin-by-bin ratios of the dijet differential cross-section at the particle level, including hadronization and underlying event effects, over that at the parton level.
The nominal corrections are calculated using \pythia 6.425 with the AUET2B tune derived for the MRST LO$^{**}$ PDF set.
The non-perturbative corrections as a function of dijet mass are shown in figure \ref{fig:npcorr:a} for the range \ystarthree.

Comparisons are also made to the \powheg~\cite{Nason:2004rx,Frixione:2007vw,Alioli:2010xd} NLO matrix element calculation\footnote{
  The folding values of \verb+foldcsi=5+, \verb+foldy=10+, \verb+foldphi=2+ associated with the spacing of the integration grid as described in ref. \cite{Nason:2007vt}, are used.
} interfaced to a parton-shower MC generator, using the CT10 PDF set.
For these calculations the factorization and renormalization scales are set to the default value of $\pt^{\mathrm{Born}}$, the transverse momentum in the $2 \to 2$ process before the hardest emission.
The parton-level configurations are passed to the \pythia generator for fragmentation and underlying event simulation.
Both the AUET2B and Perugia 2011 \cite{Skands:2010ak} tunes are considered.
The version of the program used includes an explicit calculation of all hard emission diagrams \cite{Nason:2013uba,Carli:1447073}.
This provides improved suppression of rare events, removing fluctuations that arise from the migration of jets with low values of parton-level \pt to larger values of particle-level \pt.

Corrections for electroweak tree-level effects of $O(\alpha\alphas, \alpha^2)$ as well as weak loop effects of $O(\alpha\alphas^2)$ \cite{Dittmaier:2012kx} are applied to both the \nlojet and \powheg predictions.
The calculations are derived for NLO electroweak processes on a LO QCD prediction of the observable in the phase space considered here.
In general, electroweak effects on the cross-sections are $<1\%$ for values of $\ystar \ge 0.5$, but reach up to 9\% for $\mass > 3\TeV$ in the range \ystarone.
The magnitude of the corrections on the theoretical predictions as a function of dijet mass in several ranges of \ystar is shown in figure \ref{fig:npcorr:b}.
The electroweak corrections show almost no dependence on the jet radius parameter.

\subsection{Theoretical uncertainties}\label{subsec:theoryunc}

To estimate the uncertainty due to missing higher-order terms in the fixed-order calculations, the renormalization scale is varied up and down by a factor of two.
The uncertainty due to the choice of factorization scale, specifying the separation between the short-scale hard scatter and long-scale hadronization, is also varied by a factor of two.
All permutations of these two scale choices are considered, except for cases where the renormalization and factorization scales are varied in opposite directions.
In these extreme cases, logarithmic factors in the theoretical calculations may become large, resulting in instabilities in the prediction.
The maximal deviations from the nominal prediction are taken as the uncertainty due to the scale choice.
The scale uncertainty is generally within $^{+5\%}_{-15\%}$ for the \rfour calculation, and $\pm 10\%$ for the \rsix calculation.

The uncertainties on the cross-sections due to that on $\alphas$ are estimated using two additional proton PDF sets, for which different values of $\alphas$ are assumed in the fits.
This follows the recommended prescription in ref. \cite{Lai:2010nw}, such that the effect on the PDF set as well as on the matrix elements is included.
The resulting uncertainties are approximately $\pm 4\%$ across all dijet-mass and \ystar ranges considered in this analysis.

The multiple uncorrelated uncertainty components of each PDF set, as provided by the various PDF analyses, are also propagated through the theoretical calculations.
The PDF analyses generally derive these from the experimental uncertainties on the data used in the fits.
For the results shown in section \ref{sec:results} the standard Hessian sum in quadrature \cite{Pumplin:2001ct} of the various independent components is taken.
The NNPDF2.1 and NNPDF2.3 PDF sets are exceptions, where uncertainties are expressed in terms of \emph{replicas} instead of by independent components.
These replicas represent a collection of equally likely PDF sets, where the data entering the PDF fit were varied within their experimental uncertainties.
For the plots shown in section \ref{sec:results}, the uncertainties on the NNPDF PDF sets are propagated using the RMS of the replicas, producing equivalent PDF uncertainties on the theoretical predictions.
For the frequentist method described in section \ref{sec:stattest}, these replicas are used to derive a covariance matrix for the theoretical predictions.
The eigenvector decomposition of this matrix provides a set of independent uncertainty components, which can be treated in the same way as those in the other PDF sets.

In cases where variations of the theoretical parameters are also available, these are treated as additional uncertainty components assuming a Gaussian distribution.
For the HERAPDF1.5, NNPDF2.1, and MSTW 2008 PDF sets, the results of varying the heavy-quark masses within their uncertainties are taken as additional uncertainty components.
The nominal value of the charm quark mass is taken as $m_{c} = 1.40 \GeV$, and the bottom quark mass as $m_{b} = 4.75 \GeV$.
For NNPDF2.1 the mass of the charm quark is provided with a symmetric uncertainty, while that of the bottom quark is provided with an asymmetric uncertainty.
NNPDF2.3 does not currently provide an estimate of the uncertainties due to heavy-quark masses.
For HERAPDF1.5 and MSTW 2008, asymmetric uncertainties for both the charm and bottom quark masses are available.
When considering the HERAPDF1.5 PDF set, the uncertainty components corresponding to the strange quark fraction and the $Q^2$ requirement are also included as asymmetric uncertainties.

In addition, the HERAPDF1.5 analysis provides four additional PDF sets arising from the choice of the theoretical parameters and fit functions, which are treated here as separate predictions.
These are referred to as \emph{variations} in the following, and include:
\begin{enumerate}
\item Varying the starting scale $Q^2_0$ from $1.9\GeV^2$ to $1.5\GeV^2$, while adding two additional parameters to the gluon fit function.
\item Varying the starting scale $Q^2_0$ from $1.9\GeV^2$ to $2.5\GeV^2$.
\item Including an extra parameter in the valence $\textit{u}$-quark fit function.
\item Including an extra parameter in the $\bar{\textit{u}}$-quark fit function.
\end{enumerate}
Including the maximal deviations due to these four variations with respect to the original doubles the magnitude of the theoretical uncertainties at high dijet mass, as seen in section \ref{sec:results}.
When performing the quantitative comparison described in section \ref{sec:stattest}, the four variations are treated as distinct PDF sets, since they are obtained using different parameterizations and their statistical interpretation in terms of uncertainties is not well defined.

The uncertainties on the theoretical predictions due to those on the PDFs range from $2\%$ at low dijet mass up to $20\%$ at high dijet mass for the range of smallest \ystar values.
For the largest values of \ystar, the uncertainties reach $100$--$200\%$ at high dijet mass, depending on the PDF set.

The uncertainties on the non-perturbative corrections, arising from the modelling of the fragmentation process and the underlying event, are estimated as the maximal deviations of the corrections from the nominal (see section \ref{subsec:theorycalc}) using the following three configurations:
\pythia with the AUET2B tune and CTEQ6L1 PDF set, and \herwigpp 2.5.2 with the UE-EE-3 tune using the MRST LO$^{**}$ or CTEQ6L1 PDF sets.
In addition, the statistical uncertainty due to the limited size of the sample generated using the nominal tune is included.
The uncertainty increases from $1\%$ for the range \ystar$<0.5$, up to $5\%$ for larger values of \ystar.
The statistical uncertainty significantly contributes to the total uncertainty only for the range \ystarsix.

There are several cases where the PDF sets provide uncertainties at the $90\%$ confidence level (CL);
in particular, the PDF uncertainty components for the CT10 PDF set, as well as $\alphas$ uncertainties for the CT10, HERAPDF1.5, and NNPDF2.1/2.3 PDF sets.
In these cases, the magnitudes of the uncertainties are scaled to the $68\%$ CL to match the other uncertainties.
In general, the scale uncertainties are dominant at low dijet mass, and the PDF uncertainties are dominant at high dijet mass.

\section{Trigger, jet reconstruction and data selection}
\label{sec:datasel}

This measurement uses the full data set of $pp$ collisions collected with the ATLAS detector at \onlyseventev centre-of-mass energy in 2011, corresponding to an integrated luminosity of \lumifbshort \cite{Aad:2013ucp}.
Only events collected during stable beam conditions and passing all data-quality requirements are considered in this analysis.
At least one primary vertex, reconstructed using two or more tracks, each with $\pt > 400\MeV$, must be present to reject cosmic ray events and beam background.
The primary vertex with the highest $\sum p_{\mathrm{T}}^{2}$ of associated tracks is selected as the hard-scatter vertex.

Due to the high luminosity, a suite of single-jet triggers was used to collect data, where only the trigger with the highest \pt threshold remained unprescaled.
An event must satisfy all three levels of the jet trigger system based on the transverse energy (\et) of jet-like objects.
Level-1 provides a fast, hardware decision using the combined \et of low-granularity calorimeter towers.
Level-2 performs a simple jet reconstruction in a window around the geometric region passed at Level-1, with a threshold generally 20 \GeV higher than at Level-1.
Finally, a jet reconstruction using the \AKT algorithm with \rfour is performed over the entire detector solid angle by the Event Filter (EF).
The EF requires transverse energy thresholds typically 5 \GeV higher than those used at Level-2.

\begin{figure}[htp!]
\begin{center}
  \includegraphics[width=0.75\textwidth]{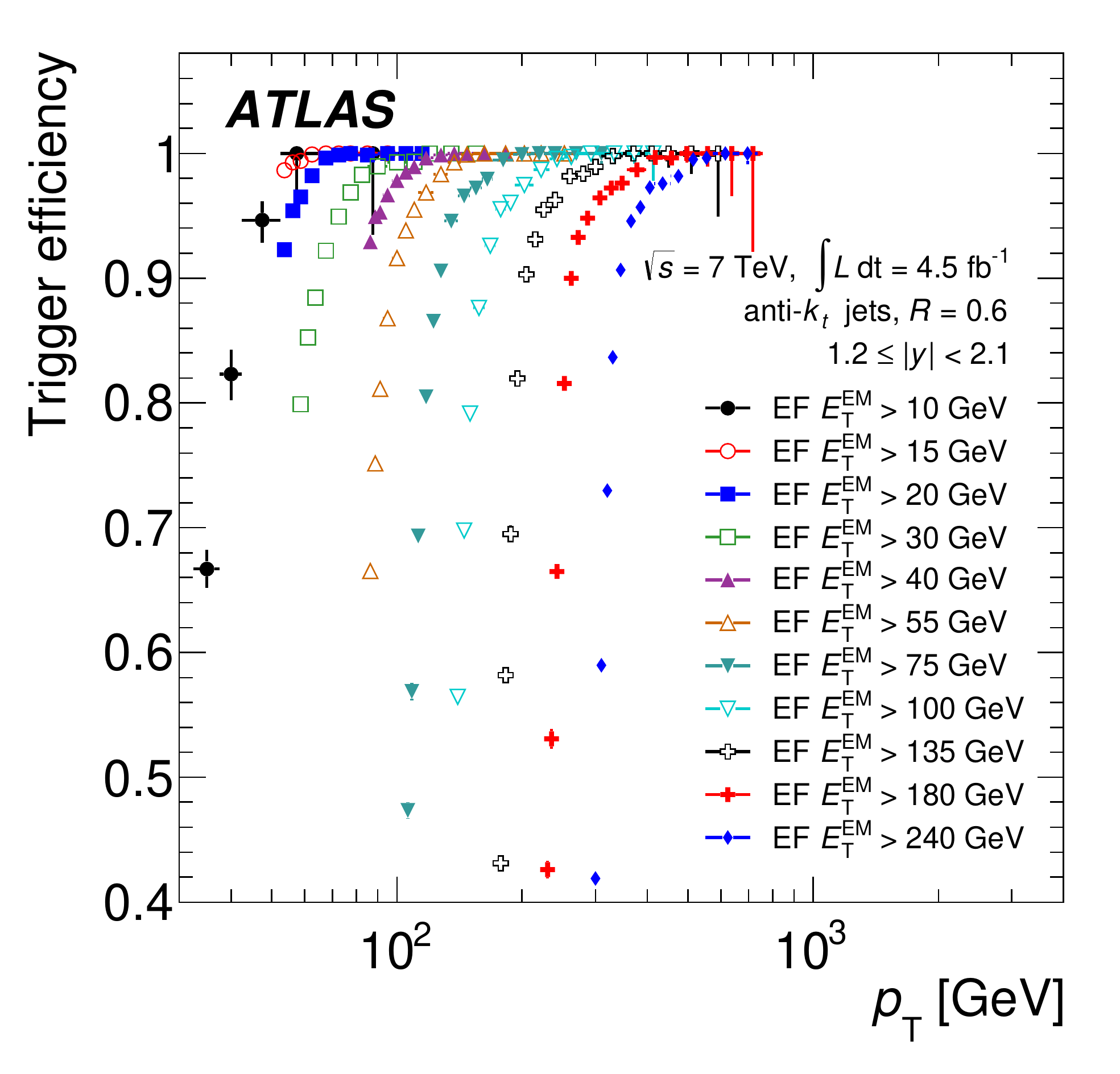}
  \caption{
    Jet trigger efficiency as a function of calibrated jet \pt for jets in the interval $1.2 \leq \absy < 2.1$ with radius parameter \rsix, shown for various trigger thresholds.
    The energy of jets in the trigger system is measured at the electromagnetic energy scale (EM scale), which correctly measures the energy deposited by electromagnetic showers.
  }
  \label{fig:TrigEff}
\end{center}
\end{figure}

The efficiencies of the central jet triggers ($\abseta < 3.2$) relevant to this analysis are shown in figure \ref{fig:TrigEff}.
They are determined using an unbiased sample known to be fully efficient in the jet-\pt range of interest.
The unbiased sample can be either a random trigger, or a single-jet trigger that has already been shown to be fully efficient in the relevant jet-\pt range.
The efficiency is presented in a representative rapidity interval ($1.2 \leq \absy < 2.1$), as a function of calibrated jet \pt, for jets with radius parameter \rsix.
Triggers are used only where the probability that a jet fires the trigger is $> 99\%$.
Because events are triggered using \rfour jets, the \pt at which calibrated \rsix jets become fully efficient is significantly higher than that for calibrated \rfour jets.
To take advantage of the lower \pt at which \rfour jets are fully efficient, distinct \pt ranges are used to collect jets for each value of the radius parameter.

A dijet event is considered in this analysis if either the leading jet, the subleading jet, or both are found to satisfy one of the jet trigger requirements.
Because of the random nature of the prescale decision, some events are taken not because of the leading jet, but instead because of the subleading jet.
This two-jet trigger strategy results in an increase in the sample size of about $10\%$.
To properly account for the combined prescale of two overlapping triggers the ``inclusion method for fully efficient combinations'' described in ref. \cite{Lendermann:2009ah} is used.

After events are selected by the trigger system, they are fully reconstructed offline.
The input objects to the jet algorithm are three-dimensional ``topological'' clusters \cite{Lampl:2008zz} corrected to the electromagnetic energy scale.

Each cluster is constructed from a seed calorimeter cell with $|E_{\rm cell}| > 4\sigma$, where $\sigma$ is the RMS of the total noise of the cell from both electronic and pileup sources.
Neighbouring cells are iteratively added to the cluster if they have $|E_{\rm cell}| > 2\sigma$.
Finally, an outer layer of all surrounding cells is added.
A calibration that accounts for dead material, out-of-cluster losses for pions, and calorimeter response, is applied to clusters identified as hadronic by their topology and energy density \cite{Barillari:1112035}.
This additional calibration serves to improve the energy resolution from the jet-constituent level through the clustering and calibration steps.
Taken as input to the \AKT jet reconstruction algorithm, each cluster is considered as a massless particle with an energy $E=\sum E_{\rm cell}$, and a direction given by the energy-weighted barycentre of the cells in the cluster with respect to the geometrical centre of the ATLAS detector.

The four-momentum of an uncalibrated jet is defined as the sum of four-momenta of the clusters making up the jet.
The jet is then calibrated in four steps:
\begin{enumerate}
\item 
  Additional energy due to pileup is subtracted using a correction
  derived from MC simulation and validated in situ as a function of \avgmu,
  the number of primary vertices (\npv) in the bunch crossing, and jet $\eta$ \cite{ATLAS:2012lla}.
\item 
  The direction of the jet is corrected such that the jet 
  originates from the selected hard-scatter vertex of the event
  instead of the geometrical centre of ATLAS.
\item 
  Using MC simulation, the energy and the position of the jet are corrected for
  instrumental effects (calorimeter non-compensation, additional dead
  material, and effects due to the magnetic field) and the jet energy scale is restored
  on average to that of the particle-level jet.
  For the calibration, the particle-level jet does not include muons and
  non-interacting particles.
\item 
  An additional in situ calibration is applied to correct for 
  residual differences between MC simulation and data, derived by combining the
  results of $\gamma$--jet, $Z$--jet, and multijet momentum balance techniques.
\end{enumerate}
The full calibration procedure is described in detail in ref. \cite{ATLAS-CONF-2013-004}.

The jet acceptance is restricted to $\absy < 3.0$ so that the trigger efficiency remains $> 99\%$.
Furthermore, the leading jet is required to have $\pt > 100 \GeV$ and the subleading jet is required to have $\pt > 50 \GeV$ to be consistent with the asymmetric cuts imposed on the theoretical predictions.

Part of the data-taking period was affected by a read-out problem in a region of the LAr calorimeter, causing jets in this region to be poorly reconstructed.
Since the same unfolding procedure is used for the entire data sample, events are rejected if either the leading or the subleading jet falls in the region $-0.88 < \phi < -0.5$, for all $\abseta$, in order to avoid a bias in the spectra.
This requirement results in a loss in acceptance of approximately 10\%.
The inefficiency is accounted for and corrected in the data unfolding procedure.

The leading and subleading jets must also fulfil the ``medium'' quality criteria as described in ref. \cite{ATLAS-CONF-2012-020}, designed to reject cosmic rays, beam halo, and detector noise.
If either jet fails these criteria, the event is not considered.
More than four (two) million dijet events are selected with these criteria using jets with radius parameter \rfour (\rsix), with the difference in sample size resulting mostly from the trigger requirements.

\section{Stability of the results under pileup conditions}
\label{sec:pileup}

\begin{figure}[ht!]
\begin{center}
  \subfigure[\ystarone]{\includegraphics[width=0.45\textwidth]{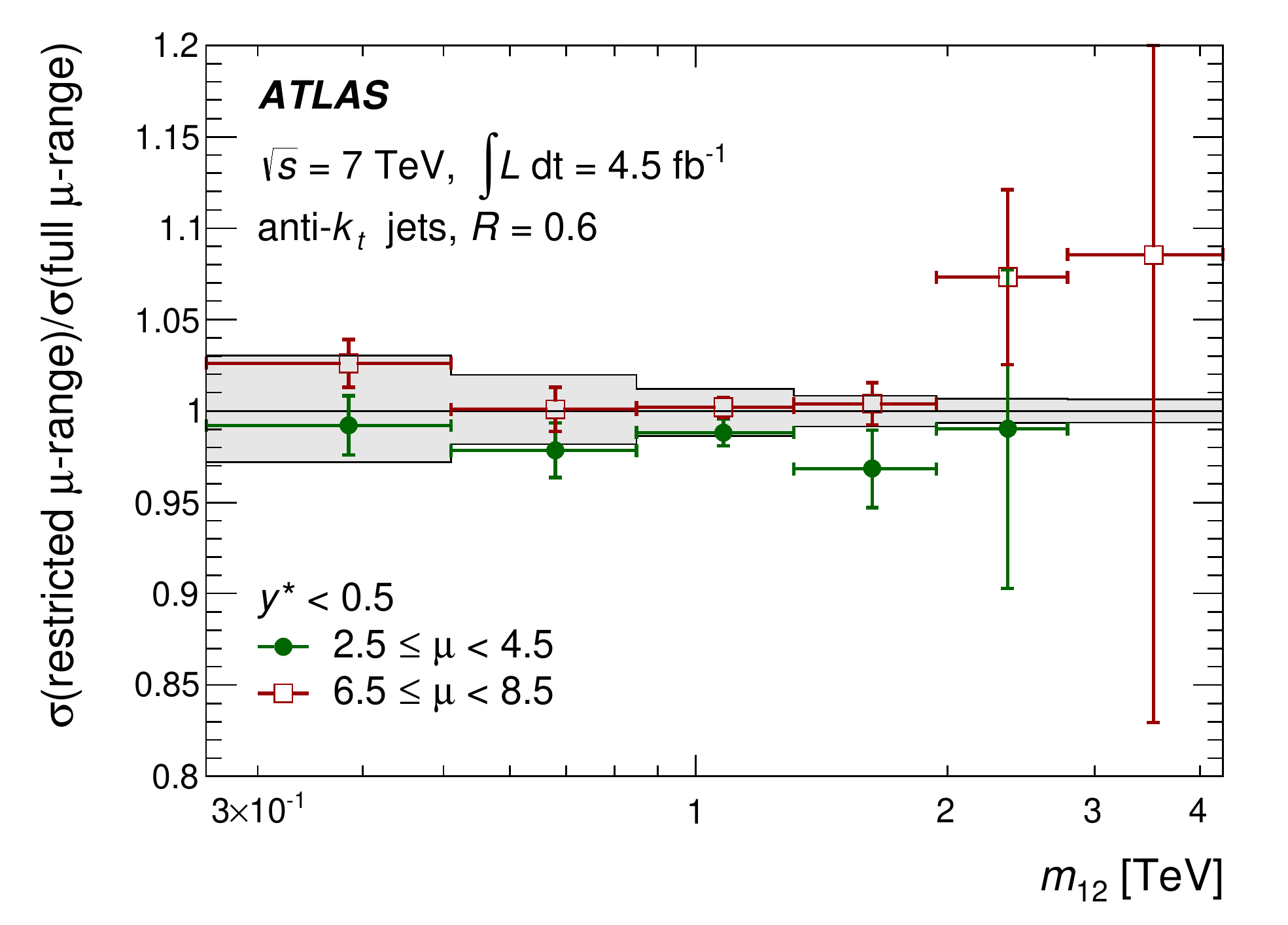}}
  \subfigure[\ystarthree]{\includegraphics[width=0.45\textwidth]{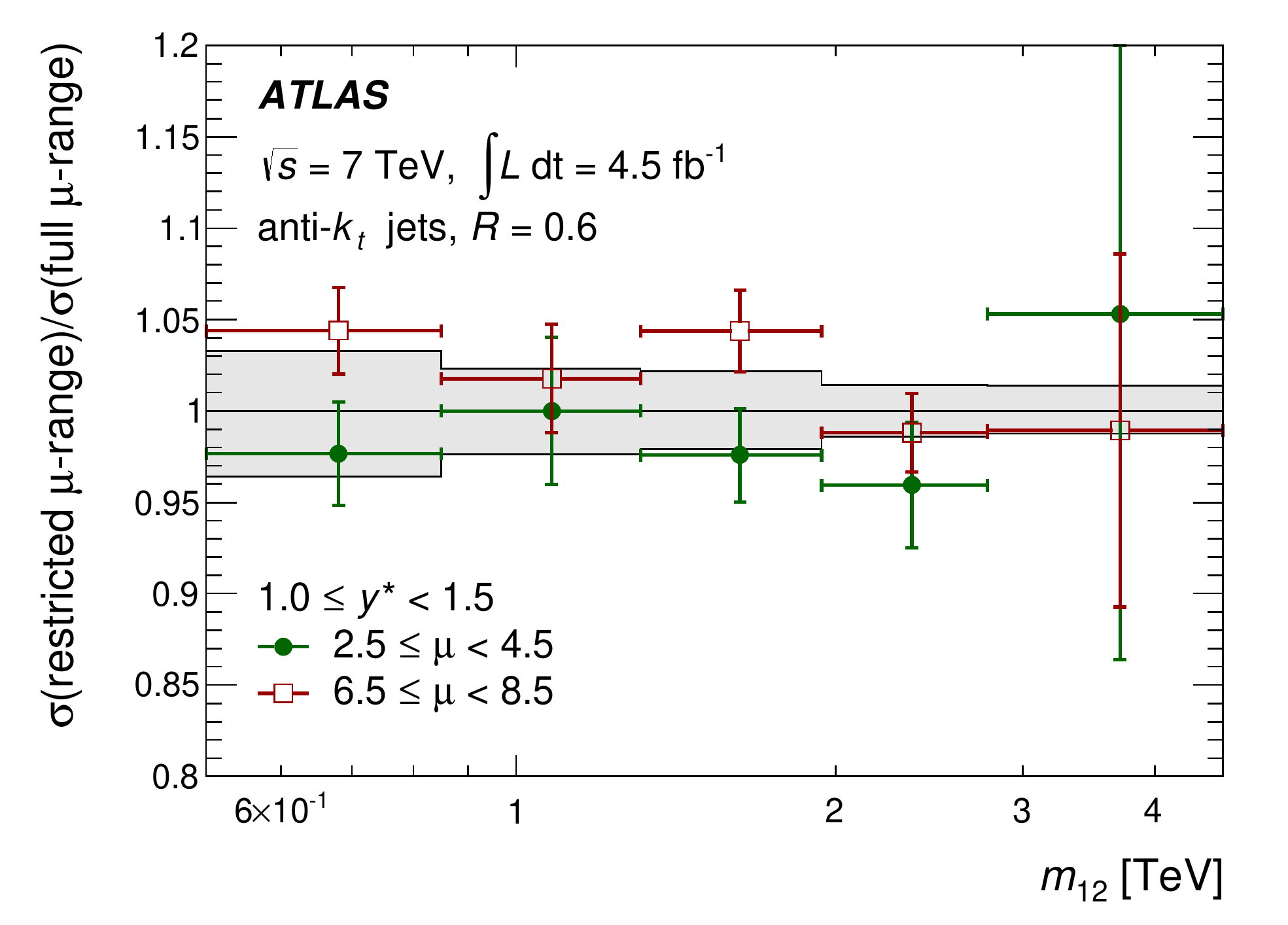}}
  \caption{
    Luminosity-normalized dijet yields as a function of dijet mass for two ranges of \actmu, in the range (a) \ystarone and (b) \ystarthree.
    The measurements are shown as ratios with respect to the full luminosity-normalized dijet yields.
    The gray bands represent the uncertainty on the jet energy calibration that accounts for the \avgmu and \npv dependence, propagated to the luminosity-normalized dijet yields.
    The statistical uncertainty shown by the error bars is propagated assuming no correlations between the samples.
    This approximation has a small impact, and does not reduce the agreement observed within the pileup uncertainties.
  }
  \label{fig:pileup}
\end{center}
\end{figure}

The dijet mass is sensitive to the effects of pileup through the energies, and to a lesser extent the directions, of the leading and subleading jets.
As such, it is important to check the stability of the measurement with respect to various pileup conditions.
The effects of pileup are removed at the per-jet level during the jet energy calibration (see section \ref{sec:datasel}).
To check for any remaining effects, the integrated luminosity delivered for several ranges of \actmu is determined separately for each trigger used in this analysis.
This information is then used to compute the luminosity-normalized dijet yields in different ranges of \actmu.
A comparison of the yields for two ranges of \actmu is shown in figure \ref{fig:pileup}.
While statistical fluctuations are present, the residual bias is covered by the uncertainties on the jet energy calibration due to \avgmu and \npv, derived through the in situ validation studies (see section \ref{sec:sysunc}) \cite{ATLAS-CONF-2013-004,ATLAS:2012lla}.

The dependence of the luminosity-normalized dijet yields on the position in the accelerator bunch train is also studied.
Because the first bunches in a train do not fully benefit from the compensation of previous bunches due to the long LAr calorimeter pulses, a bias is observed in the jet energy calibration.
This bias can be studied by defining a control region using only events collected from the middle of the bunch train, where full closure in the jet energy calibration is obtained.
By comparing the luminosity-normalized dijet yields using the full sample to that from the subsample of events collected from the middle of the bunch train, any remaining effects on the measurement due to pileup are estimated.
An increase in the luminosity-normalized dijet yields using the full sample compared to the subsample from the middle of the bunch train is observed, up to 5\% at low dijet mass.
This increase is well described in the MC simulation, so that the effect is corrected for during the unfolding step.
All remaining differences are covered by the jet energy calibration uncertainty components arising from pileup.

The stability of the luminosity-normalized dijet yields in the lowest dijet-mass bins is studied as a function of the date on which the data were collected.
Since portions of the hadronic calorimeter became non-functional during the course of running\footnote{
  Up to 5\% of the modules in the barrel and extended barrels of the hadronic calorimeter were turned off by the end of data taking.
} and pileup increased throughout the year, this provides an important check of the stability of the result.
The observed variations are consistent with those due to \actmu, which on average increased until the end of data taking. 
Pileup is found to be the dominant source of variations, whereas detector effects are small in comparison.

\section{Data unfolding}
\label{sec:unfolding}

The cross-sections as a function of dijet mass are obtained by unfolding the data distributions, correcting for detector resolutions and inefficiencies, as well as for the presence of muons and neutrinos in the particle-level jets (see section \ref{sec:xsdef}).
The same procedure as in ref. \cite{Aad:2011fc} is followed, using the iterative, dynamically stabilized (IDS) unfolding method \cite{Malaescu:2011yg}, a modified Bayesian technique.
To account for bin-to-bin migrations, a transfer matrix is built from MC simulations, relating the particle-level and reconstruction-level dijet mass, and reflecting all the effects mentioned above.
The matching is done in the \mass--\ystar plane, such that only a requirement on the presence of a dijet system is made.
Since migrations between neighbouring bins predominantly occur due to jet energy resolution smearing the dijet mass, and less frequently due to jet angular resolution, the unfolding is performed separately for each range of \ystar.

Data are unfolded to the particle level using a three-step procedure, consisting of correcting for matching inefficiency at the reconstruction level, unfolding for detector effects, and correcting for matching inefficiency at the particle level.
The final result is given by the equation:
\ifdraft \begin{linenomath} \fi
\begin{equation}
N_i^\mathrm{part} = \Sigma_j N_j^\mathrm{reco} \cdot \epsilon_j^\mathrm{reco}A_{ij}/\epsilon^\mathrm{part}_i
\end{equation}
\ifdraft \end{linenomath} \fi
where $i$ ($j$) is the particle-level (reconstruction-level) bin index, and $N_k^\mathrm{part}$ ($N_k^\mathrm{reco}$) is the number of particle-level (reconstruction-level) events in bin $k$.
The quantities $\epsilon_k^\mathrm{reco}$ ($\epsilon_k^\mathrm{part}$) represent the fraction of reconstruction-level (particle-level) events matched to particle-level (reconstruction-level) events for each bin $k$.
The element of the unfolding matrix, $A_{ij}$, provides the probability for a reconstruction-level event in bin $j$ to be associated with a particle-level event in bin $i$.
The transfer matrix is improved through a series of iterations, where the particle-level MC distribution is reweighted to the shape of the unfolded data spectrum.
The number of iterations is chosen such that the bias in the closure test (see below) is at the sub-percent level.
This is achieved after one iteration for this measurement.

The statistical uncertainties following the unfolding procedure are estimated using pseudo-experiments.
Each event in the data is fluctuated using a Poisson distribution with a mean of one before applying any additional event weights.
For the combination of this measurement with future results, the pseudo-random Poisson distribution is seeded uniquely for each event, so that the pseudo-experiments are fully reproducible.
Each resulting pseudo-experiment of the data spectrum is then unfolded using a transfer matrix and efficiency corrections obtained by fluctuating each event in the MC simulation according to a Poisson distribution.
Finally, the unfolded pseudo-experiments are used to calculate the covariance matrix.
In this way, the statistical uncertainty and bin-to-bin correlations for both the data and the MC simulation are encoded in the covariance matrix.
For neighbour and next-to-neighbour bins the statistical correlations are generally $1$--$15\%$, although the systematic uncertainties are dominant.
The level of statistical correlation decreases quickly for bins with larger dijet-mass separations.

A data-driven closure test is used to derive the bias of the spectrum shape due to mis-modelling by the MC simulation.
The particle-level MC simulation is reweighted directly in the transfer matrix by multiplying each column of the matrix by a given weight.
These weights are chosen to improve the agreement between data and reconstruction-level MC simulation.
The modified reconstruction-level MC simulation is unfolded using the original transfer matrix, and the result is compared with the modified particle-level spectrum.
The resulting bias is considered as a systematic uncertainty.\footnote{
   For this measurement the IDS method results in the smallest bias when compared with the bin-by-bin technique used in ref. \cite{Aad:2010ad} or the SVD method \cite{Hocker:1995kb}.
}

\section{Experimental uncertainties}
\label{sec:sysunc}

The uncertainty on the jet energy calibration is the dominant uncertainty for this measurement.
Complete details of its derivation can be found in ref.~\cite{ATLAS-CONF-2013-004}.
Uncertainties in the central region are determined from in situ calibration techniques, such as the transverse momentum balance in $Z/\gamma$--jet and multijet events, for which a comparison between data and MC simulation is performed.
The uncertainty in the central region is propagated to the forward region using transverse momentum balance between a central and a forward jet in dijet events.
The difference in the balance observed between MC simulation samples generated with \pythia and \herwig results in an additional large uncertainty in the forward region.
The uncertainty due to jet energy calibration on each individual jet is $1$--$4\%$ in the central region, and up to 5\% in the forward region.
The improvement of the in situ jet calibration techniques over the single-particle response used for data taken in 2010~\cite{Aad:2011he} leads to a reduction in the magnitude of the uncertainty compared to that achieved in the 2010 measurement, despite the increased level of pileup.
As a result of the different techniques employed in the 2010 and current analyses, the correlations between the two measurements are non-trivial.

The uncertainty due to the jet energy calibration is propagated to the measured cross-sections using MC simulation.
Each jet in the sample is scaled up or down by one standard deviation of a given uncertainty component, after which the luminosity-normalized dijet yield is measured from the resulting sample.
The yields from the original sample and the samples where all jets were scaled are unfolded, and the difference is taken as the uncertainty due to that component.
Since the sources of jet energy calibration uncertainty are taken as uncorrelated with each other, the corresponding uncertainty components on the cross-section are also taken as uncorrelated.
Because the correlations between the various experimental uncertainty components are not perfectly known, two additional jet energy calibration uncertainty configurations are considered.
They have \emph{stronger} and \emph{weaker} correlations with respect to the nominal configuration, depending on the number of uncertainty sources considered as fully correlated or independent of one another~\cite{ATLAS-CONF-2013-004}.

Jet energy and angular resolutions are estimated using MC simulation, after using an angular matching of particle-level and reconstruction-level jets.
The resolution is obtained from a Gaussian fit to the distribution of the ratio (difference) of reconstruction-level and particle-level jet energy (angle).
Jet energy resolutions are cross-checked in data using in situ techniques such as the bisector method in dijet events \cite{Aad:2012ag}, where good agreement is observed with MC simulation.
The uncertainty on the jet energy resolution comes from varying the selection parameters for jets, such as the amount of nearby jet activity, and depends on both jet \pt and jet $\eta$.
The jet angular bias is found to be negligible, while the resolution varies between 0.005 radians and 0.07 radians.
An uncertainty of $10\%$ on the jet angular resolution is shown to cover the observed differences in a comparison between data and MC simulation.

The resolution uncertainties are propagated to the measured cross-sections through the transfer matrix.
All jets in the MC sample are smeared according to the uncertainty on the resolution, either the jet energy or jet angular variable.
To reduce the dependence on the MC sample size, this process is repeated for each event 1000 times.
The transfer matrix resulting from this smeared sample is used to unfold the luminosity-normalized dijet yields, and the deviation from the measured cross-sections unfolded using the original transfer matrix is taken as a systematic uncertainty.

The uncertainty due to the jet reconstruction inefficiency as a function of jet \pt is estimated by comparing the efficiency for reconstructing a calorimeter jet, given the presence of an independently measured track-jet of the same radius, in data and in MC simulation.
Here, a track-jet refers to a jet reconstructed using the \AKT algorithm, considering as input all tracks in the event with $\pt > 500 \MeV$ and $\abseta < 2.5$, and which are assumed to have the mass of a pion.
Since this method relies on tracking, its application is restricted to the acceptance of the tracker for jets of $\abseta < 1.9$.
For jets with $\pt > 50 \GeV$, relevant for this analysis, the reconstruction efficiency in both the data and the MC simulation is found to be $100\%$ for this rapidity region, leading to no additional uncertainty.
The same efficiency is assumed for the forward region, where jets have more energy for a given value of \pt;
therefore, their reconstruction efficiency is likely to be as good as or better than that of jets in the central region.

Comparing the jet quality selection efficiency for jets passing the ``medium'' quality criteria in data and MC simulation, an agreement of the efficiency within $0.25\%$ is found~\cite{ATLAS-CONF-2012-020}.
Because two jets are considered for each dijet system, a $0.5\%$ systematic uncertainty on the cross-sections is assigned.

The impact of a possible mis-modelling of the spectrum shape in MC simulation, introduced through the unfolding as described in section \ref{sec:unfolding}, is also included.
The luminosity uncertainty is $1.8\%$ \cite{Aad:2013ucp} and is fully correlated between all data points.

The bootstrap method \cite{Bohm:2010tb} has been used to evaluate the statistical significance of all systematic uncertainties described above.
The individual uncertainties are treated as fully correlated in dijet mass and \ystar, but uncorrelated with each other, for the quantitative comparison described in section \ref{sec:stattest}.
The total uncertainty ranges from $10\%$ at low dijet mass up to $25\%$ at high dijet mass for the range $\ystar < 0.5$, and increases for larger $\ystar$.

\section{Frequentist method for a quantitative comparison of data and theory spectra}
\label{sec:stattest}

The comparison of data and theoretical predictions at the particle level rather than at the reconstruction level has the advantage that data can be used to test any theoretical model without the need for further detector simulation.
Because the additional uncertainties introduced by the unfolding procedure are at the sub-percent level, they do not have a significant effect on the power of the comparison.
The frequentist method described here provides quantitative statements about the ability of SM predictions to describe the measured cross-sections.
An extension, based on the \emph{CLs} technique \cite{Read:2002a}, is used to explore potential deviations in dijet production due to contributions beyond the SM (see section \ref{sec:setlimits}).

\subsection{Test statistic}

The test statistic, which contains information about the degree of deviation of one spectrum from another, is the key input for any quantitative comparison.
The use of a simple $\chi^2$ test statistic such as
\ifdraft \begin{linenomath} \fi
\begin{equation}
  \chi^2\left( \mathbf{d}; \mathbf{t} \right) = \sum_{i}{ \left( \frac{d_i - t_i}{\sigma_i(t_i)} \right)^2 } ,
\label{Eq:chi2diag}
\end{equation}
\ifdraft \end{linenomath} \fi
comparing data ($\mathbf{d}$) and theoretical predictions ($\mathbf{t}$), accounts only for the uncertainties on individual bins ($\sigma_i$).
This ignores the statistical and, even stronger, systematic correlations between bins.
Therefore, it has a reduced sensitivity to the differences between theoretical predictions and the measurements compared to other more robust test statistics.

The use of a covariance matrix ($C$) in the $\chi^2$ definition,
\ifdraft \begin{linenomath} \fi
\begin{equation}
  \chi^2\left( \mathbf{d}; \mathbf{t} \right) = \sum_{i,j} \left( d_i - t_i \right) \cdot \left[C^{-1}(\mathbf{t})\right]_{ij} \cdot \left( d_j - t_j \right),
  \label{Eq:chi2covMat}
\end{equation}
\ifdraft \end{linenomath} \fi
is a better approximation.
However, the covariance matrix is built from symmetrized uncertainties and thus cannot account for the asymmetries between positive and negative uncertainty components.

An alternative \chisq definition, based on fits of the uncertainty components, was proposed in refs.~\cite{D'Agostini:1993uj} and~\cite{Blobel:2003wa}.
For symmetric uncertainties it is equivalent to the definition in eq.~(\ref{Eq:chi2covMat}) \cite{Botje:2001fx}.
However, it allows a straightforward generalization of the \chisq statistic, accounting for asymmetric uncertainties by separating them from the symmetric ones in the test statistic, which is now:
\ifdraft \begin{linenomath} \fi
\begin{equation}
\begin{split}
  \chi^2\left( \mathbf{d}; \mathbf{t} \right) = \min_{\beta_a}
  & \left\{ \sum_{i,j} \left[ d_i - \left(1+\sum_{a}{\beta_a\cdot \left(\boldsymbol{\epsilon}_{a}^{\pm}(\beta_a)\right)_i} \right) t_i \right] \cdot \left[C_\mathrm{su}^{-1}(\mathbf{t})\right]_{ij} \right. \\
  & \left. \vphantom{\sum_{i,j}} \cdot \left[ d_j - \left(1+\sum_{a}{\beta_a\cdot \left(\boldsymbol{\epsilon}_a^{\pm}(\beta_a)\right)_j} \right) t_j \right] + \sum_{a}{\beta_a^2} \right\},
\end{split}
\label{Eq:chi2asymmSyst}
\end{equation}
\ifdraft \end{linenomath} \fi
where $C_\mathrm{su}$ is the covariance matrix built using only the \emph{symmetric} uncertainties, and $\beta_a$ are the profiled coefficients of the \emph{asymmetric} uncertainties which are varied in a fit that minimizes the \chisq.
Here $\boldsymbol{\epsilon}^\pm_a$ is the positive component of the $a^\mathrm{th}$ asymmetric relative uncertainty if the fitted value of $\beta_a$ is positive, or the negative component otherwise.
The magnitude of each $\boldsymbol{\epsilon}^\pm_a$ corresponds to the relative effect on the theory prediction of a one standard deviation shift of parameter $a$.
For this analysis, an uncertainty component is considered asymmetric when the absolute difference between the magnitudes of the positive and negative portions is larger than 1\% of the cross-section in at least one bin.
Due to the large asymmetric uncertainties on the theoretical predictions, this $\chi^2$ definition is not only a better motivated choice, but provides stronger statistical power than the two approaches mentioned above.

\begin{figure}[htp]
\begin{center}
  \subfigure[\label{fig:asymmetricSystematicsAndCorrelations_CT10_a} Asymmetric uncertainties.]{
    \includegraphics[width=0.45\linewidth]{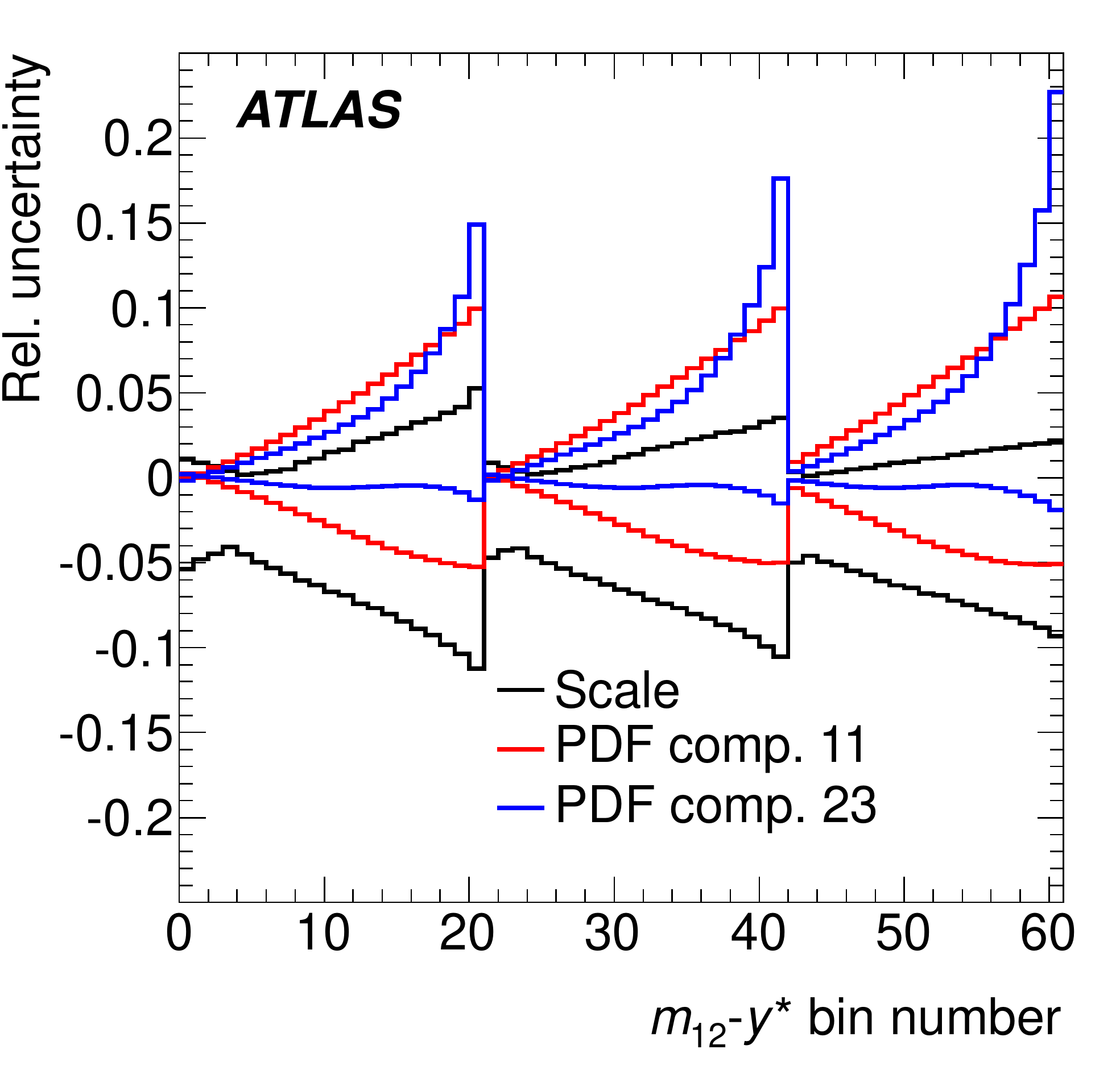}
  }\hspace{0.4cm}
  \subfigure[\label{fig:asymmetricSystematicsAndCorrelations_CT10_b} Asymmetry of the uncertainties.]{
    \includegraphics[width=0.45\linewidth]{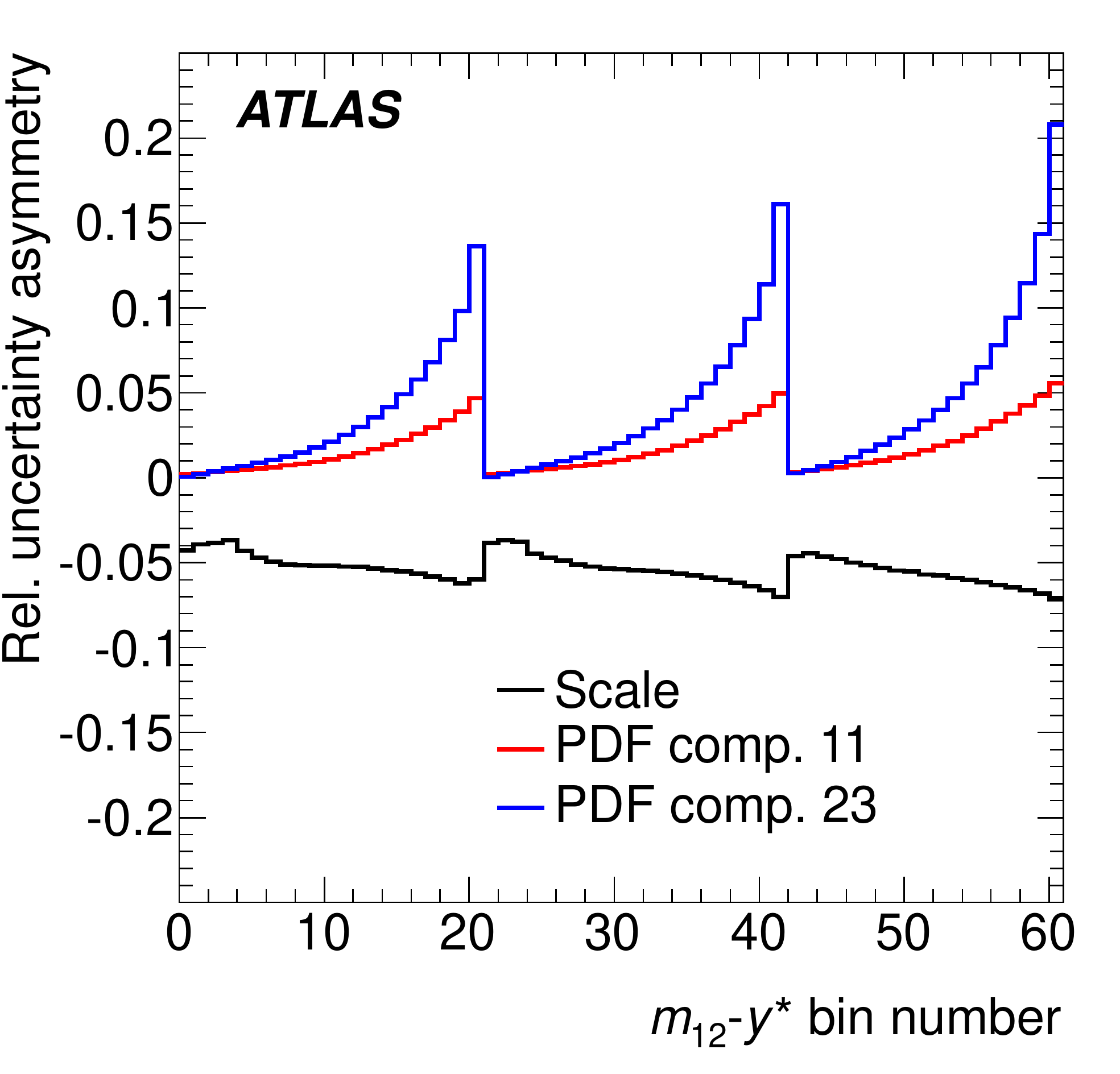}
  } \\
  \subfigure[\label{fig:asymmetricSystematicsAndCorrelations_CT10_c} Relative total uncertainty~(black) and relative total symmetric uncertainty~(blue).]{
    \includegraphics[width=0.45\linewidth]{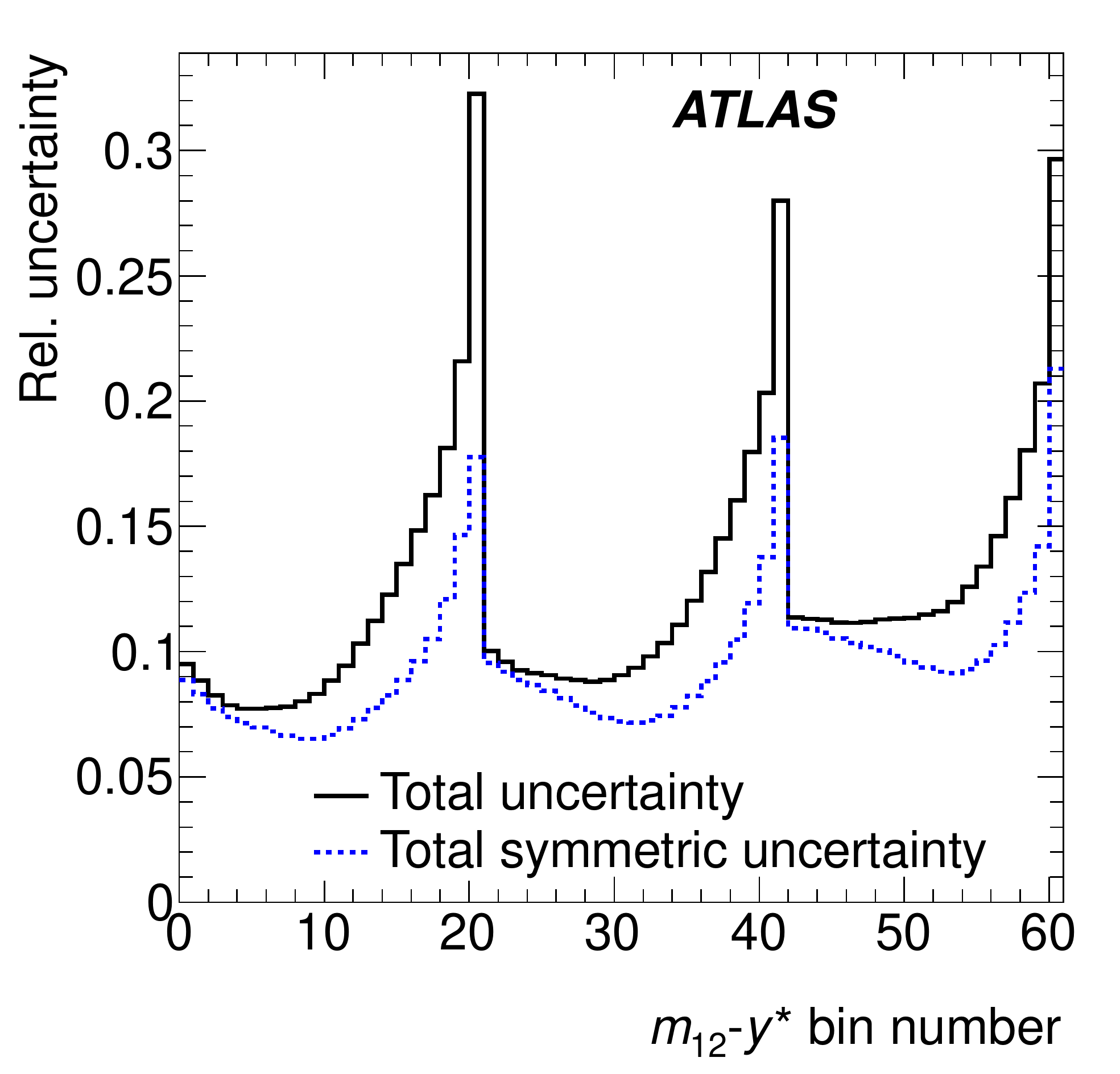}
  }\hspace{0.4cm}
  \subfigure[\label{fig:asymmetricSystematicsAndCorrelations_CT10_d} Correlation matrix for the symmetric uncertainties~(statistical and systematic).]{
    \includegraphics[width=0.45\linewidth]{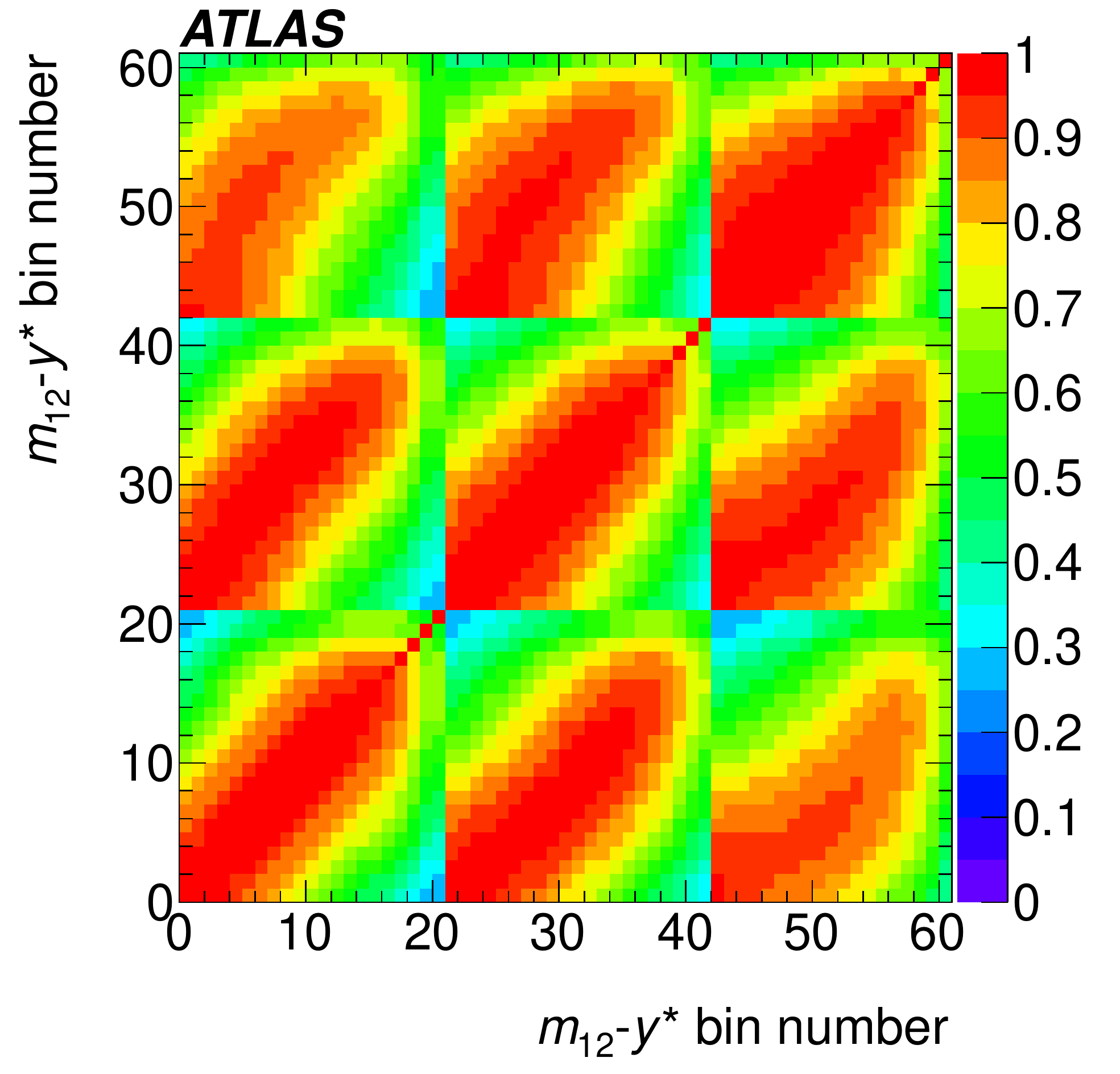}
  }
  \caption{
    Sample inputs to the asymmetric generalization of the $\chi^2$ function.
    The uncertainty components with the largest asymmetries are shown in (a), which includes the renormalization and factorization scale, and two uncertainty components (PDF comp.) of the CT10 PDF set.
    The asymmetries of the same uncertainty components are shown in (b), defined as the signed difference between the magnitudes of the positive and negative portions ($up-(-down)$).
    The relative total symmetric uncertainty (blue dashed line) and relative total uncertainty (black line) are shown in (c), along with the correlation matrix computed from the symmetric uncertainties in (d).
    These plots correspond to the theoretical prediction using the CT10 PDF set, and radius parameter \rsix. 
    For each plot, the horizontal axis covers all dijet-mass bins (ordered according to increasing dijet mass) for three ranges in \ystar{}: starting from the left end the range \ystarone, then \ystartwo~and finally \ystarthree.
    The \mass--\ystar bin numbers correspond to those in the cross-section tables in Appendix \ref{app:tables}.
  }
\label{fig:asymmetricSystematicsAndCorrelations_CT10}
\end{center}
\end{figure}

The theoretical and experimental uncertainties, including both statistical and systematic components, are included in the \chisq fit.
This relies on a detailed knowledge of the individual uncertainty components and their correlations with each other.
Sample inputs to the \chisq function using the SM predictions based on the CT10 PDF set are shown in figure~\ref{fig:asymmetricSystematicsAndCorrelations_CT10}.
To illustrate the asymmetries, examples of the most asymmetric uncertainty components are shown, namely the theoretical uncertainties due to the scale choice and two uncertainty components of the CT10 PDF set.
The positive and negative relative uncertainties are shown in figure \ref{fig:asymmetricSystematicsAndCorrelations_CT10_a} as functions of the bin number,
corresponding to those in the cross-section tables in Appendix \ref{app:tables}.
The signed difference between the magnitudes of the positive and negative relative uncertainties ($up-(-down)$) are shown in figure \ref{fig:asymmetricSystematicsAndCorrelations_CT10_b} as functions of the bin number.
The importance of the asymmetric uncertainties is highlighted in figure \ref{fig:asymmetricSystematicsAndCorrelations_CT10_c} through the comparison of the relative total symmetric uncertainty with the relative total uncertainty.
Uncertainty components exhibiting asymmetries of up to 22\% are present, and are an important fraction of the total uncertainty, in particular at high dijet mass.
The largest asymmetric uncertainties are individually comparable to the size of the total symmetric ones.
The correlation matrix computed for the symmetric uncertainties is shown in figure \ref{fig:asymmetricSystematicsAndCorrelations_CT10_d}.
Strong correlations of $90\%$ or more are observed among neighbouring bins of dijet mass, mostly due to the jet energy scale and resolution, while correlations are smaller between different ranges of \ystar.
The total symmetric uncertainty and its correlation matrix define the symmetric covariance matrix in eq.~(\ref{Eq:chi2asymmSyst}).

\subsection{Frequentist method}

The \chisq distribution expected for experiments drawn from the parent distribution of a given theory hypothesis is required in order to calculate the probability of measuring a specific \chisq value under that theory hypothesis.
This is obtained by generating a large set of pseudo-experiments that represent fluctuations of the theory hypothesis due to the full set of experimental and theoretical uncertainties.
The theory hypothesis can be the SM, or the SM with any of its extensions, depending on the study being carried out.
In the generation of pseudo-experiments, the following sources of uncertainty are considered:
\begin{itemize}
\item
  Statistical uncertainties:
  an eigenvector decomposition of the statistical covariance matrix resulting from the unfolding procedure is performed.
  The resulting eigenvectors are taken as Gaussian-distributed uncertainty components.
\item
  Systematic experimental and theoretical uncertainties:
  the symmetric components are taken as Gaussian distributed, while a two-sided Gaussian distribution is used for the asymmetric ones.
\end{itemize}
For each pseudo-experiment, the \chisq value is computed between the pseudo-data and the theory hypothesis.
In this way, the \chisq distribution that would be expected for experiments drawn from the theory hypothesis is obtained without making assumptions about its shape.

The \emph{observed \chisq} (\chiobs) value is computed using the data and the theory hypothesis.
To quantify the compatibility of the data with the theory, the ratio of the area of the \chisq distribution with $\chisq > \chiobs$ to the total area is used.
This fractional area, called a p-value, is the observed probability, under the assumption of the theory hypothesis, to find a value of \chisq with equal or lesser compatibility with the hypothesis relative to what is found with \chiobs.
If the observed p-value (\pobs) is smaller than 5\%, the theoretical prediction is considered to poorly describe the data at the 95\% CL.

The comparison of a theory hypothesis that contains an extension of the SM to the data is quantified using the \emph{CLs} technique \cite{Read:2002a}, which accounts for cases where the signal is small compared to the background.
This technique relies on the computation of two \chisq distributions:
one corresponding to the \emph{background} model (the SM) and another to the \emph{signal$+$background} model (the SM plus any extension), each with respect to the assumption of the signal$+$background model.
These distributions are calculated using the same technique described earlier, namely through the generation of large sets of pseudo-experiments:
\begin{itemize}
\item
  In the case of the background-only distribution, the \chisq values are calculated between each background pseudo-experiment and the signal$+$background prediction.
  The \emph{expected \chisq} (\chiexp) is defined as the median of this \chisq distribution.
\item
  In the case of the signal$+$background distribution, the \chisq values are calculated between each signal$+$background pseudo-experiment and the signal$+$background prediction.
\end{itemize}
These two \chisq distributions are subsequently used to calculate two p-values:
(1) the observed signal$+$background p-value ($p_{s+b}$), which is defined in the same way as the one described above, i.e. the fractional area of the signal$+$background \chisq distribution above \chiobs;
and (2) the observed background p-value ($p_b$), which is defined as the fractional area of the background-only \chisq distribution below \chiobs and measures the compatibility of the background model with the data.
In the \emph{CLs} technique, these two p-values are used to construct the quantity $\mathit{CLs} = p_{\mathrm{s}+\mathrm{b}} / (1 - p_\mathrm{b})$, from which the decision to exclude a given signal$+$background prediction is made.
The theory hypothesis is excluded at 95\% CL if the quantity \emph{CLs} is less than 0.05.
This technique has the advantage, compared to the use of $p_{s+b}$ alone, that theoretical hypotheses to which the data have little or no sensitivity are not excluded.
For comparison, the expected exclusion is calculated using the same procedure, except using the expected \chisq value instead of the observed \chisq value.

\section{Cross-section results}
\label{sec:results}

\begin{figure}[!htp]
   \begin{center}
   \subfigure[\rfour]{
      \includegraphics[width=0.6\textwidth]{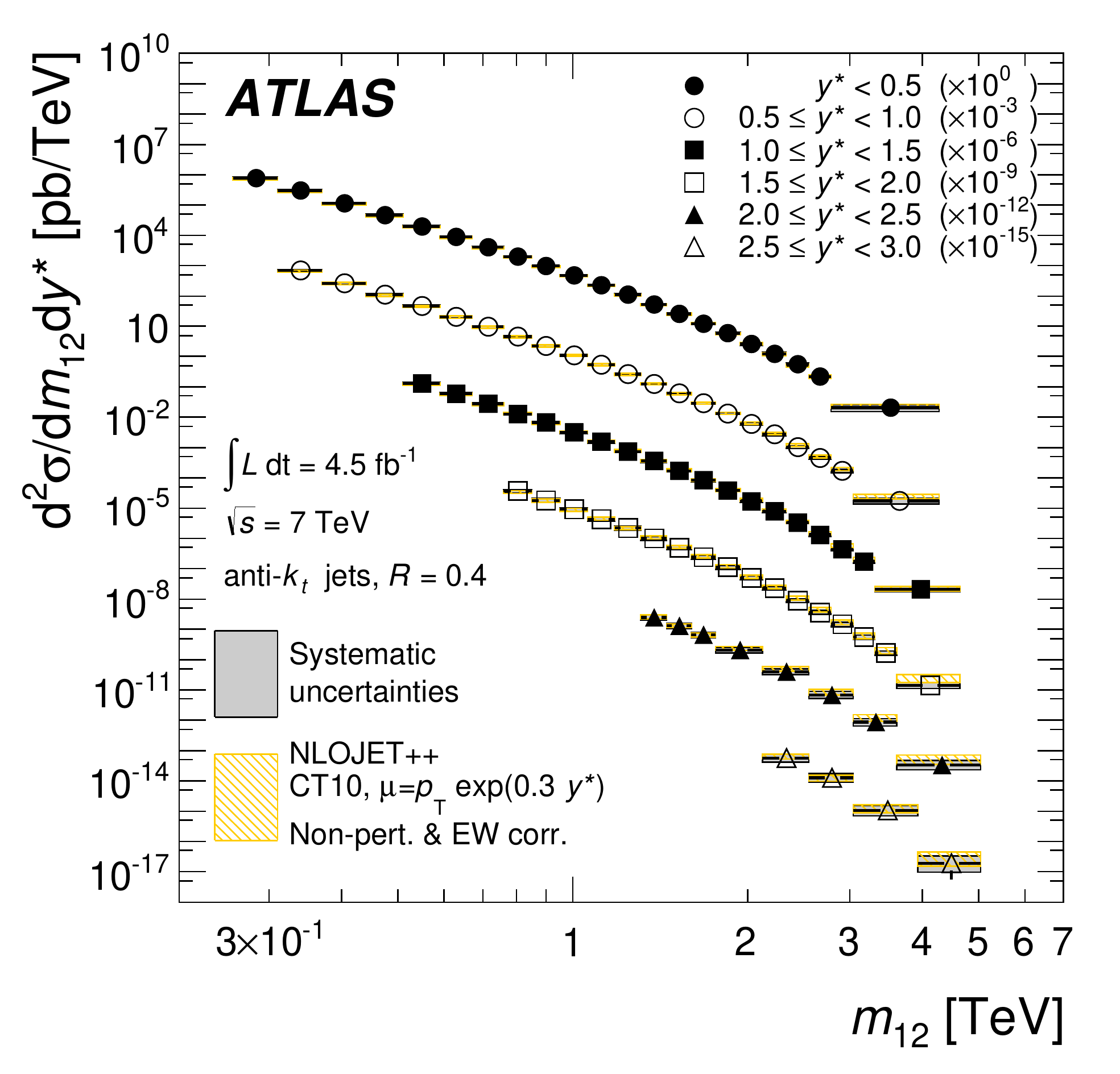}
   }
   \subfigure[\rsix]{
      \includegraphics[width=0.6\textwidth]{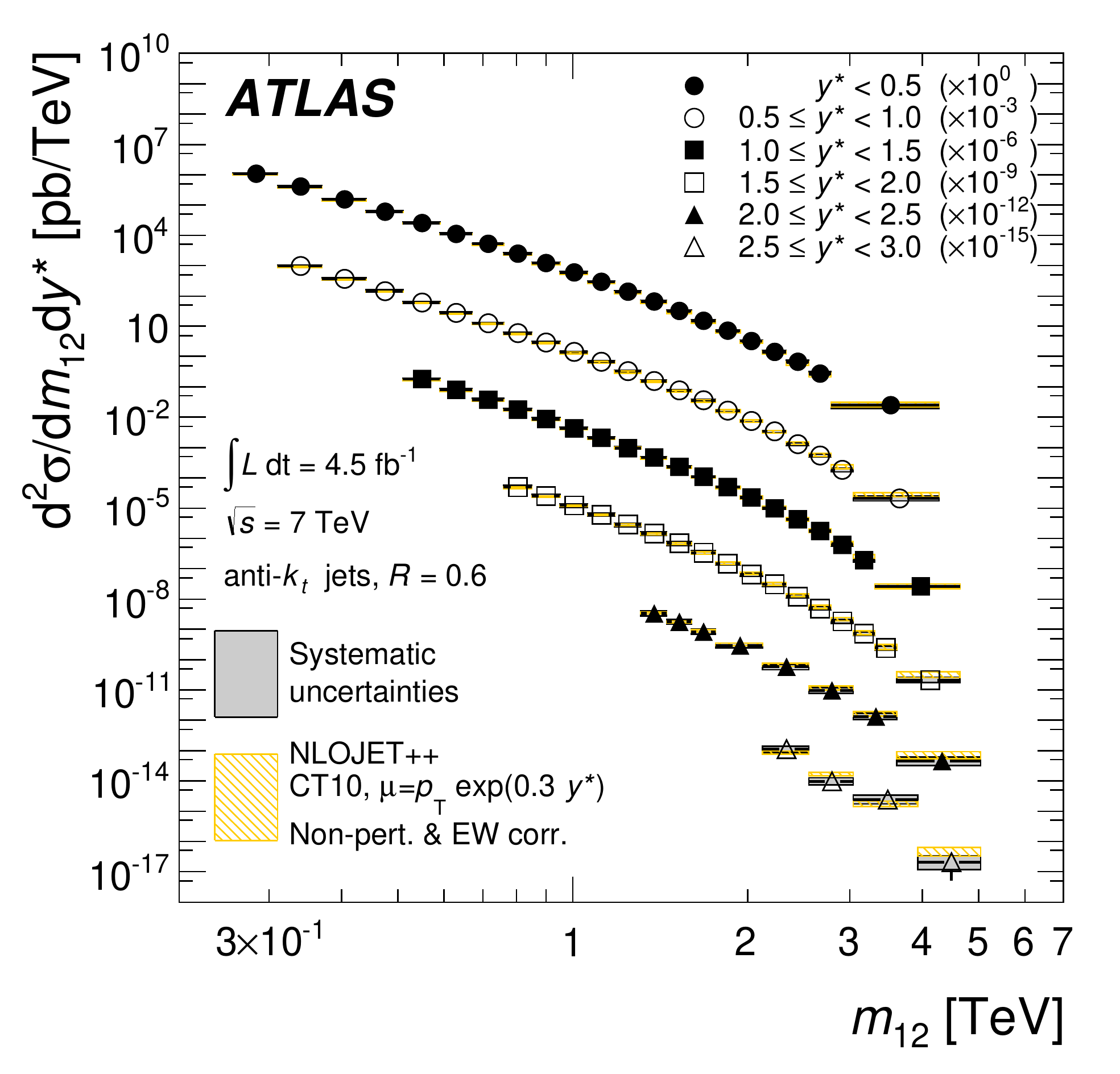}
   }
   \caption{
      Dijet double-differential cross-sections for \AKT jets with radius parameter \rfour and \rsix, shown as a function of dijet mass in different ranges of \ystar.
      To aid visibility, the cross-sections are multiplied by the factors indicated in the legend.
      The error bars indicate the statistical uncertainty on the measurement, and the dark shaded band indicates the sum in quadrature of the experimental systematic uncertainties.
      For comparison, the NLO QCD predictions of NLOJet++ using the CT10 PDF set, corrected for non-perturbative and electroweak effects, are included.
      The renormalization and factorization scale choice $\mu$ is as described in section \ref{sec:theory}.
      The hatched band shows the uncertainty associated with the theory predictions. 
      Because of the logarithmic scale on the vertical axis, the experimental and theoretical uncertainties are only visible at high dijet mass, where they are largest.
   }
   \label{fig:MassSummary04}
   \end{center}
\end{figure}

Measurements of the dijet double-differential cross-sections as a function of dijet mass in various ranges of \ystar are shown in figure \ref{fig:MassSummary04} for \AKT jets with values of the radius parameter \rfour and \rsix.
The cross-sections are measured up to a dijet mass of 5 \TeV and $\ystar$ of $3.0$, and are seen to decrease quickly with increasing dijet mass.
The NLO QCD calculations by \nlojet using the CT10 PDF set, which are corrected for non-perturbative and electroweak effects, are compared to the measured cross-sections.
No major deviation of the data from the theoretical predictions is observed over the full kinematic range, covering almost eight orders of magnitude in measured cross-section values.
More detailed quantitative comparisons are made between data and theoretical predictions in the following subsection.
At a given dijet mass, the absolute cross-sections using jets with radius parameter \rfour are smaller than those using \rsix.
This is due to the smaller value of the jet radius parameter, resulting in a smaller contribution to the jet energy from the parton shower and the underlying event.
Tables summarising the measured cross-sections are provided in Appendix \ref{app:tables}.
The quadrature sum of the uncertainties listed in the tables is the total uncertainty on the measurement, although the in situ, pileup, and flavour columns are each composed of two or more components.
The full set of cross-section values and uncertainty components, each of which is fully correlated in dijet mass and \ystar but uncorrelated with the other components, can be found in HepData \cite{Buckley:2010jn}.

\addtolength{\subfigcapskip}{-12pt}
\addtolength{\subfigcapskip}{0pt}

\subsection{Quantitative comparison with \nlojet predictions}\label{subsec:nlojetresult}

\begin{sidewaysfigure}[!htbp]
   \begin{center}
   \includegraphics[width=0.90\textwidth]{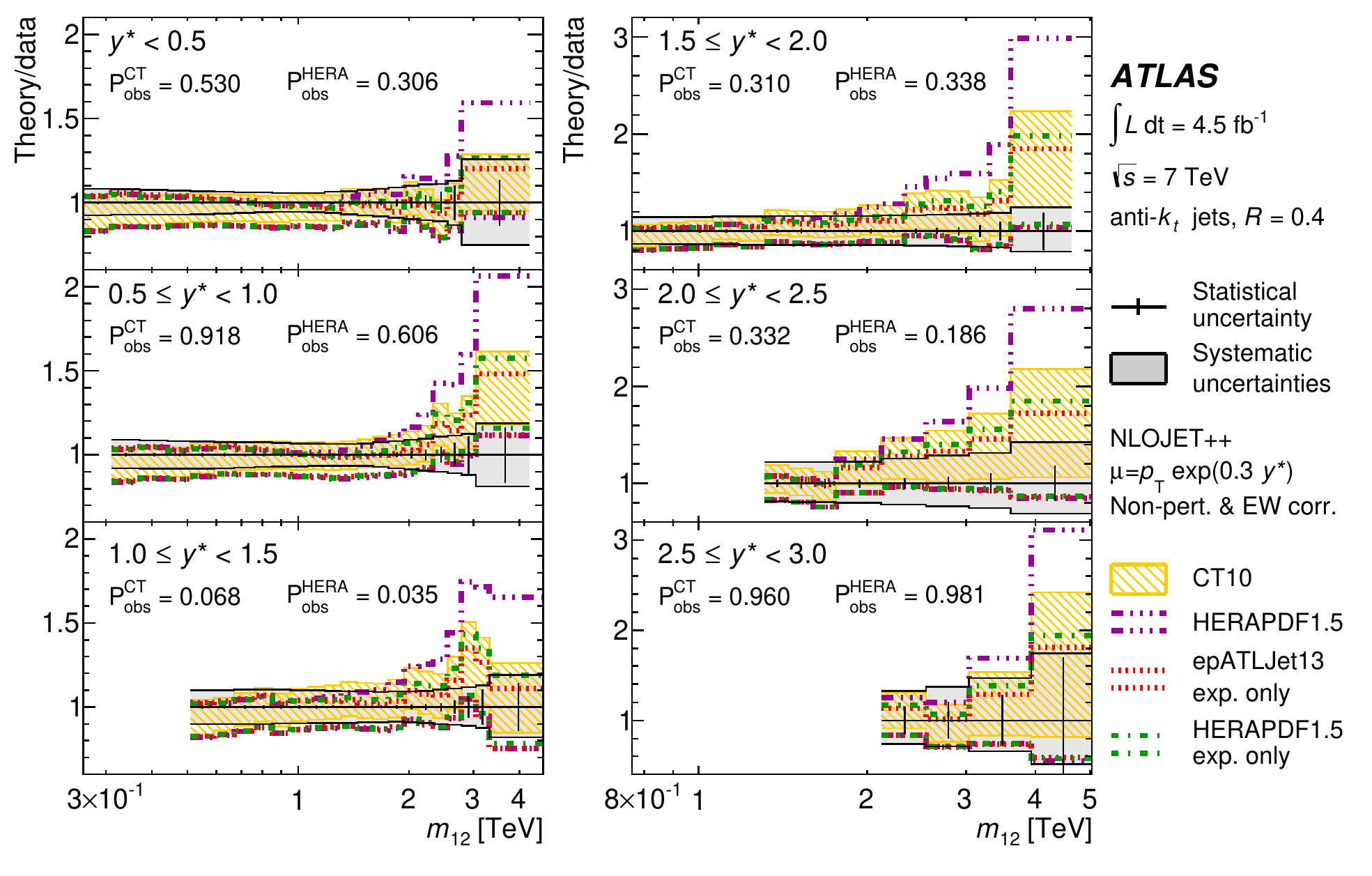}
   \caption{
      Ratio of the NLO QCD predictions of \nlojet to the measurements of the dijet double-differential cross-section as a function of dijet mass in different ranges of \ystar.
      The results are shown for jets identified using the \AKT algorithm with radius parameter \rfour.
      The predictions of \nlojet using different PDF sets (CT10, HERAPDF1.5, and epATLJet13) are shown.
      The renormalization and factorization scale choice $\mu$ is as described in section \ref{sec:theory}.
      Observed p-values resulting from the comparison of theory with data are shown considering all \mass bins in each range of \ystar separately.
      The HERAPDF1.5 analysis accounts for model and parameterization uncertainties as well as experimental uncertainties.
      The theoretical predictions are labelled with \emph{exp. only} when the model and parameterization uncertainties are not included.
   }
   \label{fig:pdf04_1}
   \end{center}
\end{sidewaysfigure}

\begin{sidewaysfigure}[!htbp]
   \begin{center}
   \includegraphics[width=0.90\textwidth]{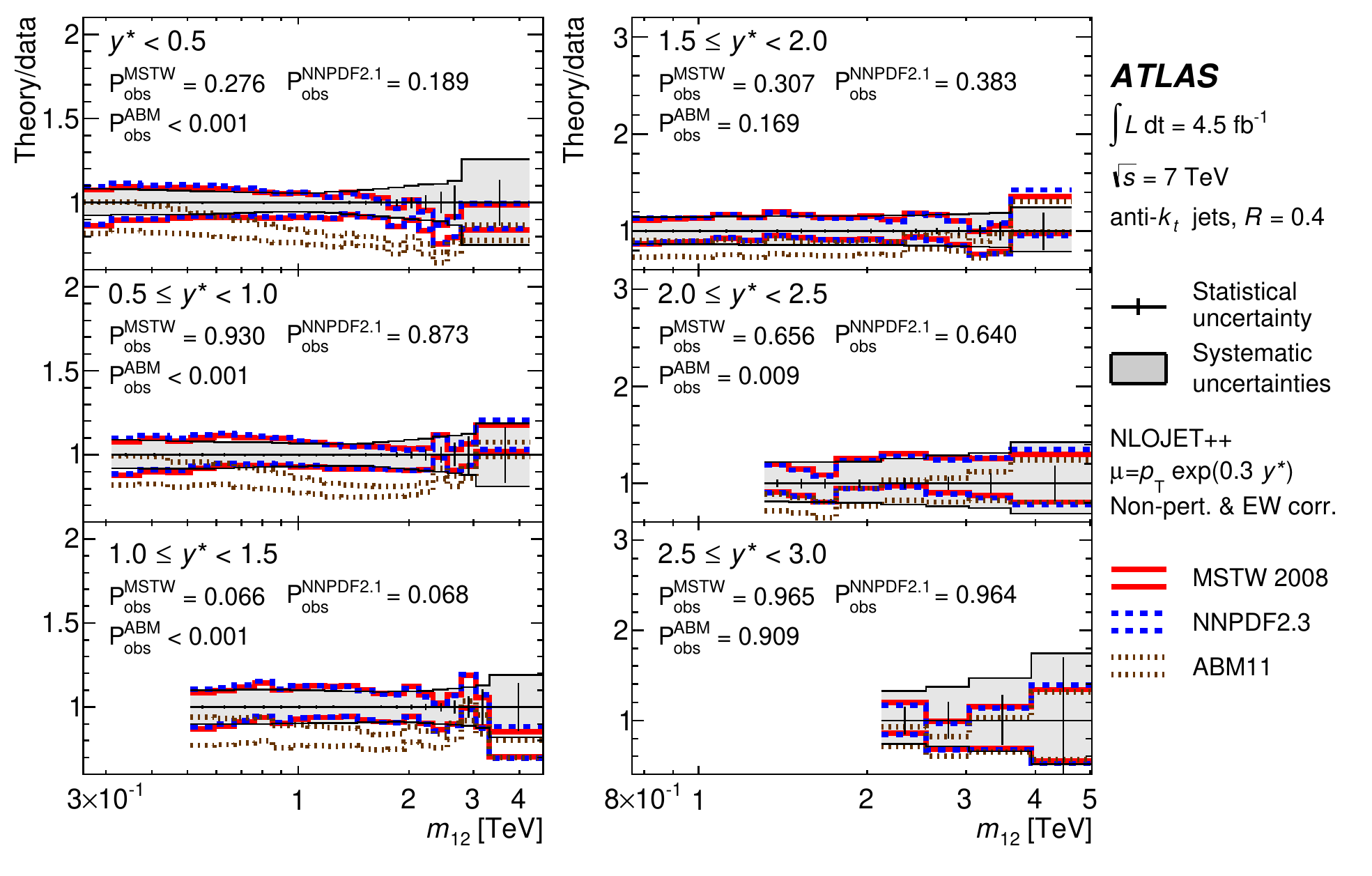}
   \caption{
      Ratio of the NLO QCD predictions of \nlojet to the measurements of the dijet double-differential cross-section as a function of dijet mass in different ranges of \ystar.
      The results are shown for jets identified using the \AKT algorithm with radius parameter \rfour.
      The predictions of \nlojet using different PDF sets (MSTW 2008, NNPDF2.3 and ABM11) are shown.
      The renormalization and factorization scale choice $\mu$ is as described in section \ref{sec:theory}.
      Observed p-values resulting from the comparison of theory with data are shown considering all \mass bins in each range of \ystar separately.
      Here, the observed p-value is shown for the NNPDF2.1 PDF set, while the lines are for the NNPDF2.3 PDF set (see text).
   }
   \label{fig:pdf04_2}
   \end{center}
\end{sidewaysfigure}

\begin{sidewaysfigure}[!htbp]
   \begin{center}
   \includegraphics[width=0.90\textwidth]{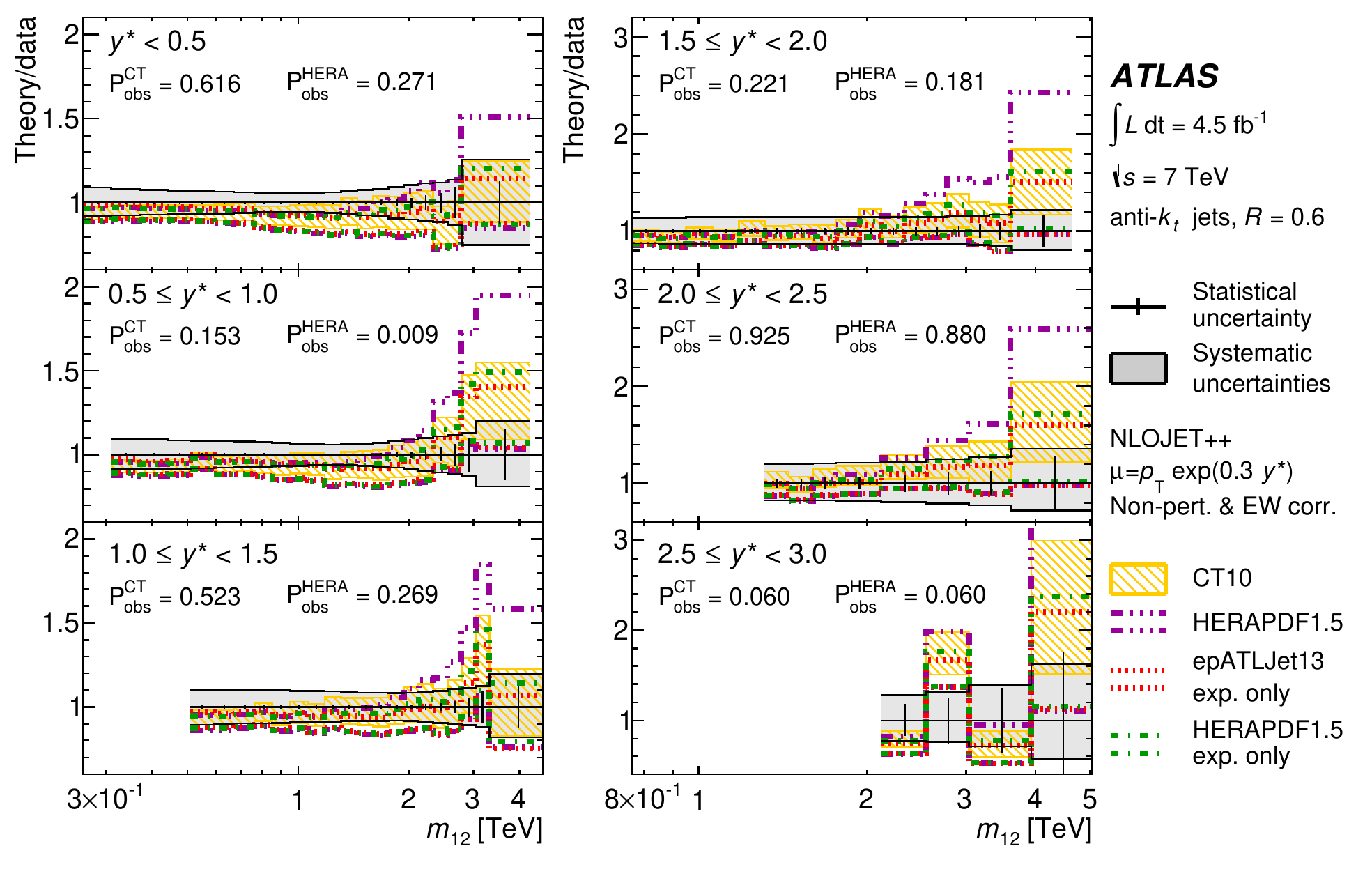}
   \caption{
      Ratio of the NLO QCD predictions of \nlojet to the measurements of the dijet double-differential cross-section as a function of dijet mass in different ranges of \ystar.
      The results are shown for jets identified using the \AKT algorithm with radius parameter \rsix.
      The predictions of \nlojet using different PDF sets (CT10, HERAPDF1.5, and epATLJet13) are shown.
      The renormalization and factorization scale choice $\mu$ is as described in section \ref{sec:theory}.
      Observed p-values resulting from the comparison of theory with data are shown considering all \mass bins in each range of \ystar separately.
      The HERAPDF1.5 analysis accounts for model and parameterization uncertainties as well as experimental uncertainties.
      The theoretical predictions are labelled with \emph{exp. only} when the model and parameterization uncertainties are not included.
   }
   \label{fig:pdf06_1}
   \end{center}
\end{sidewaysfigure}

\begin{sidewaysfigure}[!htbp]
   \begin{center}
   \includegraphics[width=0.90\textwidth]{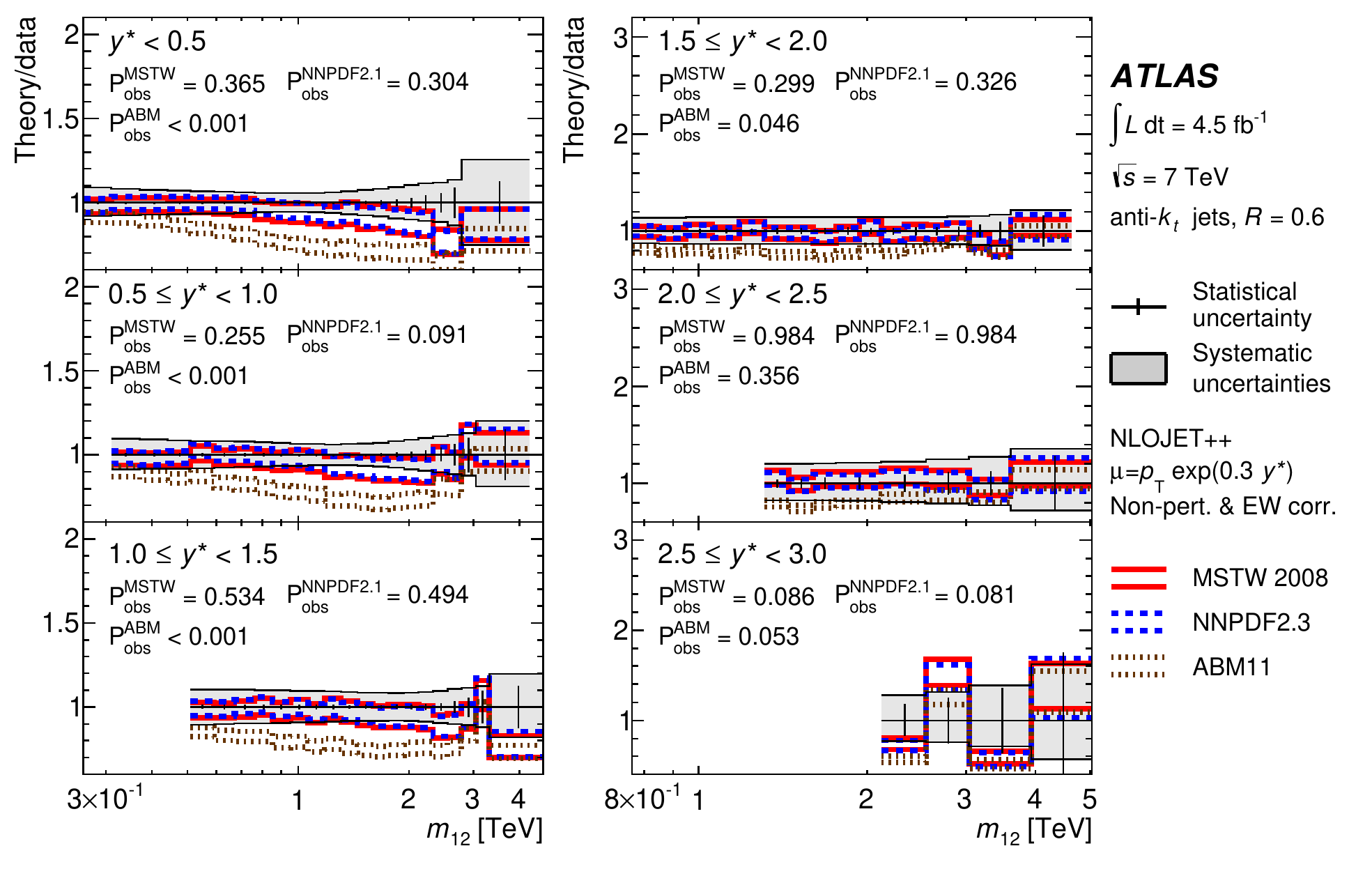}
   \caption{
      Ratio of the NLO QCD predictions of \nlojet to the measurements of the dijet double-differential cross-section as a function of dijet mass in different ranges of \ystar.
      The results are shown for jets identified using the \AKT algorithm with radius parameter \rsix.
      The predictions of \nlojet using different PDF sets (MSTW 2008, NNPDF2.3 and ABM11) are shown.
      The renormalization and factorization scale choice $\mu$ is as described in section \ref{sec:theory}.
      Observed p-values resulting from the comparison of theory with data are shown considering all \mass bins in each range of \ystar separately.
      Here, the observed p-value is shown for the NNPDF2.1 PDF set, while the lines are for the NNPDF2.3 PDF set (see text).
   }
   \label{fig:pdf06_2}
   \end{center}
\end{sidewaysfigure}

The ratio of the NLO QCD predictions from \nlojet, corrected for non-perturbative and electroweak effects, to the data is shown in figures \ref{fig:pdf04_1}--\ref{fig:pdf06_2} for various PDF sets. 
The CT10, HERAPDF1.5, epATLJet13, MSTW 2008, NNPDF2.3, and ABM11 PDF sets are used.
As discussed in section \ref{sec:sysunc}, the individual experimental and theoretical uncertainty components are fully correlated between \mass and \ystar bins.
As such, the frequentist method described in section \ref{sec:stattest} is necessary to make quantitative statements about the agreement of theoretical predictions with data.
For this measurement, the \nlojet predictions using the MSTW 2008, NNPDF2.3 and ABM11 PDF sets have smaller theoretical uncertainties than those using the CT10 and HERAPDF1.5 PDF sets.
Due to the use of ATLAS jet data in the PDF fit, the predictions using the epATLJet13 PDF set have smaller uncertainties at high dijet mass compared to those using the HERAPDF1.5 PDF set, when only considering the uncertainties due to the experimental inputs for both.

\afterpage{\clearpage}

The frequentist method is employed using the hypothesis that the SM is the underlying theory.
Here, the NNPDF2.1 PDF set is considered instead of NNPDF2.3 due to the larger number of replicas available, which allows a better determination of the uncertainty components.
The epATLJet13 PDF set is not considered since the full set of uncertainties is not available.
The resulting observed p-values are also shown in figures \ref{fig:pdf04_1}--\ref{fig:pdf06_2}, where all \mass bins are considered in each range of \ystar separately.
The agreement between the data and the various theories is good (observed p-value greater than $5\%$) in all cases, except the following.
For the predictions using HERAPDF1.5, the smallest observed p-values are seen in the range \ystarthree for jets with radius parameter \rfour, and in the range \ystartwo for jets with radius parameter \rsix, both of which are $<5\%$.
For the predictions using ABM11, the observed p-value is $<0.1\%$ for each of the first three ranges of $\ystar<1.5$, for both values of jet radius parameter.
The results using the ABM11 PDF set also show observed p-values of less than $5\%$, for the range \ystarfive for jets with radius parameter \rfour, and \ystarfour for \rsix jets.

\begin{figure}[!htbp]
\begin{center}
  \subfigure[\ystarone]{
    \includegraphics[width=0.7\linewidth]{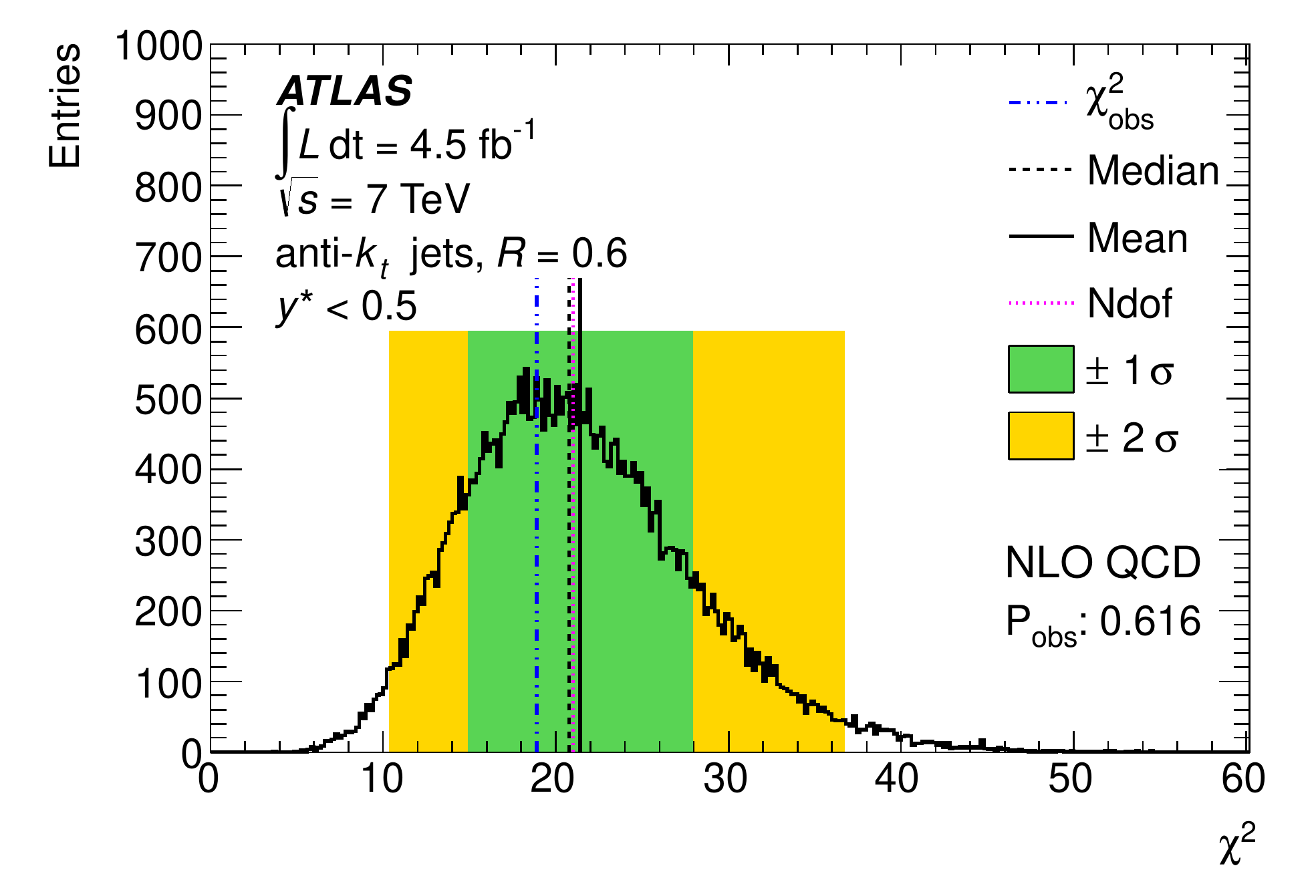}\hspace{0.0cm}
    \label{fig:Limits_NLO_chi2withCorrelationsAsymmSyst_CT10_y1}
  } \\
  \subfigure[$\ystar < 1.5$]{
    \includegraphics[width=0.7\linewidth]{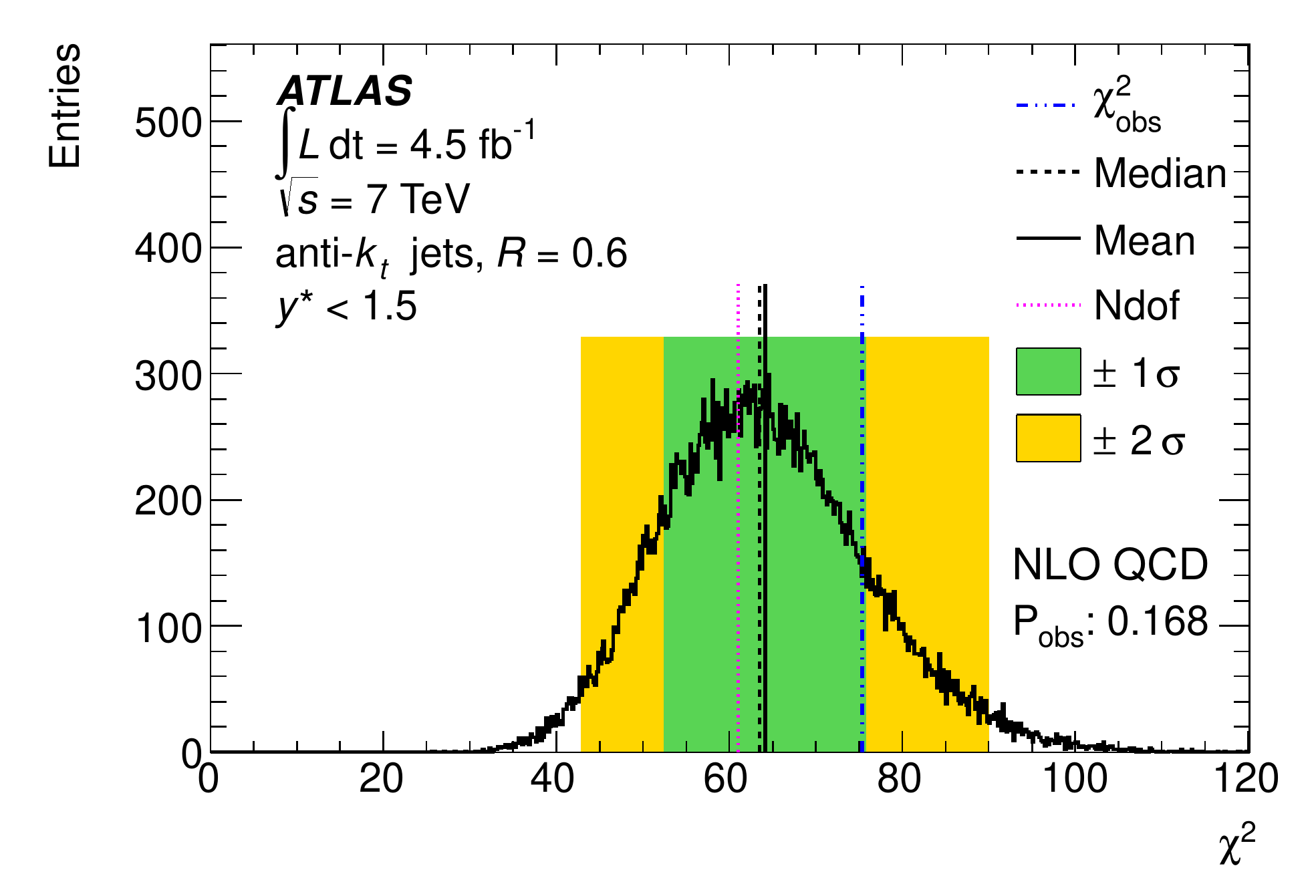}
    \label{fig:Limits_NLO_chi2withCorrelationsAsymmSyst_CT10_y4}
  } 
  \caption{
    The \chisq distribution of pseudo-experiments (black histogram) for NLO QCD using the CT10 PDF set.
    The renormalization and factorization scale choice $\mu$ is as described in section \ref{sec:theory}.
    The full information on the uncertainties, including their asymmetries and correlations, is used for both the pseudo-experiments and the \chisq calculation.
    The black vertical dashed (solid) line indicates the median (mean) of the distribution, while the green (yellow) band indicates the $\pm 1\sigma\ (\pm 2\sigma)$ region.
    The blue dot-dashed vertical lines indicate the observed \chisq, with the corresponding observed p-value given in the legend (see text).
    The pink dotted lines show the number of degrees of freedom, 21 for (a) and 61 for (b).
    The plots correspond to (a) the range \ystarone , and (b) the first three ranges of $\ystar < 1.5$ combined, for jets reconstructed with radius parameter \rsix.
  }
\label{fig:Limits_NLO_chi2withCorrelationsAsymmSyst_CT10}
\end{center}
\end{figure}

Figure~\ref{fig:Limits_NLO_chi2withCorrelationsAsymmSyst_CT10} shows the \chisq distribution from pseudo-experiments for the SM hypothesis using the CT10 PDF set, as well as the observed \chisq value.
The full range of dijet mass in the first three ranges of $\ystar < 1.5$, for jets with radius parameter \rsix, is considered.
The mean, median, $\pm 1\sigma$ and $\pm 2\sigma$ regions of the \chisq distribution from pseudo-experiments around the median, and number of degrees of freedom are also indicated.
Figure \ref{fig:Limits_NLO_chi2withCorrelationsAsymmSyst_CT10_y4} shows that a broadening of the \chisq distribution from pseudo-experiments is observed when additional degrees of freedom are included by combining multiple ranges of \ystar.
For both \chisq distributions, the mean and median are close to the number of degrees of freedom.
This shows that the choice of generalized \chisq as the test statistic in the frequentist method behaves as expected.

Because the data at larger values of \ystar are increasingly dominated by experimental uncertainties, the sensitivity to the proton PDFs is reduced.
For this reason, when considering a combination only the first three ranges of $\ystar<1.5$ are used.
While the low dijet-mass region provides constraints on the global normalization, due to the increased number of degrees of freedom it also introduces a broadening of the \chisq distribution from the pseudo-experiments that reduces sensitivity to the high dijet-mass region.
To focus on the regions where the PDF uncertainties are large while the data still provide a good constraint, the comparison between data and SM predictions is also performed using a high dijet-mass subsample.
The high dijet-mass subsample is restricted to $m_{12}> 1.31 \TeV$ for \ystarone, $m_{12}> 1.45 \TeV$ for \ystartwo, and $m_{12}> 1.60 \TeV$ for \ystarthree.

Table \ref{tab:variousPdfDataVsQCD_1} presents a reduced summary of the observed p-values for the comparison of measured cross-sections and SM predictions in both the full and high dijet-mass regions using various PDF sets, for both values of the jet radius parameter.
For the first three ranges of $\ystar < 1.5$, as well as their combination, theoretical predictions using the CT10 PDF set have an observed p-value $>6.6\%$ for all ranges of dijet mass, and are typically much larger than this.
For the HERAPDF1.5 PDF set, good agreement is found at high dijet mass in the ranges \ystarone and \ystarthree (not shown), both with observed p-values $>15\%$.
Disagreement is observed for jets with distance parameter \rsix when considering the full dijet-mass region for the first three ranges of $\ystar < 1.5$ combined, where the observed p-value is $2.5\%$.
This is due to the differences already noted for the range \ystartwo, where the observed p-value is $0.9\%$ (see figure \ref{fig:pdf06_1}).
Disagreement is also seen when limiting to the high dijet-mass region and combining the first three ranges of $\ystar < 1.5$, resulting in an observed p-value of $0.7\%$.
The observed p-values for the MSTW 2008 and NNPDF2.1 PDF sets are always $>12.5\%$ in the ranges shown in table \ref{tab:variousPdfDataVsQCD_1}.
This is particularly relevant considering these two PDF sets provide small theoretical uncertainties at high dijet mass.
A strong disagreement, where the observed p-value is generally $<0.1\%$, is observed for the ABM11 PDF set for the first three ranges of $\ystar < 1.5$ and both values of the jet radius parameter.

\clearpage

\begin{table}[!t]
\begin{center}
  \begin{tabular}{|ccccc|}
    \hline
    PDF set & \ystar    ranges & mass range  & \multicolumn{2}{c|}{${\rm P}_{\rm obs}$} \\
            &                  & (full/high) & \rfour & \rsix \\
    \hline
          & \ystarone               & high & $0.742$ & $0.785$ \\
    CT10 & $\ystar < 1.5$          & high & $0.080$ & $0.066$ \\
          & $\ystar < 1.5$          & full & $0.324$ & $0.168$ \\
    \hline
                & \ystarone               & high & $0.688$ & $0.504$ \\
    HERAPDF1.5 & $\ystar < 1.5$          & high & $0.025$ & $0.007$ \\
                & $\ystar < 1.5$          & full & $0.137$ & $0.025$ \\
    \hline      
              & \ystarone               & high & $0.328$ & $0.533$ \\
    MSTW 2008 & $\ystar < 1.5$          & high & $0.167$ & $0.183$ \\
              & $\ystar < 1.5$          & full & $0.470$ & $0.352$ \\
    \hline      
              & \ystarone               & high & $0.405$ & $0.568$ \\
    NNPDF2.1 & $\ystar < 1.5$          & high & $0.151$ & $0.125$ \\
              & $\ystar < 1.5$          & full & $0.431$ & $0.242$ \\
    \hline      
           & \ystarone               & high & $0.024$ & $<10^{-3}$ \\
    ABM11 & $\ystar < 1.5$          & high & $<10^{-3}$ & $<10^{-3}$ \\
           & $\ystar < 1.5$          & full & $<10^{-3}$ & $<10^{-3}$ \\
    \hline      
  \end{tabular}
  \caption{
    Sample of observed p-values obtained in the comparison between data and the NLO QCD predictions using the CT10, HERAPDF1.5, MSTW 2008, NNPDF2.1 and ABM11 PDF sets, with values of the jet radius parameter \rfour and \rsix.
    Results are presented for the range \ystarone, as well as the combination of the first three ranges of $\ystar < 1.5$, performing the test in the full dijet-mass range or restricting it to the high dijet-mass subsample.
    The full information on uncertainties, including their asymmetries and correlations, is used for both the pseudo-experiments and the \chisq calculation.
  }
  \label{tab:variousPdfDataVsQCD_1}
\end{center}
\end{table}

It is possible to further study the poor agreement observed at high dijet mass for the combination of the first three ranges of $\ystar<1.5$ when using the HERAPDF1.5 PDF set by exploring the four variations described in section \ref{subsec:theoryunc}.
The observed p-values using variations 1, 2, and 4 in the \nlojet predictions, shown in table \ref{tab:variousPdfDataVsQCD_2}, are generally similar to those using the default HERAPDF1.5 PDF set.
However, much smaller p-values are observed for variation 3, which has a more flexible parameterization for the valence $u$-quark contribution.
Including the present dijet measurement in the PDF analysis should provide a better constraint on the choice of parameterization.

\begin{table}[!t]
\begin{center}
  \begin{tabular}{|ccccc|}
    \hline
    PDF set       & \ystar ranges  & mass range  & \multicolumn{2}{c|}{${\rm P}_{\rm obs}$} \\
    HERAPDF1.5   &                & (full/high) & \rfour & \rsix \\
    \hline
                & \ystarone               & high & $0.692$ & $0.493$ \\
    variation 1 & $\ystar < 1.5$          & high & $0.018$ & $0.005$ \\
                & $\ystar < 1.5$          & full & $0.113$ & $0.018$ \\
    \hline
                & \ystarone               & high & $0.667$ & $0.453$ \\
    variation 2 & $\ystar < 1.5$          & high & $0.025$ & $0.008$ \\
                & $\ystar < 1.5$          & full & $0.124$ & $0.024$ \\
    \hline
                & \ystarone               & high & $0.424$ & $0.209$ \\
    variation 3 & $\ystar < 1.5$          & high & $<10^{-3}$ & $<10^{-3}$ \\
                & $\ystar < 1.5$          & full & $0.001$ & $<10^{-3}$ \\
    \hline
                & \ystarone               & high & $0.677$ & $0.525$ \\
    variation 4 & $\ystar < 1.5$          & high & $0.031$ & $0.010$ \\
                & $\ystar < 1.5$          & full & $0.160$ & $0.031$ \\
    \hline
  \end{tabular}
  \caption{
    Sample of observed p-values obtained in the comparison between data and the NLO QCD predictions based on the variations (described in section \ref{subsec:theoryunc}) of the HERAPDF1.5 PDF analysis, with values of the jet radius parameter \rfour and \rsix.
    Results are presented for the range \ystarone, as well as the combination of the first three ranges of $\ystar < 1.5$, performing the test in the full dijet-mass range or restricting it to the high dijet-mass subsample.
    The full information on uncertainties, including their asymmetries and correlations, is used for both the pseudo-experiments and the \chisq calculation.
  }
  \label{tab:variousPdfDataVsQCD_2}
\end{center}
\end{table}

The studies described above, as well as most of the previously published analyses, use test statistics that exploit the information on the sizes and bin-to-bin correlations of the uncertainties.
As such, the results are sensitive to the assumptions used to derive the uncorrelated uncertainty components, in particular those of the jet energy calibration.
To improve upon previous studies, the analysis is repeated using the two different correlation assumptions described in section \ref{sec:sysunc}.
Theoretical predictions using the CT10 PDF set in the range \ystarone for the high dijet-mass subsample, with jet radius parameter \rsix, are considered.
For both the stronger and weaker correlation assumptions the observed p-values differ by $<0.1\%$ from the value of $78.5\%$ found using the default jet energy calibration uncertainty components.
This is because the impact on the total uncertainty and its correlations is only at the few-percent level for most bins, even though these assumptions lead to a relative change of up to 8\% in the experimental uncertainty, and up to 12\% in its correlations.
Quantitative comparisons of theoretical predictions with data would benefit from alternative assumptions for the correlations of the theoretical uncertainties, in particular the large PDF uncertainty components.

\subsection{Comparison with \powheg predictions}

\begin{sidewaysfigure}[!htbp]
   \begin{center}
   \includegraphics[width=0.90\textwidth]{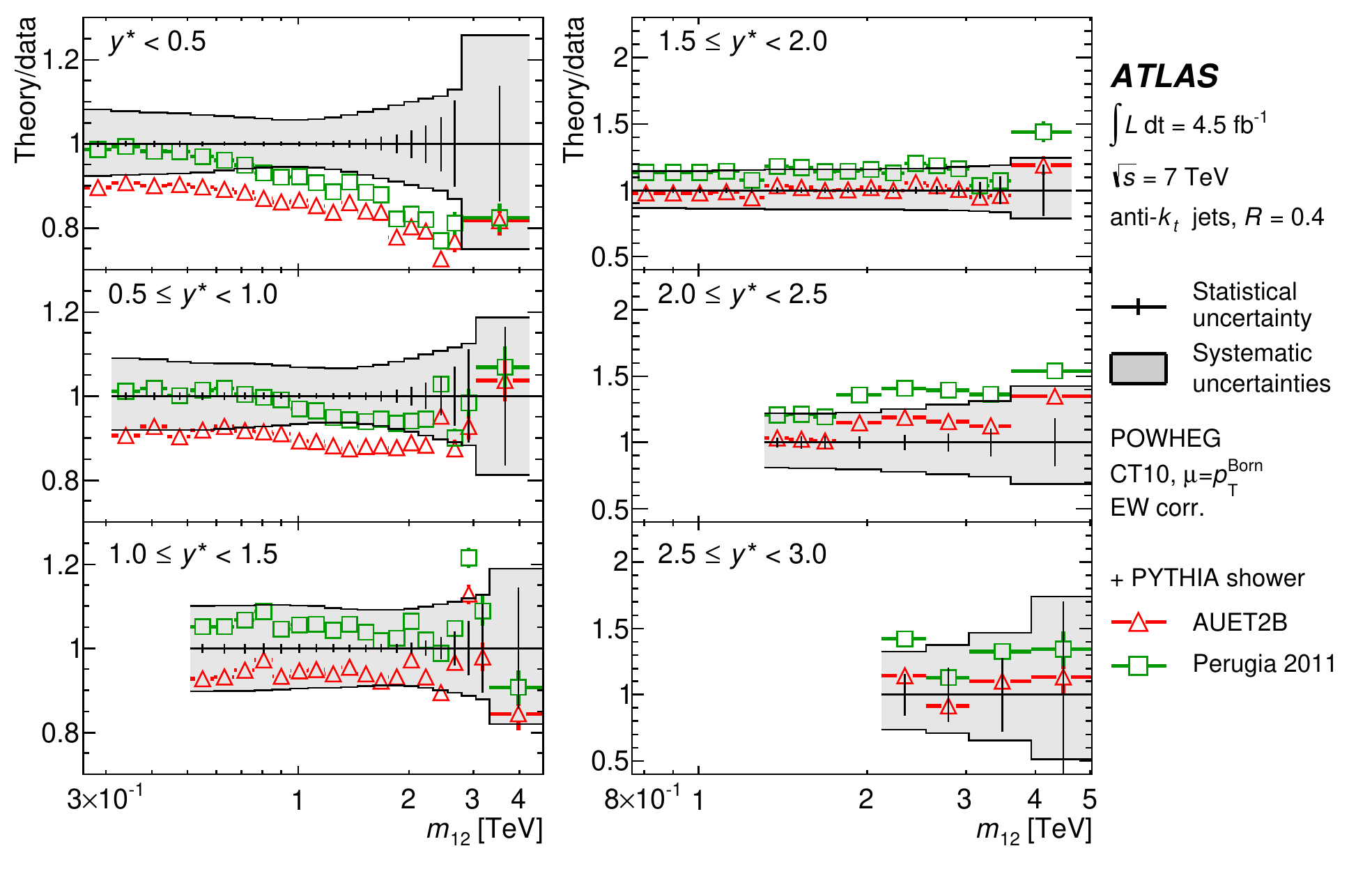}
   \caption{
      Ratio of the POWHEG predictions to the measurements of the dijet double-differential cross-sections as a function of dijet mass in different ranges of \ystar.
      The results are shown for jets identified using the \AKT algorithm with jet radius parameter \rfour.
      The predictions of \powheg with parton-shower MC simulation by \pythia are shown for the AUET2B and Perugia 2011 tunes.
      The statistical (total systematic) uncertainties of the measurements are indicated as error bars (shaded bands).
   }
   \label{fig:MassRatio04}
   \end{center}
\end{sidewaysfigure}

\begin{sidewaysfigure}[!htbp]
   \begin{center}
   \includegraphics[width=0.90\textwidth]{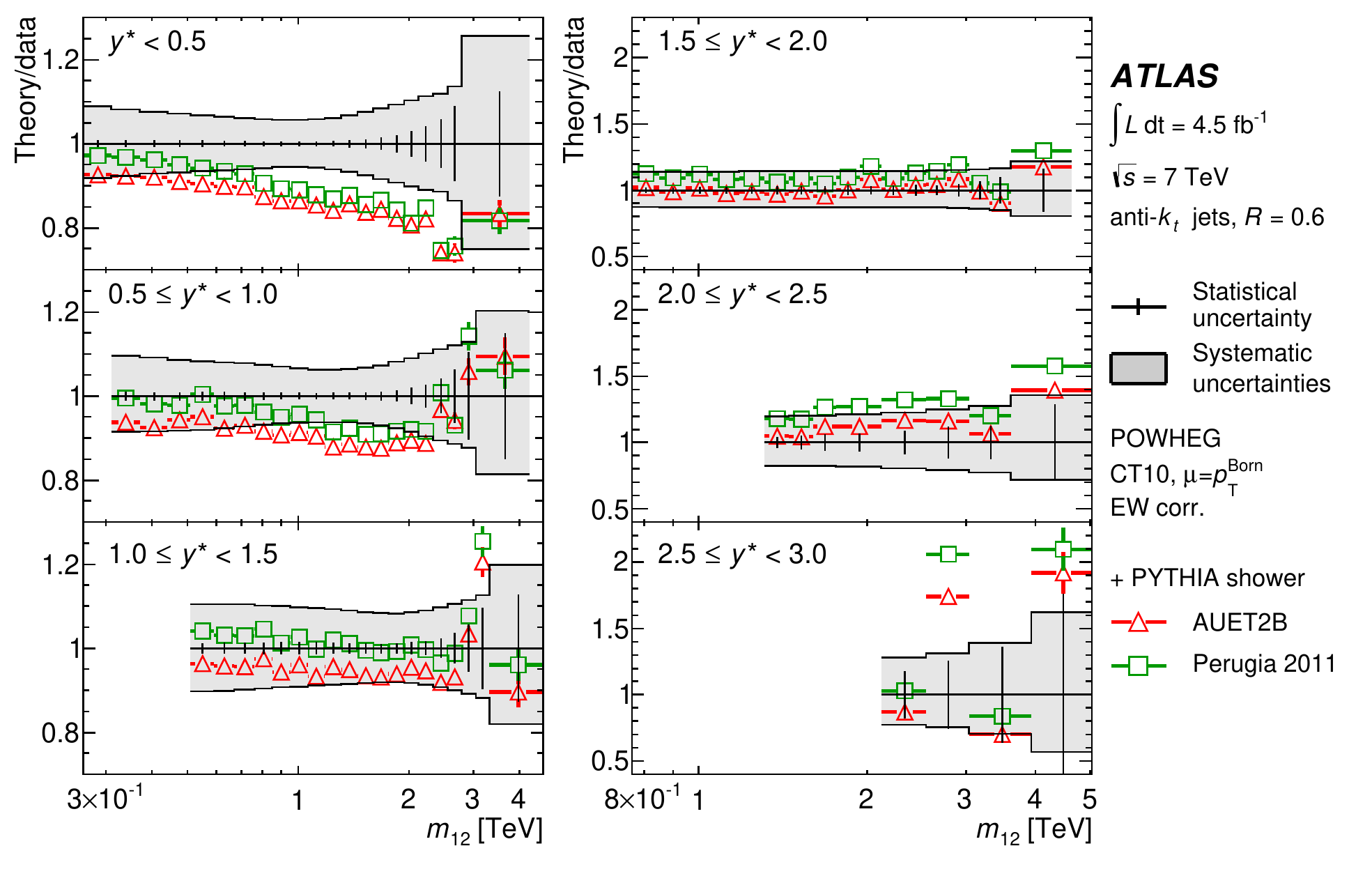}
   \caption{
      Ratio of the POWHEG predictions to the measurements of the dijet double-differential cross-sections as a function of dijet mass in different ranges of \ystar.
      The results are shown for jets identified using the \AKT algorithm with jet radius parameter \rsix.
      The predictions of \powheg with parton-shower MC simulation by \pythia are shown for the AUET2B and Perugia 2011 tunes.
      The statistical (total systematic) uncertainties of the measurements are indicated as error bars (shaded bands).
   }
   \label{fig:MassRatio06}
   \end{center}
\end{sidewaysfigure}

The ratios of the theoretical predictions from \powheg to data are shown in figures~\ref{fig:MassRatio04} and \ref{fig:MassRatio06} for values of the jet radius parameter \rfour and \rsix.
Because of the difficulty in propagating multiple PDF sets through the \powheg generation, direct calculations of the theoretical uncertainties are not made and only qualitative comparisons are provided.
However, the theoretical uncertainties are expected to be similar to those shown for the \nlojet calculations using the CT10 PDF set.

\afterpage{\clearpage}

The \powheg predictions show no major deviations from data, with the Perugia 2011 tune resulting in slightly larger cross-sections ($10$--$20\%$) than the AUET2B tune.
At high dijet mass, the difference in the theory calculations is less pronounced, decreasing to a few percent at low values of \ystar.
For the range \ystarone, theoretical predictions underestimate the measured cross-sections by up to $25\%$ at larger values of dijet mass.
In the range \ystartwo, this difference at high dijet mass is less, at around $15\%$.
At larger values of \ystar, the experimental sensitivity decreases so that no strong statements can be made.
The ratios of the theoretical predictions to the cross-section measurements using jets with radius parameter \rfour are observed to be larger than those for \rsix.
This is similar to the trend observed previously for inclusive jet cross-sections from ATLAS \cite{Aad:2011fc}.
While this trend is particularly apparent for the Perugia 2011 tune, it is almost negligible for the AUET2B tune at high dijet mass.

\section{Exploration and exclusion of contact interactions}
\label{sec:setlimits}

To illustrate the sensitivity of the measurements to physics beyond the SM, the model of QCD plus contact interactions (CIs) \cite{Gao:2011ha} with left--left coupling and destructive interference between CIs and QCD as implemented by the CIJET program \cite{Gao:2013kp} is considered.
The use of NLO QCD calculations for the CI portion unifies the treatment of the two predictions.
For this study, the measurement is restricted to the high dijet-mass subsample $\mass>1.31\TeV$ in the range \ystarone, where theory predicts the largest effect.
Because QCD predicts a more uniform \ystar distribution compared to CIs, where events are preferentially produced at smaller values of \ystar, this region provides the highest sensitivity to the CI contribution.
Furthermore, the additional QCD plus CIs matrix elements are sensitive to different quark--gluon compositions.
Given that the positive electroweak corrections and the CIs produce similar contributions, the electroweak corrections are not applied to the CI portion. 
This results in a conservative approach when setting limits for the kinematic region considered here.
Non-perturbative corrections are also not applied to the CI portion, but have a negligible size of less than 1\% at high dijet mass where CI contributions are most significant.
The PDF sets for which the SM predictions in this region provide a good description of the data are considered: CT10, HERAPDF1.5, MSTW 2008, and NNPDF2.1.
Even when considering the ABM11 PDF set in only the low dijet-mass region $\mass < 1.31\TeV$, where CIs were previously excluded \cite{Abe:1996mj,Abazov:2009ac}, the observed p-value is $<0.1\%$ in the range \ystarone.
For this reason, the ABM11 PDF set is not considered in this example.

\begin{figure}[!t]
\begin{center}
  \includegraphics[width=0.7\linewidth]{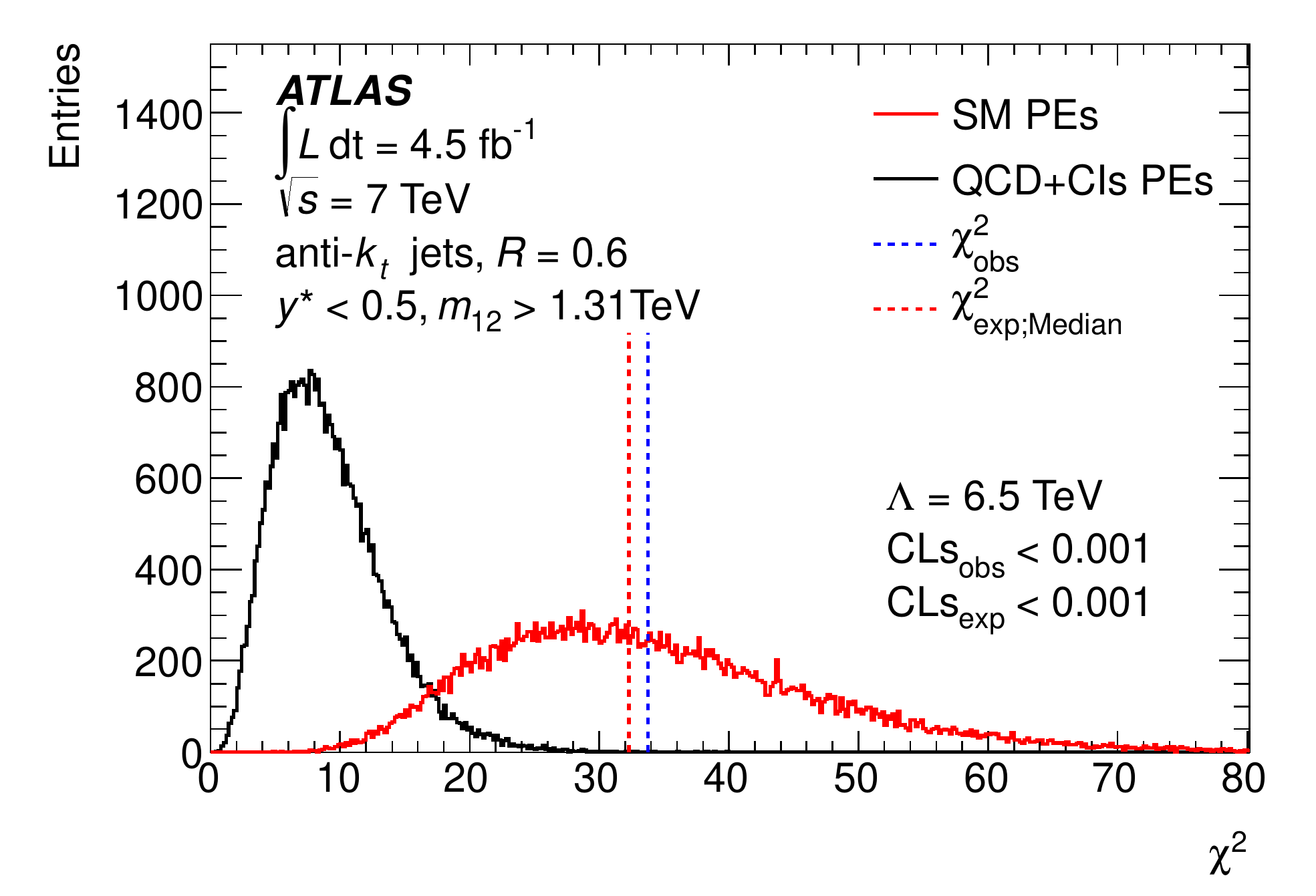}
  \caption{
    The \chisq distribution from pseudo-experiments of QCD plus CIs (black histogram) and of the SM background (red histogram) using the full information on the uncertainties, including their asymmetries and correlations, for both the pseudo-experiments (PEs) and the \chisq calculation.
    The theoretical hypothesis is the NLO QCD plus CIs prediction based on the CT10 PDF set and a compositeness scale $\Lambda=6.5\TeV$.
    The blue (red) dashed vertical line indicates the observed (expected) \chisq, with the corresponding observed (expected) \emph{CLs} value given in the legend (see text).
    The plot corresponds to the measurement in the high dijet-mass subsample for the range \ystarone and jet radius parameter \rsix.
  }
  \label{fig:Limits_QCDpCI6p5TeV_chi2withCorrelationsAsymmSyst_CT10}
\end{center}
\end{figure}

Figure \ref{fig:Limits_QCDpCI6p5TeV_chi2withCorrelationsAsymmSyst_CT10} shows the \chisq distributions from pseudo-experiments for both the SM (background) and QCD plus CIs (signal$+$background), as well as the observed and expected \chisq values.
Here, the hypothesis for the \chisq calculation is the prediction of QCD plus CIs using the CT10 PDF set with compositeness scale $\Lambda=6.5 \TeV$.
For this example, jets with radius parameter \rsix are considered.
The observed p-value for the signal$+$background model, i.e. the integral of the distribution above the observed \chisq, is significantly smaller than that obtained for the SM pseudo-data, resulting in an exclusion of the compositeness scale $\Lambda = 6.5\TeV$ with a \emph{CLs} value $<0.001$.
Similar results are obtained for jets reconstructed with radius parameter \rfour, as well as for the MSTW 2008, NNPDF2.1 and HERAPDF1.5 PDF sets.

\begin{table}[t!]
\begin{center}
  \begin{tabular}{|ccccc|}
    \hline
    PDF set       & \multicolumn{4}{c|}{\hspace{1cm} $\Lambda$ [\TeV{}]} \\
                  & \multicolumn{2}{c}{\hspace{1cm} \rfour} & \multicolumn{2}{c|}{\hspace{1cm} \rsix} \\
                  & \hspace{1cm} Exp & Obs & \hspace{1cm} Exp & Obs \\
    \hline
    CT10          & \hspace{1cm} $7.3$ & $7.2$ & \hspace{1cm} $7.1$ & $7.1$ \\
    HERAPDF1.5    & \hspace{1cm} $7.5$ & $7.7$ & \hspace{1cm} $7.3$ & $7.7$ \\
    MSTW 2008     & \hspace{1cm} $7.3$ & $7.0$ & \hspace{1cm} $7.1$ & $6.9$ \\
    NNPDF2.1      & \hspace{1cm} $7.3$ & $7.2$ & \hspace{1cm} $7.2$ & $7.0$ \\
    \hline
  \end{tabular}
  \caption{
  Expected and observed lower limits at the $95\%$ CL on the compositeness scale $\Lambda$ of the NLO QCD plus CIs model using the CT10, HERAPDF1.5, MSTW 2008, and NNPDF2.1 PDF sets, for values of the jet radius parameter \rfour and \rsix, using the measurements in the range \ystarone for the high dijet-mass subsample.
  }
  \label{tab:LambdaLimitValues}
\end{center}
\end{table}

\afterpage{\clearpage}

\begin{figure}[!t]
\begin{center}
  \subfigure[ ]{
    \includegraphics[width=0.7\linewidth]{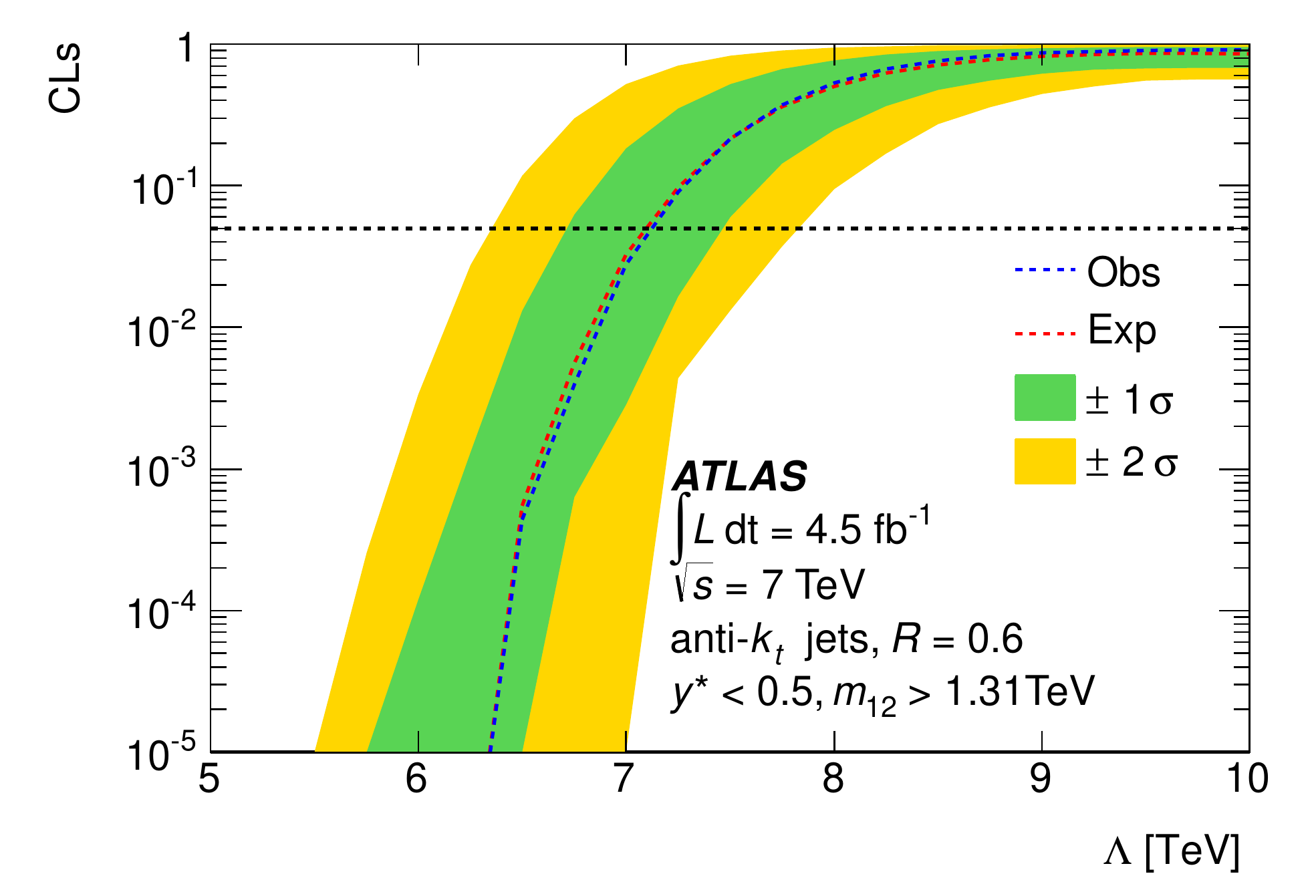}
    \label{fig:Limits_LambdaScan_chi2withCorrelationsAsymmSyst_HM_CT10:b}
  } \\
  \subfigure[ ]{
    \includegraphics[width=0.7\linewidth]{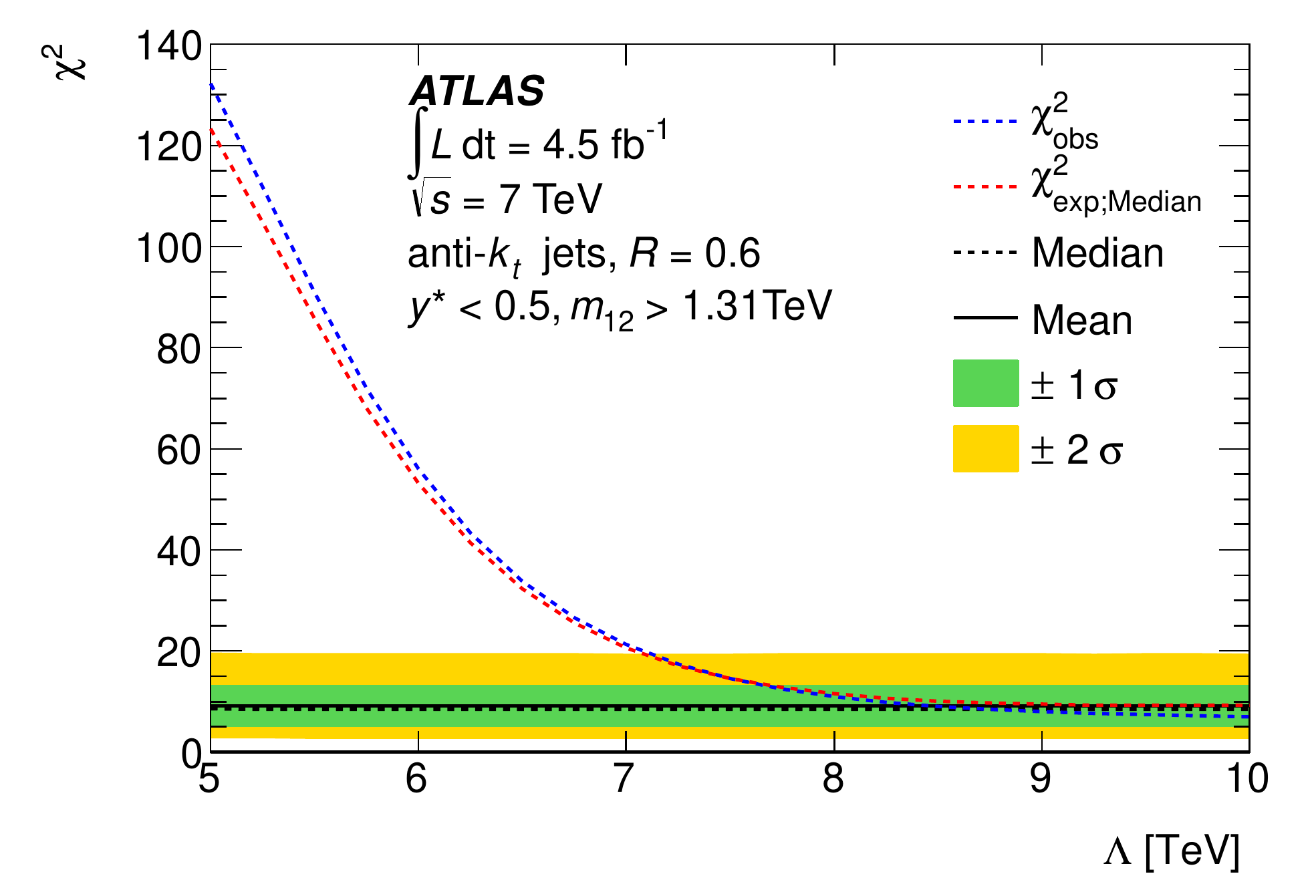}
    \label{fig:Limits_LambdaScan_chi2withCorrelationsAsymmSyst_HM_CT10:a}
  }
  \caption{
    Scan of (a) \emph{CLs} value and (b) \chisq for NLO QCD plus CIs as a function of $\Lambda$, using the CT10 PDF set.
    The green~(yellow) bands indicate the $\pm 1\sigma(\pm 2\sigma)$ regions, from pseudo-experiments of (a) the SM background and (b) the QCD plus CIs.
    The full information on uncertainties, including their asymmetries and correlations, is used for both the pseudo-experiments and the \chisq calculation.
    The dashed horizontal line in (a) indicates the $95\%$ CL exclusion, computed using the observed (expected) p-value shown by the blue (red) dashed line.
    In (b) the black dashed (solid) line indicates the median (mean) of the \chisq distribution of pseudo-experiments.
    The blue (red) dashed lines indicate the observed (expected) \chisq.
    The plots correspond to the measurement with jet radius parameter \rsix in the range \ystarone, restricted to the high dijet-mass subsample.
  }
  \label{fig:Limits_LambdaScan_chi2withCorrelationsAsymmSyst_HM_CT10}
\end{center}
\end{figure}

In order to exclude a range for the compositeness scale, a scan of the observed \emph{CLs} value as a function of $\Lambda$ is performed.
This is shown in figure \ref{fig:Limits_LambdaScan_chi2withCorrelationsAsymmSyst_HM_CT10:b}, whereas figure \ref{fig:Limits_LambdaScan_chi2withCorrelationsAsymmSyst_HM_CT10:a} shows the scan over $\Lambda$ of the observed and expected \chisq values from which the \emph{CLs} values are derived.
The observed and expected \chisq values follow each other closely for small values of $\Lambda$, converging towards a constant difference in the limit $\Lambda\to\infty$.
The scans are for the theoretical predictions with jet radius parameter \rsix, using the CT10 PDF set, and exclude the range $\Lambda < 7.1\TeV$ at the $95\%$ CL.
The granularity of the scan is 0.25\TeV, and the final value is obtained using a linear interpolation between points.
A summary of the lower limits on $\Lambda$ obtained using the different PDF sets is shown in table \ref{tab:LambdaLimitValues} for both values of the jet radius parameter.
While the expected limits are in the range $7.1$--$7.5\TeV$, the observed ones span the range $6.9$--$7.7\TeV$, the larger values being obtained for the HERAPDF1.5 PDF set.
When considering the variations of the HERAPDF1.5 PDF's, values of $\Lambda < 7.6$--$7.8\TeV$ are excluded.

\afterpage{\clearpage}

The sensitivity to the assumptions about the correlations between the jet energy calibration uncertainty components is also tested.
An effect at the level of $0.01\TeV$ on the lower limit for $\Lambda$ is observed, consistent with the small changes seen for the comparisons with the SM predictions in section \ref{sec:results}.

To compare these results with those obtained by previous studies \cite{ATLAS:2012pu} using an approximate (rather than exact) NLO QCD plus CIs calculation, the cross-sections presented here are also used to test the approximate NLO QCD plus CIs prediction.
A scaling factor computed from the ratio of the NLO to LO QCD calculations is applied to the sum of the LO QCD plus CIs predictions, together with their interference.
This factor ranges between $0.85$ and $1.20$, depending on the dijet mass and \ystar region.
Here, for the sake of consistency with previous studies, no electroweak corrections are applied to the QCD or CI contributions.
While the cross-sections used here include dijet-mass values $\mass > 1.31\TeV$ in the range \ystarone, the previous results include dijet-mass values $\mass > 2.6\TeV$ in the range $\ystar < 1.7$.
Values of $\Lambda < 6.6$--$7.3\TeV$ are excluded at the $95\%$ CL using the predictions based on the different PDF sets, with expected lower limits in the range $\Lambda < 6.6$--$6.9\TeV$.
This result is comparable to the observed (expected) lower limit of $7.6\TeV$ ($7.7\TeV$) obtained by the reconstruction-level analysis using a normalized dijet angular distribution \cite{ATLAS:2012pu}, and does not improve the previous result.

\section{Summary and conclusions}
\label{sec:conclusions}

Cross-section measurements are presented for dijet production in $pp$ collisions at \onlyseventev centre-of-mass energy as a function of dijet mass up to $5\TeV$, and for half the rapidity separation of $\ystar < 3.0$.
These measurements are based on the full data sample collected with the ATLAS detector during 2011 at the LHC, corresponding to an integrated luminosity of \lumifbshort.

Jets are reconstructed with the \AKT algorithm using two values of the radius parameter, \rfour and \rsix, in order to compare data and theory for two different regimes of hadronization, parton shower, underlying event, and higher-order QCD corrections.
The measurements are corrected for detector effects to the particle level and compared to theoretical predictions.
Fixed-order NLO QCD calculations by \nlojet, corrected for non-perturbative effects, and \powheg NLO matrix element calculations interfaced to a parton-shower MC generator are considered.
In both cases, corrections to account for electroweak effects are applied.

The statistical uncertainties are smaller than in the previous ATLAS measurement of dijet production using data collected during 2010.
In particular, the improved measurement in the high dijet-mass region can be used to constrain the PDF's at high momentum fraction.
The correlations between the two measurements, which should be accounted for in a common fit of the PDFs, are non-trivial due to the different treatments of the jet energy calibration uncertainty. 

Detailed quantitative comparisons between the measured cross-sections and theoretical predictions are presented, using a frequentist method employing a $\chi^2$ definition generalized for asymmetric uncertainties and their correlations.
Good agreement is observed when using the CT10, NNPDF2.1 and MSTW 2008 PDF sets for the theoretical predictions with values of the jet radius parameter \rfour and \rsix.
Disagreement is observed in some ranges of dijet mass and \ystar when using the HERAPDF1.5 PDF set, for both values of the jet radius parameter.
Even stronger disagreement is observed between data and the \nlojet predictions using the ABM11 PDF set, indicating the sensitivity of the measurements to the choice of PDF set.
The sensitivity of the frequentist method to assumptions about the correlations between the jet energy calibration uncertainties is tested and found to be small.
However, tests of the impact of assumptions about the correlations between the theoretical uncertainties, dominant in the comparison between data and the SM, would be desirable.

The \emph{CLs} method is employed using measured cross-sections to explore possible contributions to dijet production from physics beyond the Standard Model.
An example using a model of contact interactions is shown, computed using full NLO QCD.
An exclusion of compositeness scales $\Lambda < 6.9$--7.7\TeV is achieved, depending on the PDF set used for the calculation.
This analysis gives results similar to those from a previous ATLAS analysis using the same data set, that relied on detailed detector simulations of the contact interaction contribution, and does not improve the lower limit on $\Lambda$.
The method presented here can be used to confront other new models using measurements of dijet production in $pp$ collisions.

\acknowledgments



We thank CERN for the very successful operation of the LHC, as well as the
support staff from our institutions without whom ATLAS could not be
operated efficiently.

We acknowledge the support of ANPCyT, Argentina; YerPhI, Armenia; ARC,
Australia; BMWF and FWF, Austria; ANAS, Azerbaijan; SSTC, Belarus; CNPq and FAPESP,
Brazil; NSERC, NRC and CFI, Canada; CERN; CONICYT, Chile; CAS, MOST and NSFC,
China; COLCIENCIAS, Colombia; MSMT CR, MPO CR and VSC CR, Czech Republic;
DNRF, DNSRC and Lundbeck Foundation, Denmark; EPLANET, ERC and NSRF, European Union;
IN2P3-CNRS, CEA-DSM/IRFU, France; GNSF, Georgia; BMBF, DFG, HGF, MPG and AvH
Foundation, Germany; GSRT and NSRF, Greece; ISF, MINERVA, GIF, DIP and Benoziyo Center,
Israel; INFN, Italy; MEXT and JSPS, Japan; CNRST, Morocco; FOM and NWO,
Netherlands; BRF and RCN, Norway; MNiSW and NCN, Poland; GRICES and FCT, Portugal; MNE/IFA, Romania; MES of Russia and ROSATOM, Russian Federation; JINR; MSTD,
Serbia; MSSR, Slovakia; ARRS and MIZ\v{S}, Slovenia; DST/NRF, South Africa;
MINECO, Spain; SRC and Wallenberg Foundation, Sweden; SER, SNSF and Cantons of
Bern and Geneva, Switzerland; NSC, Taiwan; TAEK, Turkey; STFC, the Royal
Society and Leverhulme Trust, United Kingdom; DOE and NSF, United States of
America.

The crucial computing support from all WLCG partners is acknowledged
gratefully, in particular from CERN and the ATLAS Tier-1 facilities at
TRIUMF (Canada), NDGF (Denmark, Norway, Sweden), CC-IN2P3 (France),
KIT/GridKA (Germany), INFN-CNAF (Italy), NL-T1 (Netherlands), PIC (Spain),
ASGC (Taiwan), RAL (UK) and BNL (USA) and in the Tier-2 facilities
worldwide.

\appendix

\clearpage

\section{Tables of results}
\label{app:tables}

\renewcommand{\arraystretch}{1.8}

\begin{table*}[!ht]
\tiny
\centering
{\fontfamily{ptm}\selectfont
\begin{tabular}{|@{}c@{}@{}c@{}@{}c@{}@{}c@{}@{}r@{}@{}r@{}@{}r@{}@{}r@{}@{}r@{}@{}r@{}@{}r@{}@{}r@{}|@{}r@{}@{}r@{}|}
\hline
\ \ \mass--\ystar \ \ & \ \ \mass range \ \ & \ \ $\sigma$ \ \ & \ \ $\delta_{\mathrm{stat}}^{\mathrm{data}}$ \ \ & \ \ $\delta_{\mathrm{stat}}^{\mathrm{MC}}$ \ \ & \ \ $\gamma_{\mathrm{in~situ}}$ \ \ & \ \ $\gamma_{\mathrm{pileup}}$ \ \ & \ \ $\gamma_{\operatorname{close-by}}$ \ \ & \ \ $\gamma_{\mathrm{flavour}}$ \ \ & \ \ u$_{\mathrm{JER}}$ \ \ & \ \ u$_{\mathrm{JAR}}$ \ \ & \ \ u$_{\mathrm{unfold}}$ \ \ & \ \ u$_{\mathrm{qual}}$ \ \ & \ \ u$_{\mathrm{lumi}}$ \ \ \\ \relax
\ \ bin \# \ \ & \ \ [TeV] \ \ & \ \ [pb/TeV] \ \ &\ \  \% \ \ & \ \ \% \ \ & \ \  \% \ \ & \ \ \% \ \ & \ \ \% \ \ & \ \ \% \ \ & \ \ \% \ \ & \ \ \% \ \ & \ \ \% \ \ & \ \ \% \ \ & \ \ \% \ \ \\
\hline
\ \ $0$ \ \ & \ \ 0.26--0.31 \ \ & \ \ 7.77e+05 \ \ & \ \ 0.47 \ \ & \ \ 0.56\ \  & \ \ $^{+5.6}_{-5.3}$ \ \ & \ \ $^{+1.1}_{-1.0}$ \ \ & \ \ $^{+1.9}_{-1.8}$ \ \ & \ \ $^{+5.0}_{-4.9}$ \ \ & \ \ 0.9 \ \ & \ \ 0.2 \ \ & \ \ 0.2 \ \ & \multirow{21}{*}{\ \ 0.5 \ \ } & \multirow{21}{*}{\ \ 1.8 \ \ } \\
\ \ $1$ \ \ & \ \ 0.31--0.37 \ \ & \ \ 2.98e+05 \ \ & \ \ 0.45 \ \ & \ \ 0.52\ \  & \ \ $^{+5.5}_{-5.1}$ \ \ & \ \ $^{+1.0}_{-0.9}$ \ \ & \ \ $^{+1.7}_{-1.7}$ \ \ & \ \ $^{+4.7}_{-4.5}$ \ \ & \ \ 0.8 \ \ & \ \ 0.2 \ \ & \ \ 0.1 \ \ &  & \\
\ \ $2$ \ \ & \ \ 0.37--0.44 \ \ & \ \ 1.14e+05 \ \ & \ \ 0.41 \ \ & \ \ 0.55\ \  & \ \ $^{+5.4}_{-5.2}$ \ \ & \ \ $^{+1.2}_{-0.9}$ \ \ & \ \ $^{+1.8}_{-1.8}$ \ \ & \ \ $^{+4.3}_{-4.2}$ \ \ & \ \ 0.8 \ \ & \ \ 0.1 \ \ & \ \ 0.0 \ \ &  & \\
\ \ $3$ \ \ & \ \ 0.44--0.51 \ \ & \ \ 4.57e+04 \ \ & \ \ 0.43 \ \ & \ \ 0.57\ \  & \ \ $^{+5.3}_{-5.3}$ \ \ & \ \ $^{+1.2}_{-0.9}$ \ \ & \ \ $^{+1.9}_{-2.0}$ \ \ & \ \ $^{+3.9}_{-3.9}$ \ \ & \ \ 0.8 \ \ & \ \ 0.1 \ \ & \ \ 0.0 \ \ &  & \\
\ \ $4$ \ \ & \ \ 0.51--0.59 \ \ & \ \ 1.96e+04 \ \ & \ \ 0.30 \ \ & \ \ 0.48\ \  & \ \ $^{+5.2}_{-5.1}$ \ \ & \ \ $^{+1.0}_{-0.9}$ \ \ & \ \ $^{+2.0}_{-2.1}$ \ \ & \ \ $^{+3.5}_{-3.4}$ \ \ & \ \ 0.8 \ \ & \ \ 0.1 \ \ & \ \ 0.0 \ \ &  & \\
\ \ $5$ \ \ & \ \ 0.59--0.67 \ \ & \ \ 8.74e+03 \ \ & \ \ 0.30 \ \ & \ \ 0.43\ \  & \ \ $^{+5.0}_{-4.8}$ \ \ & \ \ $^{+0.8}_{-0.7}$ \ \ & \ \ $^{+2.0}_{-2.1}$ \ \ & \ \ $^{+3.1}_{-3.0}$ \ \ & \ \ 0.7 \ \ & \ \ 0.1 \ \ & \ \ 0.0 \ \ &  & \\
\ \ $6$ \ \ & \ \ 0.67--0.76 \ \ & \ \ 4.08e+03 \ \ & \ \ 0.23 \ \ & \ \ 0.44\ \  & \ \ $^{+4.9}_{-4.6}$ \ \ & \ \ $^{+0.5}_{-0.5}$ \ \ & \ \ $^{+1.9}_{-2.0}$ \ \ & \ \ $^{+2.8}_{-2.7}$ \ \ & \ \ 0.6 \ \ & \ \ 0.1 \ \ & \ \ 0.0 \ \ &  & \\
\ \ $7$ \ \ & \ \ 0.76--0.85 \ \ & \ \ 1.97e+03 \ \ & \ \ 0.27 \ \ & \ \ 0.48\ \  & \ \ $^{+4.8}_{-4.5}$ \ \ & \ \ $^{+0.3}_{-0.3}$ \ \ & \ \ $^{+1.7}_{-1.7}$ \ \ & \ \ $^{+2.5}_{-2.4}$ \ \ & \ \ 0.5 \ \ & \ \ 0.1 \ \ & \ \ 0.0 \ \ &  & \\
\ \ $8$ \ \ & \ \ 0.85--0.95 \ \ & \ \ 9.72e+02 \ \ & \ \ 0.25 \ \ & \ \ 0.47\ \  & \ \ $^{+4.7}_{-4.5}$ \ \ & \ \ $^{+0.1}_{-0.1}$ \ \ & \ \ $^{+1.4}_{-1.3}$ \ \ & \ \ $^{+2.3}_{-2.2}$ \ \ & \ \ 0.5 \ \ & \ \ 0.1 \ \ & \ \ 0.0 \ \ &  & \\
\ \ $9$ \ \ & \ \ 0.95--1.06 \ \ & \ \ 4.71e+02 \ \ & \ \ 0.31 \ \ & \ \ 0.46\ \  & \ \ $^{+4.9}_{-4.7}$ \ \ & \ \ $^{+0.1}_{-0.0}$ \ \ & \ \ $^{+0.9}_{-0.9}$ \ \ & \ \ $^{+2.0}_{-2.0}$ \ \ & \ \ 0.4 \ \ & \ \ 0.1 \ \ & \ \ 0.0 \ \ &  & \\
\ \ $10$ \ \ & \ \ 1.06--1.18 \ \ & \ \ 2.30e+02 \ \ & \ \ 0.42 \ \ & \ \ 0.42\ \  & \ \ $^{+5.2}_{-5.0}$ \ \ & \ \ $^{+0.1}_{-0.1}$ \ \ & \ \ $^{+0.6}_{-0.5}$ \ \ & \ \ $^{+1.9}_{-1.8}$ \ \ & \ \ 0.4 \ \ & \ \ 0.1 \ \ & \ \ 0.0 \ \ &  & \\
\ \ $11$ \ \ & \ \ 1.18--1.31 \ \ & \ \ 1.12e+02 \ \ & \ \ 0.60 \ \ & \ \ 0.39\ \  & \ \ $^{+5.8}_{-5.6}$ \ \ & \ \ $^{+0.1}_{-0.1}$ \ \ & \ \ $^{+0.3}_{-0.3}$ \ \ & \ \ $^{+1.8}_{-1.7}$ \ \ & \ \ 0.4 \ \ & \ \ 0.1 \ \ & \ \ 0.0 \ \ &  & \\
\ \ $12$ \ \ & \ \ 1.31--1.45 \ \ & \ \ 5.22e+01 \ \ & \ \ 0.85 \ \ & \ \ 0.45\ \  & \ \ $^{+6.4}_{-6.2}$ \ \ & \ \ $^{+0.1}_{-0.1}$ \ \ & \ \ $^{+0.1}_{-0.1}$ \ \ & \ \ $^{+1.7}_{-1.6}$ \ \ & \ \ 0.4 \ \ & \ \ 0.1 \ \ & \ \ 0.0 \ \ &  & \\
\ \ $13$ \ \ & \ \ 1.45--1.60 \ \ & \ \ 2.53e+01 \ \ & \ \ 1.2 \ \ & \ \ 0.47\ \  & \ \ $^{+7.2}_{-7.0}$ \ \ & \ \ $^{+0.1}_{-0.1}$ \ \ & \ \ $^{+0.1}_{-0.1}$ \ \ & \ \ $^{+1.6}_{-1.6}$ \ \ & \ \ 0.4 \ \ & \ \ 0.0 \ \ & \ \ 0.0 \ \ &  & \\
\ \ $14$ \ \ & \ \ 1.60--1.76 \ \ & \ \ 1.22e+01 \ \ & \ \ 1.6 \ \ & \ \ 0.46\ \  & \ \ $^{+8.0}_{-7.9}$ \ \ & \ \ $^{+0.1}_{-0.1}$ \ \ & \ \ $^{+0.0}_{-0.0}$ \ \ & \ \ $^{+1.4}_{-1.5}$ \ \ & \ \ 0.4 \ \ & \ \ 0.0 \ \ & \ \ 0.0 \ \ &  & \\
\ \ $15$ \ \ & \ \ 1.76--1.94 \ \ & \ \ 5.93e+00 \ \ & \ \ 2.2 \ \ & \ \ 0.46\ \  & \ \ $^{+8.8}_{-8.7}$ \ \ & \ \ $^{+0.1}_{-0.1}$ \ \ & \ \ $^{+0.0}_{-0.0}$ \ \ & \ \ $^{+1.3}_{-1.4}$ \ \ & \ \ 0.4 \ \ & \ \ 0.0 \ \ & \ \ 0.0 \ \ &  & \\
\ \ $16$ \ \ & \ \ 1.94--2.12 \ \ & \ \ 2.58e+00 \ \ & \ \ 3.3 \ \ & \ \ 0.50\ \  & \ \ $^{+9.7}_{-9.6}$ \ \ & \ \ $^{+0.1}_{-0.0}$ \ \ & \ \ $^{+0.0}_{-0.0}$ \ \ & \ \ $^{+1.2}_{-1.2}$ \ \ & \ \ 0.4 \ \ & \ \ 0.0 \ \ & \ \ 0.0 \ \ &  & \\
\ \ $17$ \ \ & \ \ 2.12--2.33 \ \ & \ \ 1.22e+00 \ \ & \ \ 4.4 \ \ & \ \ 0.45\ \  & \ \ $^{+11}_{-10}$ \ \ & \ \ $^{+0.0}_{-0.0}$ \ \ & \ \ $^{+0.0}_{-0.0}$ \ \ & \ \ $^{+1.1}_{-1.1}$ \ \ & \ \ 0.4 \ \ & \ \ 0.0 \ \ & \ \ 0.0 \ \ &  & \\
\ \ $18$ \ \ & \ \ 2.33--2.55 \ \ & \ \ 5.59e-01 \ \ & \ \ 6.5 \ \ & \ \ 0.39\ \  & \ \ $^{+11}_{-11}$ \ \ & \ \ $^{+0.0}_{-0.0}$ \ \ & \ \ $^{+0.0}_{-0.0}$ \ \ & \ \ $^{+1.0}_{-1.0}$ \ \ & \ \ 0.5 \ \ & \ \ 0.0 \ \ & \ \ 0.0 \ \ &  & \\
\ \ $19$ \ \ & \ \ 2.55--2.78 \ \ & \ \ 2.16e-01 \ \ & \ \ 10 \ \ & \ \ 0.44\ \  & \ \ $^{+13}_{-13}$ \ \ & \ \ $^{+0.0}_{-0.0}$ \ \ & \ \ $^{+0.0}_{-0.0}$ \ \ & \ \ $^{+1.0}_{-1.0}$ \ \ & \ \ 0.5 \ \ & \ \ 0.0 \ \ & \ \ 0.2 \ \ &  & \\
\ \ $20$ \ \ & \ \ 2.78--4.27 \ \ & \ \ 2.07e-02 \ \ & \ \ 14 \ \ & \ \ 0.30\ \  & \ \ $^{+26}_{-25}$ \ \ & \ \ $^{+0.0}_{-0.0}$ \ \ & \ \ $^{+0.0}_{-0.0}$ \ \ & \ \ $^{+0.9}_{-0.8}$ \ \ & \ \ 1.3 \ \ & \ \ 0.0 \ \ & \ \ 0.2 \ \ &  & \\
\hline
\end{tabular}
}
\caption{
Measured double-differential dijet cross-sections for the range $\ystar < 0.5$ and jet radius parameter $R=0.4$. 
Here, $\sigma$ is the measured cross-section. 
All uncertainties are given in \%. 
The variable $\delta_\mathrm{stat}^\mathrm{data}$ ($\delta_\mathrm{stat}^\mathrm{MC}$) is the statistical uncertainty from the data (MC simulation). 
The $\gamma$ components show the uncertainties due to those on the jet energy calibration from the in situ, pileup, close-by jet, and flavour components. 
The $u$ components show the uncertainty for the jet energy and angular resolution, the unfolding, the quality selection, and the integrated luminosity. 
While all columns are uncorrelated with each other, the in situ, pileup, and flavour uncertainties shown here are the sum in quadrature of multiple uncorrelated components. 
The full set of cross-section values and uncertainty components, each of which is fully correlated in dijet mass and \ystar but uncorrelated with the other components, can be found in HepData \cite{Buckley:2010jn}. 
}
\label{tab:sysunc_r04_Eta1}
\end{table*}

\renewcommand{\arraystretch}{1}

\renewcommand{\arraystretch}{1.8}

\begin{table*}[!ht]
\tiny
\centering
{\fontfamily{ptm}\selectfont
\begin{tabular}{|@{}c@{}@{}c@{}@{}c@{}@{}c@{}@{}r@{}@{}r@{}@{}r@{}@{}r@{}@{}r@{}@{}r@{}@{}r@{}@{}r@{}|@{}r@{}@{}r@{}|}
\hline
\ \ \mass--\ystar \ \ & \ \ \mass range \ \ & \ \ $\sigma$ \ \ & \ \ $\delta_{\mathrm{stat}}^{\mathrm{data}}$ \ \ & \ \ $\delta_{\mathrm{stat}}^{\mathrm{MC}}$ \ \ & \ \ $\gamma_{\mathrm{in~situ}}$ \ \ & \ \ $\gamma_{\mathrm{pileup}}$ \ \ & \ \ $\gamma_{\operatorname{close-by}}$ \ \ & \ \ $\gamma_{\mathrm{flavour}}$ \ \ & \ \ u$_{\mathrm{JER}}$ \ \ & \ \ u$_{\mathrm{JAR}}$ \ \ & \ \ u$_{\mathrm{unfold}}$ \ \ & \ \ u$_{\mathrm{qual}}$ \ \ & \ \ u$_{\mathrm{lumi}}$ \ \ \\ \relax
\ \ bin \# \ \ & \ \ [TeV] \ \ & \ \ [pb/TeV] \ \ &\ \  \% \ \ & \ \ \% \ \ & \ \  \% \ \ & \ \ \% \ \ & \ \ \% \ \ & \ \ \% \ \ & \ \ \% \ \ & \ \ \% \ \ & \ \ \% \ \ & \ \ \% \ \ & \ \ \% \ \ \\
\hline
\ \ $21$ \ \ & \ \ 0.31--0.37 \ \ & \ \ 6.92e+05 \ \ & \ \ 0.52 \ \ & \ \ 0.79\ \  & \ \ $^{+6.4}_{-5.8}$ \ \ & \ \ $^{+1.2}_{-0.9}$ \ \ & \ \ $^{+2.0}_{-1.6}$ \ \ & \ \ $^{+5.4}_{-5.0}$ \ \ & \ \ 1.4 \ \ & \ \ 0.3 \ \ & \ \ 0.2 \ \ & \multirow{21}{*}{\ \ 0.5 \ \ } & \multirow{21}{*}{\ \ 1.8 \ \ } \\
\ \ $22$ \ \ & \ \ 0.37--0.44 \ \ & \ \ 2.64e+05 \ \ & \ \ 0.54 \ \ & \ \ 0.70\ \  & \ \ $^{+6.3}_{-6.0}$ \ \ & \ \ $^{+1.2}_{-1.1}$ \ \ & \ \ $^{+2.0}_{-1.7}$ \ \ & \ \ $^{+5.0}_{-4.7}$ \ \ & \ \ 1.0 \ \ & \ \ 0.3 \ \ & \ \ 0.1 \ \ &  & \\
\ \ $23$ \ \ & \ \ 0.44--0.51 \ \ & \ \ 1.09e+05 \ \ & \ \ 0.55 \ \ & \ \ 0.72\ \  & \ \ $^{+6.2}_{-6.3}$ \ \ & \ \ $^{+1.2}_{-1.1}$ \ \ & \ \ $^{+1.8}_{-1.8}$ \ \ & \ \ $^{+4.5}_{-4.4}$ \ \ & \ \ 0.7 \ \ & \ \ 0.2 \ \ & \ \ 0.0 \ \ &  & \\
\ \ $24$ \ \ & \ \ 0.51--0.59 \ \ & \ \ 4.59e+04 \ \ & \ \ 0.53 \ \ & \ \ 0.73\ \  & \ \ $^{+6.0}_{-6.4}$ \ \ & \ \ $^{+1.2}_{-1.1}$ \ \ & \ \ $^{+1.8}_{-1.9}$ \ \ & \ \ $^{+4.0}_{-4.1}$ \ \ & \ \ 0.6 \ \ & \ \ 0.2 \ \ & \ \ 0.0 \ \ &  & \\
\ \ $25$ \ \ & \ \ 0.59--0.67 \ \ & \ \ 2.02e+04 \ \ & \ \ 0.54 \ \ & \ \ 0.77\ \  & \ \ $^{+6.0}_{-6.2}$ \ \ & \ \ $^{+1.3}_{-1.2}$ \ \ & \ \ $^{+1.9}_{-2.0}$ \ \ & \ \ $^{+3.6}_{-3.8}$ \ \ & \ \ 0.7 \ \ & \ \ 0.2 \ \ & \ \ 0.0 \ \ &  & \\
\ \ $26$ \ \ & \ \ 0.67--0.76 \ \ & \ \ 9.49e+03 \ \ & \ \ 0.41 \ \ & \ \ 0.69\ \  & \ \ $^{+6.1}_{-6.0}$ \ \ & \ \ $^{+1.2}_{-1.2}$ \ \ & \ \ $^{+2.1}_{-2.0}$ \ \ & \ \ $^{+3.4}_{-3.4}$ \ \ & \ \ 0.7 \ \ & \ \ 0.3 \ \ & \ \ 0.0 \ \ &  & \\
\ \ $27$ \ \ & \ \ 0.76--0.85 \ \ & \ \ 4.54e+03 \ \ & \ \ 0.40 \ \ & \ \ 0.65\ \  & \ \ $^{+6.1}_{-5.9}$ \ \ & \ \ $^{+1.0}_{-1.1}$ \ \ & \ \ $^{+2.1}_{-2.1}$ \ \ & \ \ $^{+3.2}_{-3.1}$ \ \ & \ \ 0.6 \ \ & \ \ 0.2 \ \ & \ \ 0.0 \ \ &  & \\
\ \ $28$ \ \ & \ \ 0.85--0.95 \ \ & \ \ 2.24e+03 \ \ & \ \ 0.35 \ \ & \ \ 0.56\ \  & \ \ $^{+6.0}_{-5.8}$ \ \ & \ \ $^{+0.7}_{-0.7}$ \ \ & \ \ $^{+1.9}_{-1.9}$ \ \ & \ \ $^{+2.9}_{-2.8}$ \ \ & \ \ 0.5 \ \ & \ \ 0.2 \ \ & \ \ 0.0 \ \ &  & \\
\ \ $29$ \ \ & \ \ 0.95--1.06 \ \ & \ \ 1.11e+03 \ \ & \ \ 0.33 \ \ & \ \ 0.56\ \  & \ \ $^{+5.8}_{-5.6}$ \ \ & \ \ $^{+0.5}_{-0.4}$ \ \ & \ \ $^{+1.7}_{-1.7}$ \ \ & \ \ $^{+2.6}_{-2.5}$ \ \ & \ \ 0.5 \ \ & \ \ 0.1 \ \ & \ \ 0.0 \ \ &  & \\
\ \ $30$ \ \ & \ \ 1.06--1.18 \ \ & \ \ 5.38e+02 \ \ & \ \ 0.36 \ \ & \ \ 0.57\ \  & \ \ $^{+5.7}_{-5.6}$ \ \ & \ \ $^{+0.2}_{-0.1}$ \ \ & \ \ $^{+1.4}_{-1.4}$ \ \ & \ \ $^{+2.4}_{-2.3}$ \ \ & \ \ 0.6 \ \ & \ \ 0.1 \ \ & \ \ 0.0 \ \ &  & \\
\ \ $31$ \ \ & \ \ 1.18--1.31 \ \ & \ \ 2.61e+02 \ \ & \ \ 0.40 \ \ & \ \ 0.57\ \  & \ \ $^{+5.8}_{-5.6}$ \ \ & \ \ $^{+0.1}_{-0.1}$ \ \ & \ \ $^{+1.1}_{-1.0}$ \ \ & \ \ $^{+2.2}_{-2.1}$ \ \ & \ \ 0.6 \ \ & \ \ 0.1 \ \ & \ \ 0.0 \ \ &  & \\
\ \ $32$ \ \ & \ \ 1.31--1.45 \ \ & \ \ 1.26e+02 \ \ & \ \ 0.54 \ \ & \ \ 0.57\ \  & \ \ $^{+6.0}_{-5.8}$ \ \ & \ \ $^{+0.1}_{-0.0}$ \ \ & \ \ $^{+0.7}_{-0.7}$ \ \ & \ \ $^{+2.0}_{-1.9}$ \ \ & \ \ 0.6 \ \ & \ \ 0.1 \ \ & \ \ 0.0 \ \ &  & \\
\ \ $33$ \ \ & \ \ 1.45--1.60 \ \ & \ \ 6.01e+01 \ \ & \ \ 0.76 \ \ & \ \ 0.56\ \  & \ \ $^{+6.4}_{-6.1}$ \ \ & \ \ $^{+0.1}_{-0.1}$ \ \ & \ \ $^{+0.4}_{-0.4}$ \ \ & \ \ $^{+1.8}_{-1.7}$ \ \ & \ \ 0.6 \ \ & \ \ 0.1 \ \ & \ \ 0.0 \ \ &  & \\
\ \ $34$ \ \ & \ \ 1.60--1.76 \ \ & \ \ 2.86e+01 \ \ & \ \ 1.1 \ \ & \ \ 0.53\ \  & \ \ $^{+7.0}_{-6.7}$ \ \ & \ \ $^{+0.1}_{-0.1}$ \ \ & \ \ $^{+0.2}_{-0.2}$ \ \ & \ \ $^{+1.7}_{-1.6}$ \ \ & \ \ 0.6 \ \ & \ \ 0.1 \ \ & \ \ 0.0 \ \ &  & \\
\ \ $35$ \ \ & \ \ 1.76--1.94 \ \ & \ \ 1.33e+01 \ \ & \ \ 1.5 \ \ & \ \ 0.52\ \  & \ \ $^{+7.7}_{-7.4}$ \ \ & \ \ $^{+0.1}_{-0.1}$ \ \ & \ \ $^{+0.1}_{-0.1}$ \ \ & \ \ $^{+1.6}_{-1.5}$ \ \ & \ \ 0.6 \ \ & \ \ 0.1 \ \ & \ \ 0.0 \ \ &  & \\
\ \ $36$ \ \ & \ \ 1.94--2.12 \ \ & \ \ 6.05e+00 \ \ & \ \ 2.1 \ \ & \ \ 0.55\ \  & \ \ $^{+8.6}_{-8.2}$ \ \ & \ \ $^{+0.1}_{-0.1}$ \ \ & \ \ $^{+0.1}_{-0.1}$ \ \ & \ \ $^{+1.4}_{-1.4}$ \ \ & \ \ 0.6 \ \ & \ \ 0.1 \ \ & \ \ 0.0 \ \ &  & \\
\ \ $37$ \ \ & \ \ 2.12--2.33 \ \ & \ \ 2.68e+00 \ \ & \ \ 3.1 \ \ & \ \ 0.54\ \  & \ \ $^{+9.5}_{-9.1}$ \ \ & \ \ $^{+0.1}_{-0.1}$ \ \ & \ \ $^{+0.0}_{-0.0}$ \ \ & \ \ $^{+1.3}_{-1.3}$ \ \ & \ \ 0.7 \ \ & \ \ 0.1 \ \ & \ \ 0.0 \ \ &  & \\
\ \ $38$ \ \ & \ \ 2.33--2.55 \ \ & \ \ 1.03e+00 \ \ & \ \ 4.7 \ \ & \ \ 0.54\ \  & \ \ $^{+11}_{-10}$ \ \ & \ \ $^{+0.1}_{-0.1}$ \ \ & \ \ $^{+0.0}_{-0.0}$ \ \ & \ \ $^{+1.2}_{-1.2}$ \ \ & \ \ 0.8 \ \ & \ \ 0.1 \ \ & \ \ 0.0 \ \ &  & \\
\ \ $39$ \ \ & \ \ 2.55--2.78 \ \ & \ \ 4.52e-01 \ \ & \ \ 7.0 \ \ & \ \ 0.49\ \  & \ \ $^{+11}_{-11}$ \ \ & \ \ $^{+0.1}_{-0.1}$ \ \ & \ \ $^{+0.0}_{-0.0}$ \ \ & \ \ $^{+1.1}_{-1.1}$ \ \ & \ \ 0.8 \ \ & \ \ 0.1 \ \ & \ \ 0.1 \ \ &  & \\
\ \ $40$ \ \ & \ \ 2.78--3.04 \ \ & \ \ 1.67e-01 \ \ & \ \ 11 \ \ & \ \ 0.47\ \  & \ \ $^{+12}_{-12}$ \ \ & \ \ $^{+0.0}_{-0.1}$ \ \ & \ \ $^{+0.0}_{-0.0}$ \ \ & \ \ $^{+1.0}_{-1.0}$ \ \ & \ \ 0.9 \ \ & \ \ 0.1 \ \ & \ \ 1.0 \ \ &  & \\
\ \ $41$ \ \ & \ \ 3.04--4.27 \ \ & \ \ 1.72e-02 \ \ & \ \ 16 \ \ & \ \ 0.28\ \  & \ \ $^{+19}_{-19}$ \ \ & \ \ $^{+0.0}_{-0.0}$ \ \ & \ \ $^{+0.0}_{-0.0}$ \ \ & \ \ $^{+0.8}_{-0.9}$ \ \ & \ \ 1.7 \ \ & \ \ 0.0 \ \ & \ \ 1.8 \ \ &  & \\
\hline
\end{tabular}
}
\caption{
Measured double-differential dijet cross-sections for the range $0.5 \leq \ystar < 1.0$ and jet radius parameter $R=0.4$. 
Here, $\sigma$ is the measured cross-section. 
All uncertainties are given in \%. 
The variable $\delta_\mathrm{stat}^\mathrm{data}$ ($\delta_\mathrm{stat}^\mathrm{MC}$) is the statistical uncertainty from the data (MC simulation). 
The $\gamma$ components show the uncertainties due to those on the jet energy calibration from the in situ, pileup, close-by jet, and flavour components. 
The $u$ components show the uncertainty for the jet energy and angular resolution, the unfolding, the quality selection, and the integrated luminosity. 
While all columns are uncorrelated with each other, the in situ, pileup, and flavour uncertainties shown here are the sum in quadrature of multiple uncorrelated components. 
The full set of cross-section values and uncertainty components, each of which is fully correlated in dijet mass and \ystar but uncorrelated with the other components, can be found in HepData \cite{Buckley:2010jn}. 
}
\label{tab:sysunc_r04_Eta2}
\end{table*}

\renewcommand{\arraystretch}{1}

\renewcommand{\arraystretch}{1.8}

\begin{table*}[!ht]
\tiny
\centering
{\fontfamily{ptm}\selectfont
\begin{tabular}{|@{}c@{}@{}c@{}@{}c@{}@{}c@{}@{}r@{}@{}r@{}@{}r@{}@{}r@{}@{}r@{}@{}r@{}@{}r@{}@{}r@{}|@{}r@{}@{}r@{}|}
\hline
\ \ \mass--\ystar \ \ & \ \ \mass range \ \ & \ \ $\sigma$ \ \ & \ \ $\delta_{\mathrm{stat}}^{\mathrm{data}}$ \ \ & \ \ $\delta_{\mathrm{stat}}^{\mathrm{MC}}$ \ \ & \ \ $\gamma_{\mathrm{in~situ}}$ \ \ & \ \ $\gamma_{\mathrm{pileup}}$ \ \ & \ \ $\gamma_{\operatorname{close-by}}$ \ \ & \ \ $\gamma_{\mathrm{flavour}}$ \ \ & \ \ u$_{\mathrm{JER}}$ \ \ & \ \ u$_{\mathrm{JAR}}$ \ \ & \ \ u$_{\mathrm{unfold}}$ \ \ & \ \ u$_{\mathrm{qual}}$ \ \ & \ \ u$_{\mathrm{lumi}}$ \ \ \\ \relax
\ \ bin \# \ \ & \ \ [TeV] \ \ & \ \ [pb/TeV] \ \ &\ \  \% \ \ & \ \ \% \ \ & \ \  \% \ \ & \ \ \% \ \ & \ \ \% \ \ & \ \ \% \ \ & \ \ \% \ \ & \ \ \% \ \ & \ \ \% \ \ & \ \ \% \ \ & \ \ \% \ \ \\
\hline
\ \ $42$ \ \ & \ \ 0.51--0.59 \ \ & \ \ 1.31e+05 \ \ & \ \ 0.82 \ \ & \ \ 0.79\ \  & \ \ $^{+8.3}_{-8.5}$ \ \ & \ \ $^{+1.2}_{-1.1}$ \ \ & \ \ $^{+1.6}_{-1.8}$ \ \ & \ \ $^{+4.8}_{-5.0}$ \ \ & \ \ 1.3 \ \ & \ \ 0.1 \ \ & \ \ 0.3 \ \ & \multirow{19}{*}{\ \ 0.5 \ \ } & \multirow{19}{*}{\ \ 1.8 \ \ } \\
\ \ $43$ \ \ & \ \ 0.59--0.67 \ \ & \ \ 5.92e+04 \ \ & \ \ 0.86 \ \ & \ \ 0.80\ \  & \ \ $^{+8.6}_{-8.6}$ \ \ & \ \ $^{+1.0}_{-1.1}$ \ \ & \ \ $^{+1.6}_{-1.9}$ \ \ & \ \ $^{+4.6}_{-4.6}$ \ \ & \ \ 1.2 \ \ & \ \ 0.1 \ \ & \ \ 0.1 \ \ &  & \\
\ \ $44$ \ \ & \ \ 0.67--0.76 \ \ & \ \ 2.73e+04 \ \ & \ \ 0.90 \ \ & \ \ 0.73\ \  & \ \ $^{+8.8}_{-8.7}$ \ \ & \ \ $^{+1.0}_{-1.2}$ \ \ & \ \ $^{+1.7}_{-1.9}$ \ \ & \ \ $^{+4.4}_{-4.3}$ \ \ & \ \ 1.0 \ \ & \ \ 0.1 \ \ & \ \ 0.0 \ \ &  & \\
\ \ $45$ \ \ & \ \ 0.76--0.85 \ \ & \ \ 1.28e+04 \ \ & \ \ 0.93 \ \ & \ \ 0.80\ \  & \ \ $^{+8.9}_{-8.7}$ \ \ & \ \ $^{+0.9}_{-1.2}$ \ \ & \ \ $^{+1.9}_{-2.0}$ \ \ & \ \ $^{+4.1}_{-4.0}$ \ \ & \ \ 0.9 \ \ & \ \ 0.1 \ \ & \ \ 0.0 \ \ &  & \\
\ \ $46$ \ \ & \ \ 0.85--0.95 \ \ & \ \ 6.60e+03 \ \ & \ \ 0.89 \ \ & \ \ 0.75\ \  & \ \ $^{+9.0}_{-8.7}$ \ \ & \ \ $^{+0.9}_{-1.1}$ \ \ & \ \ $^{+2.0}_{-2.0}$ \ \ & \ \ $^{+3.8}_{-3.6}$ \ \ & \ \ 0.9 \ \ & \ \ 0.2 \ \ & \ \ 0.0 \ \ &  & \\
\ \ $47$ \ \ & \ \ 0.95--1.06 \ \ & \ \ 3.18e+03 \ \ & \ \ 0.83 \ \ & \ \ 0.77\ \  & \ \ $^{+9.1}_{-8.6}$ \ \ & \ \ $^{+0.8}_{-1.1}$ \ \ & \ \ $^{+2.0}_{-1.9}$ \ \ & \ \ $^{+3.6}_{-3.3}$ \ \ & \ \ 0.9 \ \ & \ \ 0.2 \ \ & \ \ 0.0 \ \ &  & \\
\ \ $48$ \ \ & \ \ 1.06--1.18 \ \ & \ \ 1.53e+03 \ \ & \ \ 0.69 \ \ & \ \ 0.67\ \  & \ \ $^{+8.9}_{-8.5}$ \ \ & \ \ $^{+0.8}_{-1.0}$ \ \ & \ \ $^{+2.1}_{-2.0}$ \ \ & \ \ $^{+3.3}_{-3.0}$ \ \ & \ \ 0.9 \ \ & \ \ 0.3 \ \ & \ \ 0.0 \ \ &  & \\
\ \ $49$ \ \ & \ \ 1.18--1.31 \ \ & \ \ 7.43e+02 \ \ & \ \ 0.74 \ \ & \ \ 0.59\ \  & \ \ $^{+8.7}_{-8.5}$ \ \ & \ \ $^{+0.6}_{-0.9}$ \ \ & \ \ $^{+2.1}_{-2.0}$ \ \ & \ \ $^{+2.9}_{-2.8}$ \ \ & \ \ 0.8 \ \ & \ \ 0.3 \ \ & \ \ 0.0 \ \ &  & \\
\ \ $50$ \ \ & \ \ 1.31--1.45 \ \ & \ \ 3.54e+02 \ \ & \ \ 0.66 \ \ & \ \ 0.54\ \  & \ \ $^{+8.5}_{-8.4}$ \ \ & \ \ $^{+0.4}_{-0.7}$ \ \ & \ \ $^{+2.0}_{-1.9}$ \ \ & \ \ $^{+2.7}_{-2.6}$ \ \ & \ \ 0.8 \ \ & \ \ 0.2 \ \ & \ \ 0.0 \ \ &  & \\
\ \ $51$ \ \ & \ \ 1.45--1.60 \ \ & \ \ 1.72e+02 \ \ & \ \ 0.68 \ \ & \ \ 0.53\ \  & \ \ $^{+8.4}_{-8.3}$ \ \ & \ \ $^{+0.2}_{-0.4}$ \ \ & \ \ $^{+1.8}_{-1.7}$ \ \ & \ \ $^{+2.4}_{-2.4}$ \ \ & \ \ 0.7 \ \ & \ \ 0.2 \ \ & \ \ 0.0 \ \ &  & \\
\ \ $52$ \ \ & \ \ 1.60--1.76 \ \ & \ \ 8.30e+01 \ \ & \ \ 0.78 \ \ & \ \ 0.53\ \  & \ \ $^{+8.5}_{-8.3}$ \ \ & \ \ $^{+0.1}_{-0.2}$ \ \ & \ \ $^{+1.5}_{-1.4}$ \ \ & \ \ $^{+2.2}_{-2.2}$ \ \ & \ \ 0.7 \ \ & \ \ 0.1 \ \ & \ \ 0.0 \ \ &  & \\
\ \ $53$ \ \ & \ \ 1.76--1.94 \ \ & \ \ 3.80e+01 \ \ & \ \ 0.91 \ \ & \ \ 0.51\ \  & \ \ $^{+8.7}_{-8.3}$ \ \ & \ \ $^{+0.0}_{-0.1}$ \ \ & \ \ $^{+1.1}_{-1.1}$ \ \ & \ \ $^{+2.1}_{-2.1}$ \ \ & \ \ 0.7 \ \ & \ \ 0.1 \ \ & \ \ 0.0 \ \ &  & \\
\ \ $54$ \ \ & \ \ 1.94--2.12 \ \ & \ \ 1.67e+01 \ \ & \ \ 1.3 \ \ & \ \ 0.55\ \  & \ \ $^{+9.0}_{-8.6}$ \ \ & \ \ $^{+0.0}_{-0.1}$ \ \ & \ \ $^{+0.8}_{-0.8}$ \ \ & \ \ $^{+2.0}_{-1.9}$ \ \ & \ \ 0.8 \ \ & \ \ 0.1 \ \ & \ \ 0.0 \ \ &  & \\
\ \ $55$ \ \ & \ \ 2.12--2.33 \ \ & \ \ 7.74e+00 \ \ & \ \ 1.8 \ \ & \ \ 0.52\ \  & \ \ $^{+9.4}_{-9.0}$ \ \ & \ \ $^{+0.1}_{-0.1}$ \ \ & \ \ $^{+0.5}_{-0.5}$ \ \ & \ \ $^{+1.8}_{-1.8}$ \ \ & \ \ 0.9 \ \ & \ \ 0.1 \ \ & \ \ 0.0 \ \ &  & \\
\ \ $56$ \ \ & \ \ 2.33--2.55 \ \ & \ \ 3.35e+00 \ \ & \ \ 2.6 \ \ & \ \ 0.44\ \  & \ \ $^{+10}_{-9.6}$ \ \ & \ \ $^{+0.1}_{-0.1}$ \ \ & \ \ $^{+0.3}_{-0.3}$ \ \ & \ \ $^{+1.7}_{-1.6}$ \ \ & \ \ 0.9 \ \ & \ \ 0.1 \ \ & \ \ 0.0 \ \ &  & \\
\ \ $57$ \ \ & \ \ 2.55--2.78 \ \ & \ \ 1.29e+00 \ \ & \ \ 4.0 \ \ & \ \ 0.46\ \  & \ \ $^{+11}_{-10}$ \ \ & \ \ $^{+0.1}_{-0.1}$ \ \ & \ \ $^{+0.2}_{-0.2}$ \ \ & \ \ $^{+1.5}_{-1.5}$ \ \ & \ \ 1.0 \ \ & \ \ 0.1 \ \ & \ \ 0.0 \ \ &  & \\
\ \ $58$ \ \ & \ \ 2.78--3.04 \ \ & \ \ 4.39e-01 \ \ & \ \ 6.4 \ \ & \ \ 0.45\ \  & \ \ $^{+12}_{-11}$ \ \ & \ \ $^{+0.1}_{-0.1}$ \ \ & \ \ $^{+0.1}_{-0.1}$ \ \ & \ \ $^{+1.4}_{-1.4}$ \ \ & \ \ 1.1 \ \ & \ \ 0.1 \ \ & \ \ 0.1 \ \ &  & \\
\ \ $59$ \ \ & \ \ 3.04--3.31 \ \ & \ \ 1.72e-01 \ \ & \ \ 11 \ \ & \ \ 0.42\ \  & \ \ $^{+13}_{-12}$ \ \ & \ \ $^{+0.1}_{-0.1}$ \ \ & \ \ $^{+0.1}_{-0.1}$ \ \ & \ \ $^{+1.3}_{-1.3}$ \ \ & \ \ 1.2 \ \ & \ \ 0.1 \ \ & \ \ 0.9 \ \ &  & \\
\ \ $60$ \ \ & \ \ 3.31--4.64 \ \ & \ \ 2.11e-02 \ \ & \ \ 14 \ \ & \ \ 0.28\ \  & \ \ $^{+19}_{-18}$ \ \ & \ \ $^{+0.0}_{-0.0}$ \ \ & \ \ $^{+0.0}_{-0.0}$ \ \ & \ \ $^{+1.0}_{-0.9}$ \ \ & \ \ 1.7 \ \ & \ \ 0.0 \ \ & \ \ 1.5 \ \ &  & \\
\hline
\end{tabular}
}
\caption{
Measured double-differential dijet cross-sections for the range $1.0 \leq \ystar < 1.5$ and jet radius parameter $R=0.4$. 
Here, $\sigma$ is the measured cross-section. 
All uncertainties are given in \%. 
The variable $\delta_\mathrm{stat}^\mathrm{data}$ ($\delta_\mathrm{stat}^\mathrm{MC}$) is the statistical uncertainty from the data (MC simulation). 
The $\gamma$ components show the uncertainties due to those on the jet energy calibration from the in situ, pileup, close-by jet, and flavour components. 
The $u$ components show the uncertainty for the jet energy and angular resolution, the unfolding, the quality selection, and the integrated luminosity. 
While all columns are uncorrelated with each other, the in situ, pileup, and flavour uncertainties shown here are the sum in quadrature of multiple uncorrelated components. 
The full set of cross-section values and uncertainty components, each of which is fully correlated in dijet mass and \ystar but uncorrelated with the other components, can be found in HepData \cite{Buckley:2010jn}. 
}
\label{tab:sysunc_r04_Eta3}
\end{table*}

\renewcommand{\arraystretch}{1}

\renewcommand{\arraystretch}{1.8}

\begin{table*}[!ht]
\tiny
\centering
{\fontfamily{ptm}\selectfont
\begin{tabular}{|@{}c@{}@{}c@{}@{}c@{}@{}c@{}@{}r@{}@{}r@{}@{}r@{}@{}r@{}@{}r@{}@{}r@{}@{}r@{}@{}r@{}|@{}r@{}@{}r@{}|}
\hline
\ \ \mass--\ystar \ \ & \ \ \mass range \ \ & \ \ $\sigma$ \ \ & \ \ $\delta_{\mathrm{stat}}^{\mathrm{data}}$ \ \ & \ \ $\delta_{\mathrm{stat}}^{\mathrm{MC}}$ \ \ & \ \ $\gamma_{\mathrm{in~situ}}$ \ \ & \ \ $\gamma_{\mathrm{pileup}}$ \ \ & \ \ $\gamma_{\operatorname{close-by}}$ \ \ & \ \ $\gamma_{\mathrm{flavour}}$ \ \ & \ \ u$_{\mathrm{JER}}$ \ \ & \ \ u$_{\mathrm{JAR}}$ \ \ & \ \ u$_{\mathrm{unfold}}$ \ \ & \ \ u$_{\mathrm{qual}}$ \ \ & \ \ u$_{\mathrm{lumi}}$ \ \ \\ \relax
\ \ bin \# \ \ & \ \ [TeV] \ \ & \ \ [pb/TeV] \ \ &\ \  \% \ \ & \ \ \% \ \ & \ \  \% \ \ & \ \ \% \ \ & \ \ \% \ \ & \ \ \% \ \ & \ \ \% \ \ & \ \ \% \ \ & \ \ \% \ \ & \ \ \% \ \ & \ \ \% \ \ \\
\hline
\ \ $61$ \ \ & \ \ 0.76--0.85 \ \ & \ \ 3.60e+04 \ \ & \ \ 1.4 \ \ & \ \ 0.97\ \  & \ \ $^{+13}_{-12}$ \ \ & \ \ $^{+1.4}_{-1.7}$ \ \ & \ \ $^{+1.8}_{-2.1}$ \ \ & \ \ $^{+4.9}_{-4.8}$ \ \ & \ \ 1.5 \ \ & \ \ 0.3 \ \ & \ \ 0.2 \ \ & \multirow{17}{*}{\ \ 0.5 \ \ } & \multirow{17}{*}{\ \ 1.8 \ \ } \\
\ \ $62$ \ \ & \ \ 0.85--0.95 \ \ & \ \ 1.82e+04 \ \ & \ \ 1.7 \ \ & \ \ 0.99\ \  & \ \ $^{+13}_{-12}$ \ \ & \ \ $^{+1.5}_{-1.5}$ \ \ & \ \ $^{+1.8}_{-2.0}$ \ \ & \ \ $^{+4.6}_{-4.6}$ \ \ & \ \ 1.4 \ \ & \ \ 0.3 \ \ & \ \ 0.1 \ \ &  & \\
\ \ $63$ \ \ & \ \ 0.95--1.06 \ \ & \ \ 9.01e+03 \ \ & \ \ 1.8 \ \ & \ \ 0.87\ \  & \ \ $^{+13}_{-13}$ \ \ & \ \ $^{+1.5}_{-1.3}$ \ \ & \ \ $^{+1.8}_{-1.9}$ \ \ & \ \ $^{+4.4}_{-4.3}$ \ \ & \ \ 1.4 \ \ & \ \ 0.3 \ \ & \ \ 0.0 \ \ &  & \\
\ \ $64$ \ \ & \ \ 1.06--1.18 \ \ & \ \ 4.32e+03 \ \ & \ \ 1.9 \ \ & \ \ 0.87\ \  & \ \ $^{+14}_{-13}$ \ \ & \ \ $^{+1.4}_{-1.3}$ \ \ & \ \ $^{+1.8}_{-1.8}$ \ \ & \ \ $^{+4.2}_{-4.0}$ \ \ & \ \ 1.3 \ \ & \ \ 0.3 \ \ & \ \ 0.0 \ \ &  & \\
\ \ $65$ \ \ & \ \ 1.18--1.31 \ \ & \ \ 2.17e+03 \ \ & \ \ 2.0 \ \ & \ \ 0.78\ \  & \ \ $^{+14}_{-13}$ \ \ & \ \ $^{+1.3}_{-1.3}$ \ \ & \ \ $^{+1.9}_{-1.9}$ \ \ & \ \ $^{+3.9}_{-3.7}$ \ \ & \ \ 1.3 \ \ & \ \ 0.3 \ \ & \ \ 0.0 \ \ &  & \\
\ \ $66$ \ \ & \ \ 1.31--1.45 \ \ & \ \ 9.98e+02 \ \ & \ \ 1.9 \ \ & \ \ 0.72\ \  & \ \ $^{+15}_{-13}$ \ \ & \ \ $^{+1.3}_{-1.4}$ \ \ & \ \ $^{+2.0}_{-2.0}$ \ \ & \ \ $^{+3.7}_{-3.4}$ \ \ & \ \ 1.3 \ \ & \ \ 0.2 \ \ & \ \ 0.0 \ \ &  & \\
\ \ $67$ \ \ & \ \ 1.45--1.60 \ \ & \ \ 4.94e+02 \ \ & \ \ 2.2 \ \ & \ \ 0.69\ \  & \ \ $^{+15}_{-14}$ \ \ & \ \ $^{+1.3}_{-1.4}$ \ \ & \ \ $^{+2.1}_{-2.1}$ \ \ & \ \ $^{+3.4}_{-3.2}$ \ \ & \ \ 1.3 \ \ & \ \ 0.2 \ \ & \ \ 0.0 \ \ &  & \\
\ \ $68$ \ \ & \ \ 1.60--1.76 \ \ & \ \ 2.42e+02 \ \ & \ \ 1.9 \ \ & \ \ 0.65\ \  & \ \ $^{+15}_{-14}$ \ \ & \ \ $^{+1.3}_{-1.3}$ \ \ & \ \ $^{+2.2}_{-2.1}$ \ \ & \ \ $^{+3.1}_{-3.0}$ \ \ & \ \ 1.3 \ \ & \ \ 0.2 \ \ & \ \ 0.0 \ \ &  & \\
\ \ $69$ \ \ & \ \ 1.76--1.94 \ \ & \ \ 1.11e+02 \ \ & \ \ 1.9 \ \ & \ \ 0.62\ \  & \ \ $^{+15}_{-14}$ \ \ & \ \ $^{+1.1}_{-1.1}$ \ \ & \ \ $^{+2.3}_{-2.1}$ \ \ & \ \ $^{+2.9}_{-2.8}$ \ \ & \ \ 1.3 \ \ & \ \ 0.1 \ \ & \ \ 0.0 \ \ &  & \\
\ \ $70$ \ \ & \ \ 1.94--2.12 \ \ & \ \ 4.99e+01 \ \ & \ \ 2.2 \ \ & \ \ 0.66\ \  & \ \ $^{+15}_{-14}$ \ \ & \ \ $^{+0.9}_{-0.9}$ \ \ & \ \ $^{+2.2}_{-2.0}$ \ \ & \ \ $^{+2.8}_{-2.7}$ \ \ & \ \ 1.4 \ \ & \ \ 0.1 \ \ & \ \ 0.0 \ \ &  & \\
\ \ $71$ \ \ & \ \ 2.12--2.33 \ \ & \ \ 2.28e+01 \ \ & \ \ 2.0 \ \ & \ \ 0.60\ \  & \ \ $^{+15}_{-14}$ \ \ & \ \ $^{+0.7}_{-0.6}$ \ \ & \ \ $^{+2.1}_{-1.9}$ \ \ & \ \ $^{+2.6}_{-2.6}$ \ \ & \ \ 1.5 \ \ & \ \ 0.1 \ \ & \ \ 0.0 \ \ &  & \\
\ \ $72$ \ \ & \ \ 2.33--2.55 \ \ & \ \ 8.88e+00 \ \ & \ \ 2.4 \ \ & \ \ 0.65\ \  & \ \ $^{+16}_{-14}$ \ \ & \ \ $^{+0.5}_{-0.4}$ \ \ & \ \ $^{+1.9}_{-1.7}$ \ \ & \ \ $^{+2.6}_{-2.5}$ \ \ & \ \ 1.5 \ \ & \ \ 0.2 \ \ & \ \ 0.0 \ \ &  & \\
\ \ $73$ \ \ & \ \ 2.55--2.78 \ \ & \ \ 3.64e+00 \ \ & \ \ 3.1 \ \ & \ \ 0.64\ \  & \ \ $^{+16}_{-15}$ \ \ & \ \ $^{+0.3}_{-0.2}$ \ \ & \ \ $^{+1.6}_{-1.4}$ \ \ & \ \ $^{+2.5}_{-2.4}$ \ \ & \ \ 1.6 \ \ & \ \ 0.2 \ \ & \ \ 0.0 \ \ &  & \\
\ \ $74$ \ \ & \ \ 2.78--3.04 \ \ & \ \ 1.45e+00 \ \ & \ \ 4.1 \ \ & \ \ 0.56\ \  & \ \ $^{+17}_{-15}$ \ \ & \ \ $^{+0.1}_{-0.1}$ \ \ & \ \ $^{+1.3}_{-1.1}$ \ \ & \ \ $^{+2.4}_{-2.3}$ \ \ & \ \ 1.7 \ \ & \ \ 0.2 \ \ & \ \ 0.0 \ \ &  & \\
\ \ $75$ \ \ & \ \ 3.04--3.31 \ \ & \ \ 5.47e-01 \ \ & \ \ 6.0 \ \ & \ \ 0.51\ \  & \ \ $^{+18}_{-16}$ \ \ & \ \ $^{+0.1}_{-0.1}$ \ \ & \ \ $^{+1.0}_{-0.9}$ \ \ & \ \ $^{+2.3}_{-2.2}$ \ \ & \ \ 1.9 \ \ & \ \ 0.2 \ \ & \ \ 0.1 \ \ &  & \\
\ \ $76$ \ \ & \ \ 3.31--3.61 \ \ & \ \ 1.64e-01 \ \ & \ \ 10 \ \ & \ \ 0.59\ \  & \ \ $^{+19}_{-16}$ \ \ & \ \ $^{+0.1}_{-0.1}$ \ \ & \ \ $^{+0.8}_{-0.7}$ \ \ & \ \ $^{+2.2}_{-2.2}$ \ \ & \ \ 2.1 \ \ & \ \ 0.2 \ \ & \ \ 0.5 \ \ &  & \\
\ \ $77$ \ \ & \ \ 3.61--4.64 \ \ & \ \ 1.42e-02 \ \ & \ \ 20 \ \ & \ \ 0.51\ \  & \ \ $^{+24}_{-21}$ \ \ & \ \ $^{+0.1}_{-0.1}$ \ \ & \ \ $^{+0.0}_{-0.0}$ \ \ & \ \ $^{+1.7}_{-1.9}$ \ \ & \ \ 3.1 \ \ & \ \ 0.1 \ \ & \ \ 1.4 \ \ &  & \\
\hline
\end{tabular}
}
\caption{
Measured double-differential dijet cross-sections for the range $1.5 \leq \ystar < 2.0$ and jet radius parameter $R=0.4$. 
Here, $\sigma$ is the measured cross-section. 
All uncertainties are given in \%. 
The variable $\delta_\mathrm{stat}^\mathrm{data}$ ($\delta_\mathrm{stat}^\mathrm{MC}$) is the statistical uncertainty from the data (MC simulation). 
The $\gamma$ components show the uncertainties due to those on the jet energy calibration from the in situ, pileup, close-by jet, and flavour components. 
The $u$ components show the uncertainty for the jet energy and angular resolution, the unfolding, the quality selection, and the integrated luminosity. 
While all columns are uncorrelated with each other, the in situ, pileup, and flavour uncertainties shown here are the sum in quadrature of multiple uncorrelated components. 
The full set of cross-section values and uncertainty components, each of which is fully correlated in dijet mass and \ystar but uncorrelated with the other components, can be found in HepData \cite{Buckley:2010jn}. 
}
\label{tab:sysunc_r04_Eta4}
\end{table*}

\renewcommand{\arraystretch}{1}

\renewcommand{\arraystretch}{1.8}

\begin{table*}[!ht]
\tiny
\centering
{\fontfamily{ptm}\selectfont
\begin{tabular}{|@{}c@{}@{}c@{}@{}c@{}@{}c@{}@{}r@{}@{}r@{}@{}r@{}@{}r@{}@{}r@{}@{}r@{}@{}r@{}@{}r@{}|@{}r@{}@{}r@{}|}
\hline
\ \ \mass--\ystar \ \ & \ \ \mass range \ \ & \ \ $\sigma$ \ \ & \ \ $\delta_{\mathrm{stat}}^{\mathrm{data}}$ \ \ & \ \ $\delta_{\mathrm{stat}}^{\mathrm{MC}}$ \ \ & \ \ $\gamma_{\mathrm{in~situ}}$ \ \ & \ \ $\gamma_{\mathrm{pileup}}$ \ \ & \ \ $\gamma_{\operatorname{close-by}}$ \ \ & \ \ $\gamma_{\mathrm{flavour}}$ \ \ & \ \ u$_{\mathrm{JER}}$ \ \ & \ \ u$_{\mathrm{JAR}}$ \ \ & \ \ u$_{\mathrm{unfold}}$ \ \ & \ \ u$_{\mathrm{qual}}$ \ \ & \ \ u$_{\mathrm{lumi}}$ \ \ \\ \relax
\ \ bin \# \ \ & \ \ [TeV] \ \ & \ \ [pb/TeV] \ \ &\ \  \% \ \ & \ \ \% \ \ & \ \  \% \ \ & \ \ \% \ \ & \ \ \% \ \ & \ \ \% \ \ & \ \ \% \ \ & \ \ \% \ \ & \ \ \% \ \ & \ \ \% \ \ & \ \ \% \ \ \\
\hline
\ \ $78$ \ \ & \ \ 1.31--1.45 \ \ & \ \ 2.48e+03 \ \ & \ \ 3.9 \ \ & \ \ 1.1\ \  & \ \ $^{+21}_{-18}$ \ \ & \ \ $^{+1.8}_{-1.9}$ \ \ & \ \ $^{+1.9}_{-1.8}$ \ \ & \ \ $^{+4.2}_{-3.8}$ \ \ & \ \ 2.2 \ \ & \ \ 0.5 \ \ & \ \ 0.1 \ \ & \multirow{8}{*}{\ \ 0.5 \ \ } & \multirow{8}{*}{\ \ 1.8 \ \ } \\
\ \ $79$ \ \ & \ \ 1.45--1.60 \ \ & \ \ 1.29e+03 \ \ & \ \ 4.6 \ \ & \ \ 1.2\ \  & \ \ $^{+21}_{-19}$ \ \ & \ \ $^{+1.7}_{-1.9}$ \ \ & \ \ $^{+1.8}_{-1.8}$ \ \ & \ \ $^{+3.9}_{-3.6}$ \ \ & \ \ 2.0 \ \ & \ \ 0.6 \ \ & \ \ 0.1 \ \ &  & \\
\ \ $80$ \ \ & \ \ 1.60--1.76 \ \ & \ \ 6.61e+02 \ \ & \ \ 4.8 \ \ & \ \ 1.0\ \  & \ \ $^{+21}_{-19}$ \ \ & \ \ $^{+1.7}_{-1.8}$ \ \ & \ \ $^{+1.8}_{-1.8}$ \ \ & \ \ $^{+3.6}_{-3.3}$ \ \ & \ \ 1.9 \ \ & \ \ 0.5 \ \ & \ \ 0.1 \ \ &  & \\
\ \ $81$ \ \ & \ \ 1.76--2.12 \ \ & \ \ 2.05e+02 \ \ & \ \ 4.5 \ \ & \ \ 0.69\ \  & \ \ $^{+22}_{-20}$ \ \ & \ \ $^{+1.8}_{-1.7}$ \ \ & \ \ $^{+1.9}_{-1.9}$ \ \ & \ \ $^{+3.2}_{-3.0}$ \ \ & \ \ 1.8 \ \ & \ \ 0.4 \ \ & \ \ 0.0 \ \ &  & \\
\ \ $82$ \ \ & \ \ 2.12--2.55 \ \ & \ \ 4.03e+01 \ \ & \ \ 5.8 \ \ & \ \ 0.65\ \  & \ \ $^{+25}_{-22}$ \ \ & \ \ $^{+2.1}_{-1.9}$ \ \ & \ \ $^{+2.2}_{-2.0}$ \ \ & \ \ $^{+2.9}_{-2.6}$ \ \ & \ \ 1.9 \ \ & \ \ 0.2 \ \ & \ \ 0.1 \ \ &  & \\
\ \ $83$ \ \ & \ \ 2.55--3.04 \ \ & \ \ 6.83e+00 \ \ & \ \ 6.8 \ \ & \ \ 0.53\ \  & \ \ $^{+28}_{-24}$ \ \ & \ \ $^{+2.2}_{-2.1}$ \ \ & \ \ $^{+2.4}_{-2.3}$ \ \ & \ \ $^{+2.7}_{-2.5}$ \ \ & \ \ 2.4 \ \ & \ \ 0.4 \ \ & \ \ 0.1 \ \ &  & \\
\ \ $84$ \ \ & \ \ 3.04--3.61 \ \ & \ \ 8.77e-01 \ \ & \ \ 10 \ \ & \ \ 0.65\ \  & \ \ $^{+31}_{-25}$ \ \ & \ \ $^{+2.1}_{-1.8}$ \ \ & \ \ $^{+2.5}_{-2.3}$ \ \ & \ \ $^{+2.6}_{-2.5}$ \ \ & \ \ 3.2 \ \ & \ \ 0.7 \ \ & \ \ 0.0 \ \ &  & \\
\ \ $85$ \ \ & \ \ 3.61--5.04 \ \ & \ \ 3.31e-02 \ \ & \ \ 18 \ \ & \ \ 1.2\ \  & \ \ $^{+42}_{-31}$ \ \ & \ \ $^{+0.9}_{-0.9}$ \ \ & \ \ $^{+2.1}_{-1.1}$ \ \ & \ \ $^{+2.5}_{-2.9}$ \ \ & \ \ 6.2 \ \ & \ \ 0.9 \ \ & \ \ 2.3 \ \ &  & \\
\hline
\end{tabular}
}
\caption{
Measured double-differential dijet cross-sections for the range $2.0 \leq \ystar < 2.5$ and jet radius parameter $R=0.4$. 
Here, $\sigma$ is the measured cross-section. 
All uncertainties are given in \%. 
The variable $\delta_\mathrm{stat}^\mathrm{data}$ ($\delta_\mathrm{stat}^\mathrm{MC}$) is the statistical uncertainty from the data (MC simulation). 
The $\gamma$ components show the uncertainties due to those on the jet energy calibration from the in situ, pileup, close-by jet, and flavour components. 
The $u$ components show the uncertainty for the jet energy and angular resolution, the unfolding, the quality selection, and the integrated luminosity. 
While all columns are uncorrelated with each other, the in situ, pileup, and flavour uncertainties shown here are the sum in quadrature of multiple uncorrelated components. 
The full set of cross-section values and uncertainty components, each of which is fully correlated in dijet mass and \ystar but uncorrelated with the other components, can be found in HepData \cite{Buckley:2010jn}. 
}
\label{tab:sysunc_r04_Eta5}
\end{table*}

\renewcommand{\arraystretch}{1}

\renewcommand{\arraystretch}{1.8}

\begin{table*}[!ht]
\tiny
\centering
{\fontfamily{ptm}\selectfont
\begin{tabular}{|@{}c@{}@{}c@{}@{}c@{}@{}c@{}@{}r@{}@{}r@{}@{}r@{}@{}r@{}@{}r@{}@{}r@{}@{}r@{}@{}r@{}|@{}r@{}@{}r@{}|}
\hline
\ \ \mass--\ystar \ \ & \ \ \mass range \ \ & \ \ $\sigma$ \ \ & \ \ $\delta_{\mathrm{stat}}^{\mathrm{data}}$ \ \ & \ \ $\delta_{\mathrm{stat}}^{\mathrm{MC}}$ \ \ & \ \ $\gamma_{\mathrm{in~situ}}$ \ \ & \ \ $\gamma_{\mathrm{pileup}}$ \ \ & \ \ $\gamma_{\operatorname{close-by}}$ \ \ & \ \ $\gamma_{\mathrm{flavour}}$ \ \ & \ \ u$_{\mathrm{JER}}$ \ \ & \ \ u$_{\mathrm{JAR}}$ \ \ & \ \ u$_{\mathrm{unfold}}$ \ \ & \ \ u$_{\mathrm{qual}}$ \ \ & \ \ u$_{\mathrm{lumi}}$ \ \ \\ \relax
\ \ bin \# \ \ & \ \ [TeV] \ \ & \ \ [pb/TeV] \ \ &\ \  \% \ \ & \ \ \% \ \ & \ \  \% \ \ & \ \ \% \ \ & \ \ \% \ \ & \ \ \% \ \ & \ \ \% \ \ & \ \ \% \ \ & \ \ \% \ \ & \ \ \% \ \ & \ \ \% \ \ \\
\hline
\ \ $86$ \ \ & \ \ 2.12--2.55 \ \ & \ \ 5.73e+01 \ \ & \ \ 16 \ \ & \ \ 1.2\ \  & \ \ $^{+32}_{-26}$ \ \ & \ \ $^{+2.1}_{-1.8}$ \ \ & \ \ $^{+1.8}_{-1.3}$ \ \ & \ \ $^{+3.0}_{-2.5}$ \ \ & \ \ 2.8 \ \ & \ \ 0.3 \ \ & \ \ 0.2 \ \ & \multirow{4}{*}{\ \ 0.5 \ \ } & \multirow{4}{*}{\ \ 1.8 \ \ } \\
\ \ $87$ \ \ & \ \ 2.55--3.04 \ \ & \ \ 1.28e+01 \ \ & \ \ 20 \ \ & \ \ 1.3\ \  & \ \ $^{+37}_{-29}$ \ \ & \ \ $^{+2.9}_{-2.2}$ \ \ & \ \ $^{+2.1}_{-1.8}$ \ \ & \ \ $^{+2.8}_{-2.0}$ \ \ & \ \ 3.5 \ \ & \ \ 0.3 \ \ & \ \ 0.1 \ \ &  & \\
\ \ $88$ \ \ & \ \ 3.04--3.93 \ \ & \ \ 1.06e+00 \ \ & \ \ 28 \ \ & \ \ 1.4\ \  & \ \ $^{+46}_{-34}$ \ \ & \ \ $^{+4.5}_{-3.0}$ \ \ & \ \ $^{+2.8}_{-2.5}$ \ \ & \ \ $^{+2.4}_{-1.9}$ \ \ & \ \ 5.0 \ \ & \ \ 0.2 \ \ & \ \ 0.4 \ \ &  & \\
\ \ $89$ \ \ & \ \ 3.93--5.04 \ \ & \ \ 1.90e-02 \ \ & \ \ 70 \ \ & \ \ 2.1\ \  & \ \ $^{+73}_{-48}$ \ \ & \ \ $^{+7.0}_{-5.1}$ \ \ & \ \ $^{+4.2}_{-3.1}$ \ \ & \ \ $^{+2.3}_{-2.4}$ \ \ & \ \ 10 \ \ & \ \ 0.5 \ \ & \ \ 1.7 \ \ &  & \\
\hline
\end{tabular}
}
\caption{
Measured double-differential dijet cross-sections for the range $2.5 \leq \ystar < 3.0$ and jet radius parameter $R=0.4$. 
Here, $\sigma$ is the measured cross-section. 
All uncertainties are given in \%. 
The variable $\delta_\mathrm{stat}^\mathrm{data}$ ($\delta_\mathrm{stat}^\mathrm{MC}$) is the statistical uncertainty from the data (MC simulation). 
The $\gamma$ components show the uncertainties due to those on the jet energy calibration from the in situ, pileup, close-by jet, and flavour components. 
The $u$ components show the uncertainty for the jet energy and angular resolution, the unfolding, the quality selection, and the integrated luminosity. 
While all columns are uncorrelated with each other, the in situ, pileup, and flavour uncertainties shown here are the sum in quadrature of multiple uncorrelated components. 
The full set of cross-section values and uncertainty components, each of which is fully correlated in dijet mass and \ystar but uncorrelated with the other components, can be found in HepData \cite{Buckley:2010jn}. 
}
\label{tab:sysunc_r04_Eta6}
\end{table*}

\renewcommand{\arraystretch}{1}

\renewcommand{\arraystretch}{1.8}

\begin{table*}[!ht]
\tiny
\centering
{\fontfamily{ptm}\selectfont
\begin{tabular}{|@{}c@{}@{}c@{}@{}c@{}@{}c@{}@{}r@{}@{}r@{}@{}r@{}@{}r@{}@{}r@{}@{}r@{}@{}r@{}@{}r@{}|@{}r@{}@{}r@{}|}
\hline
\ \ \mass--\ystar \ \ & \ \ \mass range \ \ & \ \ $\sigma$ \ \ & \ \ $\delta_{\mathrm{stat}}^{\mathrm{data}}$ \ \ & \ \ $\delta_{\mathrm{stat}}^{\mathrm{MC}}$ \ \ & \ \ $\gamma_{\mathrm{in~situ}}$ \ \ & \ \ $\gamma_{\mathrm{pileup}}$ \ \ & \ \ $\gamma_{\operatorname{close-by}}$ \ \ & \ \ $\gamma_{\mathrm{flavour}}$ \ \ & \ \ u$_{\mathrm{JER}}$ \ \ & \ \ u$_{\mathrm{JAR}}$ \ \ & \ \ u$_{\mathrm{unfold}}$ \ \ & \ \ u$_{\mathrm{qual}}$ \ \ & \ \ u$_{\mathrm{lumi}}$ \ \ \\ \relax
\ \ bin \# \ \ & \ \ [TeV] \ \ & \ \ [pb/TeV] \ \ &\ \  \% \ \ & \ \ \% \ \ & \ \  \% \ \ & \ \ \% \ \ & \ \ \% \ \ & \ \ \% \ \ & \ \ \% \ \ & \ \ \% \ \ & \ \ \% \ \ & \ \ \% \ \ & \ \ \% \ \ \\
\hline
\ \ $0$ \ \ & \ \ 0.26--0.31 \ \ & \ \ 1.07e+06 \ \ & \ \ 0.55 \ \ & \ \ 0.61\ \  & \ \ $^{+5.6}_{-5.3}$ \ \ & \ \ $^{+1.6}_{-1.3}$ \ \ & \ \ $^{+2.9}_{-2.3}$ \ \ & \ \ $^{+5.5}_{-5.1}$ \ \ & \ \ 2.0 \ \ & \ \ 0.2 \ \ & \ \ 0.3 \ \ & \multirow{21}{*}{\ \ 0.5 \ \ } & \multirow{21}{*}{\ \ 1.8 \ \ } \\
\ \ $1$ \ \ & \ \ 0.31--0.37 \ \ & \ \ 4.06e+05 \ \ & \ \ 0.60 \ \ & \ \ 0.52\ \  & \ \ $^{+5.4}_{-5.1}$ \ \ & \ \ $^{+1.3}_{-1.1}$ \ \ & \ \ $^{+2.4}_{-2.0}$ \ \ & \ \ $^{+5.1}_{-4.8}$ \ \ & \ \ 1.5 \ \ & \ \ 0.2 \ \ & \ \ 0.1 \ \ &  & \\
\ \ $2$ \ \ & \ \ 0.37--0.44 \ \ & \ \ 1.52e+05 \ \ & \ \ 0.57 \ \ & \ \ 0.55\ \  & \ \ $^{+5.2}_{-4.9}$ \ \ & \ \ $^{+1.1}_{-1.1}$ \ \ & \ \ $^{+2.0}_{-1.9}$ \ \ & \ \ $^{+4.5}_{-4.3}$ \ \ & \ \ 1.0 \ \ & \ \ 0.2 \ \ & \ \ 0.1 \ \ &  & \\
\ \ $3$ \ \ & \ \ 0.44--0.51 \ \ & \ \ 6.06e+04 \ \ & \ \ 0.57 \ \ & \ \ 0.58\ \  & \ \ $^{+5.1}_{-4.8}$ \ \ & \ \ $^{+1.1}_{-1.1}$ \ \ & \ \ $^{+1.8}_{-1.9}$ \ \ & \ \ $^{+4.0}_{-3.8}$ \ \ & \ \ 0.8 \ \ & \ \ 0.2 \ \ & \ \ 0.0 \ \ &  & \\
\ \ $4$ \ \ & \ \ 0.51--0.59 \ \ & \ \ 2.57e+04 \ \ & \ \ 0.58 \ \ & \ \ 0.51\ \  & \ \ $^{+4.9}_{-4.8}$ \ \ & \ \ $^{+1.1}_{-1.1}$ \ \ & \ \ $^{+1.8}_{-1.9}$ \ \ & \ \ $^{+3.5}_{-3.4}$ \ \ & \ \ 0.7 \ \ & \ \ 0.2 \ \ & \ \ 0.0 \ \ &  & \\
\ \ $5$ \ \ & \ \ 0.59--0.67 \ \ & \ \ 1.13e+04 \ \ & \ \ 0.42 \ \ & \ \ 0.44\ \  & \ \ $^{+4.8}_{-4.7}$ \ \ & \ \ $^{+1.0}_{-1.2}$ \ \ & \ \ $^{+1.8}_{-1.8}$ \ \ & \ \ $^{+3.1}_{-3.0}$ \ \ & \ \ 0.7 \ \ & \ \ 0.2 \ \ & \ \ 0.0 \ \ &  & \\
\ \ $6$ \ \ & \ \ 0.67--0.76 \ \ & \ \ 5.21e+03 \ \ & \ \ 0.46 \ \ & \ \ 0.43\ \  & \ \ $^{+4.8}_{-4.7}$ \ \ & \ \ $^{+0.9}_{-1.2}$ \ \ & \ \ $^{+1.6}_{-1.6}$ \ \ & \ \ $^{+2.8}_{-2.7}$ \ \ & \ \ 0.7 \ \ & \ \ 0.2 \ \ & \ \ 0.0 \ \ &  & \\
\ \ $7$ \ \ & \ \ 0.76--0.85 \ \ & \ \ 2.51e+03 \ \ & \ \ 0.36 \ \ & \ \ 0.47\ \  & \ \ $^{+4.8}_{-4.6}$ \ \ & \ \ $^{+0.7}_{-1.0}$ \ \ & \ \ $^{+1.3}_{-1.3}$ \ \ & \ \ $^{+2.5}_{-2.4}$ \ \ & \ \ 0.5 \ \ & \ \ 0.2 \ \ & \ \ 0.0 \ \ &  & \\
\ \ $8$ \ \ & \ \ 0.85--0.95 \ \ & \ \ 1.24e+03 \ \ & \ \ 0.36 \ \ & \ \ 0.45\ \  & \ \ $^{+4.8}_{-4.6}$ \ \ & \ \ $^{+0.5}_{-0.7}$ \ \ & \ \ $^{+0.9}_{-1.0}$ \ \ & \ \ $^{+2.2}_{-2.1}$ \ \ & \ \ 0.4 \ \ & \ \ 0.1 \ \ & \ \ 0.0 \ \ &  & \\
\ \ $9$ \ \ & \ \ 0.95--1.06 \ \ & \ \ 5.99e+02 \ \ & \ \ 0.37 \ \ & \ \ 0.45\ \  & \ \ $^{+5.0}_{-4.8}$ \ \ & \ \ $^{+0.3}_{-0.4}$ \ \ & \ \ $^{+0.6}_{-0.7}$ \ \ & \ \ $^{+1.9}_{-1.8}$ \ \ & \ \ 0.3 \ \ & \ \ 0.1 \ \ & \ \ 0.0 \ \ &  & \\
\ \ $10$ \ \ & \ \ 1.06--1.18 \ \ & \ \ 2.90e+02 \ \ & \ \ 0.38 \ \ & \ \ 0.43\ \  & \ \ $^{+5.2}_{-5.1}$ \ \ & \ \ $^{+0.1}_{-0.2}$ \ \ & \ \ $^{+0.4}_{-0.4}$ \ \ & \ \ $^{+1.7}_{-1.6}$ \ \ & \ \ 0.3 \ \ & \ \ 0.1 \ \ & \ \ 0.0 \ \ &  & \\
\ \ $11$ \ \ & \ \ 1.18--1.31 \ \ & \ \ 1.40e+02 \ \ & \ \ 0.52 \ \ & \ \ 0.38\ \  & \ \ $^{+5.7}_{-5.6}$ \ \ & \ \ $^{+0.1}_{-0.1}$ \ \ & \ \ $^{+0.2}_{-0.2}$ \ \ & \ \ $^{+1.4}_{-1.4}$ \ \ & \ \ 0.3 \ \ & \ \ 0.1 \ \ & \ \ 0.0 \ \ &  & \\
\ \ $12$ \ \ & \ \ 1.31--1.45 \ \ & \ \ 6.56e+01 \ \ & \ \ 0.74 \ \ & \ \ 0.41\ \  & \ \ $^{+6.4}_{-6.3}$ \ \ & \ \ $^{+0.1}_{-0.1}$ \ \ & \ \ $^{+0.1}_{-0.1}$ \ \ & \ \ $^{+1.3}_{-1.3}$ \ \ & \ \ 0.3 \ \ & \ \ 0.1 \ \ & \ \ 0.0 \ \ &  & \\
\ \ $13$ \ \ & \ \ 1.45--1.60 \ \ & \ \ 3.19e+01 \ \ & \ \ 1.0 \ \ & \ \ 0.43\ \  & \ \ $^{+7.2}_{-7.2}$ \ \ & \ \ $^{+0.1}_{-0.1}$ \ \ & \ \ $^{+0.0}_{-0.0}$ \ \ & \ \ $^{+1.2}_{-1.2}$ \ \ & \ \ 0.4 \ \ & \ \ 0.1 \ \ & \ \ 0.0 \ \ &  & \\
\ \ $14$ \ \ & \ \ 1.60--1.76 \ \ & \ \ 1.49e+01 \ \ & \ \ 1.4 \ \ & \ \ 0.45\ \  & \ \ $^{+8.1}_{-8.1}$ \ \ & \ \ $^{+0.1}_{-0.1}$ \ \ & \ \ $^{+0.0}_{-0.0}$ \ \ & \ \ $^{+1.1}_{-1.1}$ \ \ & \ \ 0.3 \ \ & \ \ 0.1 \ \ & \ \ 0.0 \ \ &  & \\
\ \ $15$ \ \ & \ \ 1.76--1.94 \ \ & \ \ 7.05e+00 \ \ & \ \ 2.0 \ \ & \ \ 0.46\ \  & \ \ $^{+9.1}_{-9.0}$ \ \ & \ \ $^{+0.1}_{-0.1}$ \ \ & \ \ $^{+0.0}_{-0.0}$ \ \ & \ \ $^{+1.0}_{-1.0}$ \ \ & \ \ 0.3 \ \ & \ \ 0.1 \ \ & \ \ 0.0 \ \ &  & \\
\ \ $16$ \ \ & \ \ 1.94--2.12 \ \ & \ \ 3.23e+00 \ \ & \ \ 3.0 \ \ & \ \ 0.52\ \  & \ \ $^{+10}_{-9.9}$ \ \ & \ \ $^{+0.1}_{-0.1}$ \ \ & \ \ $^{+0.0}_{-0.0}$ \ \ & \ \ $^{+1.0}_{-1.0}$ \ \ & \ \ 0.4 \ \ & \ \ 0.1 \ \ & \ \ 0.0 \ \ &  & \\
\ \ $17$ \ \ & \ \ 2.12--2.33 \ \ & \ \ 1.44e+00 \ \ & \ \ 4.1 \ \ & \ \ 0.47\ \  & \ \ $^{+11}_{-11}$ \ \ & \ \ $^{+0.1}_{-0.1}$ \ \ & \ \ $^{+0.0}_{-0.0}$ \ \ & \ \ $^{+0.9}_{-0.9}$ \ \ & \ \ 0.4 \ \ & \ \ 0.1 \ \ & \ \ 0.0 \ \ &  & \\
\ \ $18$ \ \ & \ \ 2.33--2.55 \ \ & \ \ 6.81e-01 \ \ & \ \ 5.8 \ \ & \ \ 0.39\ \  & \ \ $^{+12}_{-11}$ \ \ & \ \ $^{+0.1}_{-0.1}$ \ \ & \ \ $^{+0.0}_{-0.0}$ \ \ & \ \ $^{+0.9}_{-0.9}$ \ \ & \ \ 0.4 \ \ & \ \ 0.0 \ \ & \ \ 0.0 \ \ &  & \\
\ \ $19$ \ \ & \ \ 2.55--2.78 \ \ & \ \ 2.79e-01 \ \ & \ \ 9.0 \ \ & \ \ 0.40\ \  & \ \ $^{+13}_{-13}$ \ \ & \ \ $^{+0.1}_{-0.1}$ \ \ & \ \ $^{+0.0}_{-0.0}$ \ \ & \ \ $^{+0.9}_{-0.8}$ \ \ & \ \ 0.5 \ \ & \ \ 0.0 \ \ & \ \ 0.0 \ \ &  & \\
\ \ $20$ \ \ & \ \ 2.78--4.27 \ \ & \ \ 2.52e-02 \ \ & \ \ 12 \ \ & \ \ 0.29\ \  & \ \ $^{+26}_{-25}$ \ \ & \ \ $^{+0.0}_{-0.0}$ \ \ & \ \ $^{+0.0}_{-0.0}$ \ \ & \ \ $^{+0.7}_{-0.7}$ \ \ & \ \ 0.8 \ \ & \ \ 0.0 \ \ & \ \ 0.0 \ \ &  & \\
\hline
\end{tabular}
}
\caption{
Measured double-differential dijet cross-sections for the range $\ystar < 0.5$ and jet radius parameter $R=0.6$. 
Here, $\sigma$ is the measured cross-section. 
All uncertainties are given in \%. 
The variable $\delta_\mathrm{stat}^\mathrm{data}$ ($\delta_\mathrm{stat}^\mathrm{MC}$) is the statistical uncertainty from the data (MC simulation). 
The $\gamma$ components show the uncertainties due to those on the jet energy calibration from the in situ, pileup, close-by jet, and flavour components. 
The $u$ components show the uncertainty for the jet energy and angular resolution, the unfolding, the quality selection, and the integrated luminosity. 
While all columns are uncorrelated with each other, the in situ, pileup, and flavour uncertainties shown here are the sum in quadrature of multiple uncorrelated components. 
The full set of cross-section values and uncertainty components, each of which is fully correlated in dijet mass and \ystar but uncorrelated with the other components, can be found in HepData \cite{Buckley:2010jn}. 
}
\label{tab:sysunc_r06_Eta1}
\end{table*}

\renewcommand{\arraystretch}{1}

\renewcommand{\arraystretch}{1.8}

\begin{table*}[!ht]
\tiny
\centering
{\fontfamily{ptm}\selectfont
\begin{tabular}{|@{}c@{}@{}c@{}@{}c@{}@{}c@{}@{}r@{}@{}r@{}@{}r@{}@{}r@{}@{}r@{}@{}r@{}@{}r@{}@{}r@{}|@{}r@{}@{}r@{}|}
\hline
\ \ \mass--\ystar \ \ & \ \ \mass range \ \ & \ \ $\sigma$ \ \ & \ \ $\delta_{\mathrm{stat}}^{\mathrm{data}}$ \ \ & \ \ $\delta_{\mathrm{stat}}^{\mathrm{MC}}$ \ \ & \ \ $\gamma_{\mathrm{in~situ}}$ \ \ & \ \ $\gamma_{\mathrm{pileup}}$ \ \ & \ \ $\gamma_{\operatorname{close-by}}$ \ \ & \ \ $\gamma_{\mathrm{flavour}}$ \ \ & \ \ u$_{\mathrm{JER}}$ \ \ & \ \ u$_{\mathrm{JAR}}$ \ \ & \ \ u$_{\mathrm{unfold}}$ \ \ & \ \ u$_{\mathrm{qual}}$ \ \ & \ \ u$_{\mathrm{lumi}}$ \ \ \\ \relax
\ \ bin \# \ \ & \ \ [TeV] \ \ & \ \ [pb/TeV] \ \ &\ \  \% \ \ & \ \ \% \ \ & \ \  \% \ \ & \ \ \% \ \ & \ \ \% \ \ & \ \ \% \ \ & \ \ \% \ \ & \ \ \% \ \ & \ \ \% \ \ & \ \ \% \ \ & \ \ \% \ \ \\
\hline
\ \ $21$ \ \ & \ \ 0.31--0.37 \ \ & \ \ 9.61e+05 \ \ & \ \ 0.66 \ \ & \ \ 0.86\ \  & \ \ $^{+6.4}_{-5.9}$ \ \ & \ \ $^{+1.4}_{-1.3}$ \ \ & \ \ $^{+2.6}_{-2.4}$ \ \ & \ \ $^{+5.8}_{-5.2}$ \ \ & \ \ 2.5 \ \ & \ \ 0.3 \ \ & \ \ 0.4 \ \ & \multirow{21}{*}{\ \ 0.5 \ \ } & \multirow{21}{*}{\ \ 1.8 \ \ } \\
\ \ $22$ \ \ & \ \ 0.37--0.44 \ \ & \ \ 3.68e+05 \ \ & \ \ 0.66 \ \ & \ \ 0.76\ \  & \ \ $^{+6.4}_{-6.0}$ \ \ & \ \ $^{+1.2}_{-1.4}$ \ \ & \ \ $^{+2.5}_{-2.1}$ \ \ & \ \ $^{+5.5}_{-4.9}$ \ \ & \ \ 1.8 \ \ & \ \ 0.4 \ \ & \ \ 0.2 \ \ &  & \\
\ \ $23$ \ \ & \ \ 0.44--0.51 \ \ & \ \ 1.46e+05 \ \ & \ \ 0.71 \ \ & \ \ 0.70\ \  & \ \ $^{+6.2}_{-6.1}$ \ \ & \ \ $^{+1.1}_{-1.4}$ \ \ & \ \ $^{+2.3}_{-2.0}$ \ \ & \ \ $^{+5.0}_{-4.6}$ \ \ & \ \ 1.2 \ \ & \ \ 0.3 \ \ & \ \ 0.0 \ \ &  & \\
\ \ $24$ \ \ & \ \ 0.51--0.59 \ \ & \ \ 6.03e+04 \ \ & \ \ 0.74 \ \ & \ \ 0.76\ \  & \ \ $^{+6.1}_{-6.1}$ \ \ & \ \ $^{+1.1}_{-1.4}$ \ \ & \ \ $^{+2.2}_{-2.0}$ \ \ & \ \ $^{+4.5}_{-4.3}$ \ \ & \ \ 1.0 \ \ & \ \ 0.2 \ \ & \ \ 0.0 \ \ &  & \\
\ \ $25$ \ \ & \ \ 0.59--0.67 \ \ & \ \ 2.73e+04 \ \ & \ \ 0.74 \ \ & \ \ 0.77\ \  & \ \ $^{+6.2}_{-5.9}$ \ \ & \ \ $^{+1.3}_{-1.4}$ \ \ & \ \ $^{+2.2}_{-2.0}$ \ \ & \ \ $^{+4.1}_{-3.8}$ \ \ & \ \ 0.9 \ \ & \ \ 0.2 \ \ & \ \ 0.0 \ \ &  & \\
\ \ $26$ \ \ & \ \ 0.67--0.76 \ \ & \ \ 1.24e+04 \ \ & \ \ 0.74 \ \ & \ \ 0.72\ \  & \ \ $^{+6.1}_{-5.6}$ \ \ & \ \ $^{+1.4}_{-1.2}$ \ \ & \ \ $^{+2.2}_{-2.0}$ \ \ & \ \ $^{+3.7}_{-3.3}$ \ \ & \ \ 0.9 \ \ & \ \ 0.2 \ \ & \ \ 0.0 \ \ &  & \\
\ \ $27$ \ \ & \ \ 0.76--0.85 \ \ & \ \ 5.96e+03 \ \ & \ \ 0.67 \ \ & \ \ 0.67\ \  & \ \ $^{+5.8}_{-5.4}$ \ \ & \ \ $^{+1.4}_{-1.1}$ \ \ & \ \ $^{+2.1}_{-1.9}$ \ \ & \ \ $^{+3.3}_{-3.0}$ \ \ & \ \ 0.9 \ \ & \ \ 0.2 \ \ & \ \ 0.0 \ \ &  & \\
\ \ $28$ \ \ & \ \ 0.85--0.95 \ \ & \ \ 2.94e+03 \ \ & \ \ 0.57 \ \ & \ \ 0.57\ \  & \ \ $^{+5.6}_{-5.3}$ \ \ & \ \ $^{+1.2}_{-0.9}$ \ \ & \ \ $^{+1.9}_{-1.7}$ \ \ & \ \ $^{+2.9}_{-2.7}$ \ \ & \ \ 0.9 \ \ & \ \ 0.2 \ \ & \ \ 0.0 \ \ &  & \\
\ \ $29$ \ \ & \ \ 0.95--1.06 \ \ & \ \ 1.41e+03 \ \ & \ \ 0.58 \ \ & \ \ 0.56\ \  & \ \ $^{+5.5}_{-5.3}$ \ \ & \ \ $^{+0.9}_{-0.7}$ \ \ & \ \ $^{+1.5}_{-1.4}$ \ \ & \ \ $^{+2.5}_{-2.5}$ \ \ & \ \ 0.8 \ \ & \ \ 0.1 \ \ & \ \ 0.0 \ \ &  & \\
\ \ $30$ \ \ & \ \ 1.06--1.18 \ \ & \ \ 6.83e+02 \ \ & \ \ 0.49 \ \ & \ \ 0.55\ \  & \ \ $^{+5.4}_{-5.4}$ \ \ & \ \ $^{+0.5}_{-0.5}$ \ \ & \ \ $^{+1.1}_{-1.1}$ \ \ & \ \ $^{+2.2}_{-2.2}$ \ \ & \ \ 0.7 \ \ & \ \ 0.1 \ \ & \ \ 0.0 \ \ &  & \\
\ \ $31$ \ \ & \ \ 1.18--1.31 \ \ & \ \ 3.36e+02 \ \ & \ \ 0.51 \ \ & \ \ 0.56\ \  & \ \ $^{+5.5}_{-5.5}$ \ \ & \ \ $^{+0.3}_{-0.3}$ \ \ & \ \ $^{+0.8}_{-0.8}$ \ \ & \ \ $^{+1.9}_{-2.0}$ \ \ & \ \ 0.7 \ \ & \ \ 0.1 \ \ & \ \ 0.0 \ \ &  & \\
\ \ $32$ \ \ & \ \ 1.31--1.45 \ \ & \ \ 1.59e+02 \ \ & \ \ 0.56 \ \ & \ \ 0.57\ \  & \ \ $^{+5.8}_{-5.7}$ \ \ & \ \ $^{+0.1}_{-0.2}$ \ \ & \ \ $^{+0.5}_{-0.5}$ \ \ & \ \ $^{+1.7}_{-1.7}$ \ \ & \ \ 0.7 \ \ & \ \ 0.1 \ \ & \ \ 0.0 \ \ &  & \\
\ \ $33$ \ \ & \ \ 1.45--1.60 \ \ & \ \ 7.64e+01 \ \ & \ \ 0.67 \ \ & \ \ 0.54\ \  & \ \ $^{+6.2}_{-6.0}$ \ \ & \ \ $^{+0.0}_{-0.1}$ \ \ & \ \ $^{+0.3}_{-0.3}$ \ \ & \ \ $^{+1.5}_{-1.5}$ \ \ & \ \ 0.7 \ \ & \ \ 0.1 \ \ & \ \ 0.0 \ \ &  & \\
\ \ $34$ \ \ & \ \ 1.60--1.76 \ \ & \ \ 3.64e+01 \ \ & \ \ 0.92 \ \ & \ \ 0.51\ \  & \ \ $^{+6.9}_{-6.6}$ \ \ & \ \ $^{+0.1}_{-0.1}$ \ \ & \ \ $^{+0.2}_{-0.2}$ \ \ & \ \ $^{+1.4}_{-1.3}$ \ \ & \ \ 0.7 \ \ & \ \ 0.1 \ \ & \ \ 0.0 \ \ &  & \\
\ \ $35$ \ \ & \ \ 1.76--1.94 \ \ & \ \ 1.67e+01 \ \ & \ \ 1.3 \ \ & \ \ 0.51\ \  & \ \ $^{+7.7}_{-7.3}$ \ \ & \ \ $^{+0.1}_{-0.1}$ \ \ & \ \ $^{+0.1}_{-0.1}$ \ \ & \ \ $^{+1.2}_{-1.1}$ \ \ & \ \ 0.8 \ \ & \ \ 0.1 \ \ & \ \ 0.0 \ \ &  & \\
\ \ $36$ \ \ & \ \ 1.94--2.12 \ \ & \ \ 7.57e+00 \ \ & \ \ 1.9 \ \ & \ \ 0.54\ \  & \ \ $^{+8.8}_{-8.2}$ \ \ & \ \ $^{+0.1}_{-0.1}$ \ \ & \ \ $^{+0.0}_{-0.0}$ \ \ & \ \ $^{+1.1}_{-1.0}$ \ \ & \ \ 0.8 \ \ & \ \ 0.1 \ \ & \ \ 0.0 \ \ &  & \\
\ \ $37$ \ \ & \ \ 2.12--2.33 \ \ & \ \ 3.36e+00 \ \ & \ \ 2.7 \ \ & \ \ 0.51\ \  & \ \ $^{+9.8}_{-9.3}$ \ \ & \ \ $^{+0.1}_{-0.1}$ \ \ & \ \ $^{+0.0}_{-0.0}$ \ \ & \ \ $^{+1.0}_{-1.0}$ \ \ & \ \ 0.8 \ \ & \ \ 0.1 \ \ & \ \ 0.0 \ \ &  & \\
\ \ $38$ \ \ & \ \ 2.33--2.55 \ \ & \ \ 1.29e+00 \ \ & \ \ 4.1 \ \ & \ \ 0.53\ \  & \ \ $^{+11}_{-10}$ \ \ & \ \ $^{+0.1}_{-0.1}$ \ \ & \ \ $^{+0.0}_{-0.0}$ \ \ & \ \ $^{+1.0}_{-0.9}$ \ \ & \ \ 0.8 \ \ & \ \ 0.0 \ \ & \ \ 0.0 \ \ &  & \\
\ \ $39$ \ \ & \ \ 2.55--2.78 \ \ & \ \ 5.44e-01 \ \ & \ \ 6.4 \ \ & \ \ 0.49\ \  & \ \ $^{+12}_{-11}$ \ \ & \ \ $^{+0.1}_{-0.1}$ \ \ & \ \ $^{+0.0}_{-0.0}$ \ \ & \ \ $^{+1.0}_{-0.9}$ \ \ & \ \ 0.9 \ \ & \ \ 0.0 \ \ & \ \ 0.1 \ \ &  & \\
\ \ $40$ \ \ & \ \ 2.78--3.04 \ \ & \ \ 1.80e-01 \ \ & \ \ 10 \ \ & \ \ 0.44\ \  & \ \ $^{+13}_{-12}$ \ \ & \ \ $^{+0.1}_{-0.1}$ \ \ & \ \ $^{+0.0}_{-0.0}$ \ \ & \ \ $^{+1.0}_{-1.0}$ \ \ & \ \ 0.9 \ \ & \ \ 0.0 \ \ & \ \ 0.6 \ \ &  & \\
\ \ $41$ \ \ & \ \ 3.04--4.27 \ \ & \ \ 2.11e-02 \ \ & \ \ 15 \ \ & \ \ 0.27\ \  & \ \ $^{+20}_{-19}$ \ \ & \ \ $^{+0.1}_{-0.1}$ \ \ & \ \ $^{+0.0}_{-0.0}$ \ \ & \ \ $^{+1.3}_{-1.1}$ \ \ & \ \ 1.9 \ \ & \ \ 0.0 \ \ & \ \ 1.1 \ \ &  & \\
\hline
\end{tabular}
}
\caption{
Measured double-differential dijet cross-sections for the range $0.5 \leq \ystar < 1.0$ and jet radius parameter $R=0.6$. 
Here, $\sigma$ is the measured cross-section. 
All uncertainties are given in \%. 
The variable $\delta_\mathrm{stat}^\mathrm{data}$ ($\delta_\mathrm{stat}^\mathrm{MC}$) is the statistical uncertainty from the data (MC simulation). 
The $\gamma$ components show the uncertainties due to those on the jet energy calibration from the in situ, pileup, close-by jet, and flavour components. 
The $u$ components show the uncertainty for the jet energy and angular resolution, the unfolding, the quality selection, and the integrated luminosity. 
While all columns are uncorrelated with each other, the in situ, pileup, and flavour uncertainties shown here are the sum in quadrature of multiple uncorrelated components. 
The full set of cross-section values and uncertainty components, each of which is fully correlated in dijet mass and \ystar but uncorrelated with the other components, can be found in HepData \cite{Buckley:2010jn}. 
}
\label{tab:sysunc_r06_Eta2}
\end{table*}

\renewcommand{\arraystretch}{1}

\renewcommand{\arraystretch}{1.8}

\begin{table*}[!ht]
\tiny
\centering
{\fontfamily{ptm}\selectfont
\begin{tabular}{|@{}c@{}@{}c@{}@{}c@{}@{}c@{}@{}r@{}@{}r@{}@{}r@{}@{}r@{}@{}r@{}@{}r@{}@{}r@{}@{}r@{}|@{}r@{}@{}r@{}|}
\hline
\ \ \mass--\ystar \ \ & \ \ \mass range \ \ & \ \ $\sigma$ \ \ & \ \ $\delta_{\mathrm{stat}}^{\mathrm{data}}$ \ \ & \ \ $\delta_{\mathrm{stat}}^{\mathrm{MC}}$ \ \ & \ \ $\gamma_{\mathrm{in~situ}}$ \ \ & \ \ $\gamma_{\mathrm{pileup}}$ \ \ & \ \ $\gamma_{\operatorname{close-by}}$ \ \ & \ \ $\gamma_{\mathrm{flavour}}$ \ \ & \ \ u$_{\mathrm{JER}}$ \ \ & \ \ u$_{\mathrm{JAR}}$ \ \ & \ \ u$_{\mathrm{unfold}}$ \ \ & \ \ u$_{\mathrm{qual}}$ \ \ & \ \ u$_{\mathrm{lumi}}$ \ \ \\ \relax
\ \ bin \# \ \ & \ \ [TeV] \ \ & \ \ [pb/TeV] \ \ &\ \  \% \ \ & \ \ \% \ \ & \ \  \% \ \ & \ \ \% \ \ & \ \ \% \ \ & \ \ \% \ \ & \ \ \% \ \ & \ \ \% \ \ & \ \ \% \ \ & \ \ \% \ \ & \ \ \% \ \ \\
\hline
\ \ $42$ \ \ & \ \ 0.51--0.59 \ \ & \ \ 1.80e+05 \ \ & \ \ 1.0 \ \ & \ \ 0.84\ \  & \ \ $^{+8.0}_{-8.1}$ \ \ & \ \ $^{+1.5}_{-1.8}$ \ \ & \ \ $^{+2.6}_{-2.6}$ \ \ & \ \ $^{+5.3}_{-5.2}$ \ \ & \ \ 2.3 \ \ & \ \ 0.2 \ \ & \ \ 0.4 \ \ & \multirow{19}{*}{\ \ 0.5 \ \ } & \multirow{19}{*}{\ \ 1.8 \ \ } \\
\ \ $43$ \ \ & \ \ 0.59--0.67 \ \ & \ \ 8.07e+04 \ \ & \ \ 1.1 \ \ & \ \ 0.86\ \  & \ \ $^{+8.3}_{-8.2}$ \ \ & \ \ $^{+1.5}_{-1.6}$ \ \ & \ \ $^{+2.4}_{-2.3}$ \ \ & \ \ $^{+5.0}_{-5.0}$ \ \ & \ \ 2.0 \ \ & \ \ 0.2 \ \ & \ \ 0.2 \ \ &  & \\
\ \ $44$ \ \ & \ \ 0.67--0.76 \ \ & \ \ 3.74e+04 \ \ & \ \ 1.2 \ \ & \ \ 0.76\ \  & \ \ $^{+8.5}_{-8.2}$ \ \ & \ \ $^{+1.5}_{-1.4}$ \ \ & \ \ $^{+2.3}_{-2.1}$ \ \ & \ \ $^{+4.7}_{-4.6}$ \ \ & \ \ 1.8 \ \ & \ \ 0.2 \ \ & \ \ 0.0 \ \ &  & \\
\ \ $45$ \ \ & \ \ 0.76--0.85 \ \ & \ \ 1.75e+04 \ \ & \ \ 1.2 \ \ & \ \ 0.81\ \  & \ \ $^{+8.5}_{-8.1}$ \ \ & \ \ $^{+1.4}_{-1.3}$ \ \ & \ \ $^{+2.2}_{-2.0}$ \ \ & \ \ $^{+4.4}_{-4.1}$ \ \ & \ \ 1.6 \ \ & \ \ 0.3 \ \ & \ \ 0.0 \ \ &  & \\
\ \ $46$ \ \ & \ \ 0.85--0.95 \ \ & \ \ 8.84e+03 \ \ & \ \ 1.3 \ \ & \ \ 0.76\ \  & \ \ $^{+8.4}_{-8.0}$ \ \ & \ \ $^{+1.3}_{-1.3}$ \ \ & \ \ $^{+2.0}_{-2.0}$ \ \ & \ \ $^{+4.0}_{-3.8}$ \ \ & \ \ 1.4 \ \ & \ \ 0.2 \ \ & \ \ 0.0 \ \ &  & \\
\ \ $47$ \ \ & \ \ 0.95--1.06 \ \ & \ \ 4.21e+03 \ \ & \ \ 1.2 \ \ & \ \ 0.77\ \  & \ \ $^{+8.4}_{-7.9}$ \ \ & \ \ $^{+1.3}_{-1.3}$ \ \ & \ \ $^{+2.0}_{-2.0}$ \ \ & \ \ $^{+3.7}_{-3.5}$ \ \ & \ \ 1.4 \ \ & \ \ 0.2 \ \ & \ \ 0.0 \ \ &  & \\
\ \ $48$ \ \ & \ \ 1.06--1.18 \ \ & \ \ 2.08e+03 \ \ & \ \ 1.3 \ \ & \ \ 0.72\ \  & \ \ $^{+8.3}_{-7.8}$ \ \ & \ \ $^{+1.3}_{-1.3}$ \ \ & \ \ $^{+2.0}_{-1.9}$ \ \ & \ \ $^{+3.4}_{-3.2}$ \ \ & \ \ 1.5 \ \ & \ \ 0.2 \ \ & \ \ 0.0 \ \ &  & \\
\ \ $49$ \ \ & \ \ 1.18--1.31 \ \ & \ \ 9.64e+02 \ \ & \ \ 1.2 \ \ & \ \ 0.62\ \  & \ \ $^{+8.1}_{-7.6}$ \ \ & \ \ $^{+1.4}_{-1.3}$ \ \ & \ \ $^{+1.9}_{-1.9}$ \ \ & \ \ $^{+3.1}_{-2.9}$ \ \ & \ \ 1.5 \ \ & \ \ 0.1 \ \ & \ \ 0.0 \ \ &  & \\
\ \ $50$ \ \ & \ \ 1.31--1.45 \ \ & \ \ 4.66e+02 \ \ & \ \ 1.1 \ \ & \ \ 0.53\ \  & \ \ $^{+7.9}_{-7.5}$ \ \ & \ \ $^{+1.3}_{-1.3}$ \ \ & \ \ $^{+1.8}_{-1.7}$ \ \ & \ \ $^{+2.7}_{-2.7}$ \ \ & \ \ 1.5 \ \ & \ \ 0.1 \ \ & \ \ 0.0 \ \ &  & \\
\ \ $51$ \ \ & \ \ 1.45--1.60 \ \ & \ \ 2.26e+02 \ \ & \ \ 1.2 \ \ & \ \ 0.51\ \  & \ \ $^{+7.7}_{-7.5}$ \ \ & \ \ $^{+1.1}_{-1.0}$ \ \ & \ \ $^{+1.6}_{-1.4}$ \ \ & \ \ $^{+2.4}_{-2.4}$ \ \ & \ \ 1.5 \ \ & \ \ 0.1 \ \ & \ \ 0.0 \ \ &  & \\
\ \ $52$ \ \ & \ \ 1.60--1.76 \ \ & \ \ 1.07e+02 \ \ & \ \ 1.0 \ \ & \ \ 0.52\ \  & \ \ $^{+7.7}_{-7.5}$ \ \ & \ \ $^{+0.8}_{-0.7}$ \ \ & \ \ $^{+1.2}_{-1.1}$ \ \ & \ \ $^{+2.2}_{-2.2}$ \ \ & \ \ 1.4 \ \ & \ \ 0.1 \ \ & \ \ 0.0 \ \ &  & \\
\ \ $53$ \ \ & \ \ 1.76--1.94 \ \ & \ \ 4.91e+01 \ \ & \ \ 1.1 \ \ & \ \ 0.50\ \  & \ \ $^{+7.7}_{-7.6}$ \ \ & \ \ $^{+0.5}_{-0.4}$ \ \ & \ \ $^{+0.9}_{-0.8}$ \ \ & \ \ $^{+1.9}_{-1.9}$ \ \ & \ \ 1.3 \ \ & \ \ 0.1 \ \ & \ \ 0.0 \ \ &  & \\
\ \ $54$ \ \ & \ \ 1.94--2.12 \ \ & \ \ 2.21e+01 \ \ & \ \ 1.5 \ \ & \ \ 0.56\ \  & \ \ $^{+8.0}_{-7.8}$ \ \ & \ \ $^{+0.3}_{-0.2}$ \ \ & \ \ $^{+0.6}_{-0.6}$ \ \ & \ \ $^{+1.7}_{-1.7}$ \ \ & \ \ 1.3 \ \ & \ \ 0.1 \ \ & \ \ 0.0 \ \ &  & \\
\ \ $55$ \ \ & \ \ 2.12--2.33 \ \ & \ \ 9.86e+00 \ \ & \ \ 1.6 \ \ & \ \ 0.49\ \  & \ \ $^{+8.6}_{-8.3}$ \ \ & \ \ $^{+0.1}_{-0.0}$ \ \ & \ \ $^{+0.4}_{-0.4}$ \ \ & \ \ $^{+1.6}_{-1.6}$ \ \ & \ \ 1.4 \ \ & \ \ 0.1 \ \ & \ \ 0.0 \ \ &  & \\
\ \ $56$ \ \ & \ \ 2.33--2.55 \ \ & \ \ 4.30e+00 \ \ & \ \ 2.3 \ \ & \ \ 0.45\ \  & \ \ $^{+9.4}_{-8.9}$ \ \ & \ \ $^{+0.1}_{-0.0}$ \ \ & \ \ $^{+0.2}_{-0.2}$ \ \ & \ \ $^{+1.5}_{-1.4}$ \ \ & \ \ 1.5 \ \ & \ \ 0.1 \ \ & \ \ 0.0 \ \ &  & \\
\ \ $57$ \ \ & \ \ 2.55--2.78 \ \ & \ \ 1.73e+00 \ \ & \ \ 3.6 \ \ & \ \ 0.47\ \  & \ \ $^{+10}_{-9.7}$ \ \ & \ \ $^{+0.1}_{-0.1}$ \ \ & \ \ $^{+0.1}_{-0.1}$ \ \ & \ \ $^{+1.4}_{-1.4}$ \ \ & \ \ 1.7 \ \ & \ \ 0.1 \ \ & \ \ 0.0 \ \ &  & \\
\ \ $58$ \ \ & \ \ 2.78--3.04 \ \ & \ \ 6.16e-01 \ \ & \ \ 5.7 \ \ & \ \ 0.45\ \  & \ \ $^{+11}_{-11}$ \ \ & \ \ $^{+0.1}_{-0.1}$ \ \ & \ \ $^{+0.1}_{-0.1}$ \ \ & \ \ $^{+1.3}_{-1.3}$ \ \ & \ \ 1.8 \ \ & \ \ 0.1 \ \ & \ \ 0.1 \ \ &  & \\
\ \ $59$ \ \ & \ \ 3.04--3.31 \ \ & \ \ 1.88e-01 \ \ & \ \ 9.7 \ \ & \ \ 0.42\ \  & \ \ $^{+12}_{-12}$ \ \ & \ \ $^{+0.2}_{-0.1}$ \ \ & \ \ $^{+0.0}_{-0.0}$ \ \ & \ \ $^{+1.3}_{-1.3}$ \ \ & \ \ 1.9 \ \ & \ \ 0.1 \ \ & \ \ 0.9 \ \ &  & \\
\ \ $60$ \ \ & \ \ 3.31--4.64 \ \ & \ \ 2.61e-02 \ \ & \ \ 13 \ \ & \ \ 0.25\ \  & \ \ $^{+19}_{-18}$ \ \ & \ \ $^{+0.2}_{-0.1}$ \ \ & \ \ $^{+0.0}_{-0.0}$ \ \ & \ \ $^{+1.4}_{-1.3}$ \ \ & \ \ 3.4 \ \ & \ \ 0.0 \ \ & \ \ 1.7 \ \ &  & \\
\hline
\end{tabular}
}
\caption{
Measured double-differential dijet cross-sections for the range $1.0 \leq \ystar < 1.5$ and jet radius parameter $R=0.6$. 
Here, $\sigma$ is the measured cross-section. 
All uncertainties are given in \%. 
The variable $\delta_\mathrm{stat}^\mathrm{data}$ ($\delta_\mathrm{stat}^\mathrm{MC}$) is the statistical uncertainty from the data (MC simulation). 
The $\gamma$ components show the uncertainties due to those on the jet energy calibration from the in situ, pileup, close-by jet, and flavour components. 
The $u$ components show the uncertainty for the jet energy and angular resolution, the unfolding, the quality selection, and the integrated luminosity. 
While all columns are uncorrelated with each other, the in situ, pileup, and flavour uncertainties shown here are the sum in quadrature of multiple uncorrelated components. 
The full set of cross-section values and uncertainty components, each of which is fully correlated in dijet mass and \ystar but uncorrelated with the other components, can be found in HepData \cite{Buckley:2010jn}. 
}
\label{tab:sysunc_r06_Eta3}
\end{table*}

\renewcommand{\arraystretch}{1}

\renewcommand{\arraystretch}{1.8}

\begin{table*}[!ht]
\tiny
\centering
{\fontfamily{ptm}\selectfont
\begin{tabular}{|@{}c@{}@{}c@{}@{}c@{}@{}c@{}@{}r@{}@{}r@{}@{}r@{}@{}r@{}@{}r@{}@{}r@{}@{}r@{}@{}r@{}|@{}r@{}@{}r@{}|}
\hline
\ \ \mass--\ystar \ \ & \ \ \mass range \ \ & \ \ $\sigma$ \ \ & \ \ $\delta_{\mathrm{stat}}^{\mathrm{data}}$ \ \ & \ \ $\delta_{\mathrm{stat}}^{\mathrm{MC}}$ \ \ & \ \ $\gamma_{\mathrm{in~situ}}$ \ \ & \ \ $\gamma_{\mathrm{pileup}}$ \ \ & \ \ $\gamma_{\operatorname{close-by}}$ \ \ & \ \ $\gamma_{\mathrm{flavour}}$ \ \ & \ \ u$_{\mathrm{JER}}$ \ \ & \ \ u$_{\mathrm{JAR}}$ \ \ & \ \ u$_{\mathrm{unfold}}$ \ \ & \ \ u$_{\mathrm{qual}}$ \ \ & \ \ u$_{\mathrm{lumi}}$ \ \ \\ \relax
\ \ bin \# \ \ & \ \ [TeV] \ \ & \ \ [pb/TeV] \ \ &\ \  \% \ \ & \ \ \% \ \ & \ \  \% \ \ & \ \ \% \ \ & \ \ \% \ \ & \ \ \% \ \ & \ \ \% \ \ & \ \ \% \ \ & \ \ \% \ \ & \ \ \% \ \ & \ \ \% \ \ \\
\hline
\ \ $61$ \ \ & \ \ 0.76--0.85 \ \ & \ \ 4.96e+04 \ \ & \ \ 1.8 \ \ & \ \ 0.93\ \  & \ \ $^{+12}_{-11}$ \ \ & \ \ $^{+1.4}_{-1.4}$ \ \ & \ \ $^{+2.9}_{-2.6}$ \ \ & \ \ $^{+5.1}_{-4.9}$ \ \ & \ \ 3.8 \ \ & \ \ 0.4 \ \ & \ \ 0.2 \ \ & \multirow{17}{*}{\ \ 0.5 \ \ } & \multirow{17}{*}{\ \ 1.8 \ \ } \\
\ \ $62$ \ \ & \ \ 0.85--0.95 \ \ & \ \ 2.55e+04 \ \ & \ \ 2.1 \ \ & \ \ 0.99\ \  & \ \ $^{+12}_{-11}$ \ \ & \ \ $^{+1.4}_{-1.4}$ \ \ & \ \ $^{+2.6}_{-2.5}$ \ \ & \ \ $^{+5.0}_{-4.7}$ \ \ & \ \ 3.4 \ \ & \ \ 0.4 \ \ & \ \ 0.1 \ \ &  & \\
\ \ $63$ \ \ & \ \ 0.95--1.06 \ \ & \ \ 1.22e+04 \ \ & \ \ 2.1 \ \ & \ \ 0.93\ \  & \ \ $^{+12}_{-12}$ \ \ & \ \ $^{+1.6}_{-1.5}$ \ \ & \ \ $^{+2.4}_{-2.4}$ \ \ & \ \ $^{+4.7}_{-4.5}$ \ \ & \ \ 3.0 \ \ & \ \ 0.3 \ \ & \ \ 0.0 \ \ &  & \\
\ \ $64$ \ \ & \ \ 1.06--1.18 \ \ & \ \ 6.14e+03 \ \ & \ \ 2.4 \ \ & \ \ 0.87\ \  & \ \ $^{+13}_{-12}$ \ \ & \ \ $^{+1.7}_{-1.5}$ \ \ & \ \ $^{+2.3}_{-2.2}$ \ \ & \ \ $^{+4.4}_{-4.3}$ \ \ & \ \ 2.7 \ \ & \ \ 0.3 \ \ & \ \ 0.0 \ \ &  & \\
\ \ $65$ \ \ & \ \ 1.18--1.31 \ \ & \ \ 2.82e+03 \ \ & \ \ 2.5 \ \ & \ \ 0.85\ \  & \ \ $^{+13}_{-12}$ \ \ & \ \ $^{+1.8}_{-1.5}$ \ \ & \ \ $^{+2.4}_{-2.1}$ \ \ & \ \ $^{+4.1}_{-4.0}$ \ \ & \ \ 2.5 \ \ & \ \ 0.2 \ \ & \ \ 0.0 \ \ &  & \\
\ \ $66$ \ \ & \ \ 1.31--1.45 \ \ & \ \ 1.44e+03 \ \ & \ \ 2.8 \ \ & \ \ 0.74\ \  & \ \ $^{+13}_{-12}$ \ \ & \ \ $^{+1.9}_{-1.5}$ \ \ & \ \ $^{+2.4}_{-2.1}$ \ \ & \ \ $^{+3.9}_{-3.7}$ \ \ & \ \ 2.4 \ \ & \ \ 0.2 \ \ & \ \ 0.0 \ \ &  & \\
\ \ $67$ \ \ & \ \ 1.45--1.60 \ \ & \ \ 6.87e+02 \ \ & \ \ 3.0 \ \ & \ \ 0.66\ \  & \ \ $^{+13}_{-12}$ \ \ & \ \ $^{+1.9}_{-1.5}$ \ \ & \ \ $^{+2.4}_{-2.1}$ \ \ & \ \ $^{+3.7}_{-3.5}$ \ \ & \ \ 2.3 \ \ & \ \ 0.2 \ \ & \ \ 0.0 \ \ &  & \\
\ \ $68$ \ \ & \ \ 1.60--1.76 \ \ & \ \ 3.41e+02 \ \ & \ \ 3.1 \ \ & \ \ 0.66\ \  & \ \ $^{+13}_{-12}$ \ \ & \ \ $^{+1.8}_{-1.7}$ \ \ & \ \ $^{+2.2}_{-2.2}$ \ \ & \ \ $^{+3.4}_{-3.3}$ \ \ & \ \ 2.2 \ \ & \ \ 0.2 \ \ & \ \ 0.0 \ \ &  & \\
\ \ $69$ \ \ & \ \ 1.76--1.94 \ \ & \ \ 1.50e+02 \ \ & \ \ 3.6 \ \ & \ \ 0.68\ \  & \ \ $^{+13}_{-12}$ \ \ & \ \ $^{+1.7}_{-1.7}$ \ \ & \ \ $^{+2.0}_{-2.2}$ \ \ & \ \ $^{+3.2}_{-3.1}$ \ \ & \ \ 2.2 \ \ & \ \ 0.2 \ \ & \ \ 0.0 \ \ &  & \\
\ \ $70$ \ \ & \ \ 1.94--2.12 \ \ & \ \ 6.33e+01 \ \ & \ \ 2.9 \ \ & \ \ 0.64\ \  & \ \ $^{+13}_{-12}$ \ \ & \ \ $^{+1.5}_{-1.7}$ \ \ & \ \ $^{+1.8}_{-2.0}$ \ \ & \ \ $^{+2.9}_{-2.9}$ \ \ & \ \ 2.4 \ \ & \ \ 0.2 \ \ & \ \ 0.0 \ \ &  & \\
\ \ $71$ \ \ & \ \ 2.12--2.33 \ \ & \ \ 3.04e+01 \ \ & \ \ 3.4 \ \ & \ \ 0.56\ \  & \ \ $^{+13}_{-12}$ \ \ & \ \ $^{+1.4}_{-1.6}$ \ \ & \ \ $^{+1.5}_{-1.8}$ \ \ & \ \ $^{+2.7}_{-2.7}$ \ \ & \ \ 2.6 \ \ & \ \ 0.2 \ \ & \ \ 0.0 \ \ &  & \\
\ \ $72$ \ \ & \ \ 2.33--2.55 \ \ & \ \ 1.21e+01 \ \ & \ \ 4.3 \ \ & \ \ 0.64\ \  & \ \ $^{+14}_{-12}$ \ \ & \ \ $^{+1.2}_{-1.3}$ \ \ & \ \ $^{+1.3}_{-1.6}$ \ \ & \ \ $^{+2.6}_{-2.6}$ \ \ & \ \ 2.8 \ \ & \ \ 0.2 \ \ & \ \ 0.0 \ \ &  & \\
\ \ $73$ \ \ & \ \ 2.55--2.78 \ \ & \ \ 4.90e+00 \ \ & \ \ 4.0 \ \ & \ \ 0.64\ \  & \ \ $^{+14}_{-13}$ \ \ & \ \ $^{+1.0}_{-1.1}$ \ \ & \ \ $^{+1.1}_{-1.3}$ \ \ & \ \ $^{+2.4}_{-2.5}$ \ \ & \ \ 3.0 \ \ & \ \ 0.2 \ \ & \ \ 0.0 \ \ &  & \\
\ \ $74$ \ \ & \ \ 2.78--3.04 \ \ & \ \ 1.82e+00 \ \ & \ \ 4.8 \ \ & \ \ 0.48\ \  & \ \ $^{+14}_{-13}$ \ \ & \ \ $^{+0.8}_{-0.8}$ \ \ & \ \ $^{+0.9}_{-1.0}$ \ \ & \ \ $^{+2.3}_{-2.4}$ \ \ & \ \ 3.4 \ \ & \ \ 0.2 \ \ & \ \ 0.0 \ \ &  & \\
\ \ $75$ \ \ & \ \ 3.04--3.31 \ \ & \ \ 7.06e-01 \ \ & \ \ 7.1 \ \ & \ \ 0.50\ \  & \ \ $^{+15}_{-14}$ \ \ & \ \ $^{+0.5}_{-0.5}$ \ \ & \ \ $^{+0.7}_{-0.8}$ \ \ & \ \ $^{+2.3}_{-2.3}$ \ \ & \ \ 3.8 \ \ & \ \ 0.2 \ \ & \ \ 0.1 \ \ &  & \\
\ \ $76$ \ \ & \ \ 3.31--3.61 \ \ & \ \ 2.44e-01 \ \ & \ \ 9.7 \ \ & \ \ 0.53\ \  & \ \ $^{+16}_{-14}$ \ \ & \ \ $^{+0.3}_{-0.3}$ \ \ & \ \ $^{+0.6}_{-0.6}$ \ \ & \ \ $^{+2.2}_{-2.2}$ \ \ & \ \ 4.2 \ \ & \ \ 0.2 \ \ & \ \ 0.4 \ \ &  & \\
\ \ $77$ \ \ & \ \ 3.61--4.64 \ \ & \ \ 2.15e-02 \ \ & \ \ 16 \ \ & \ \ 0.49\ \  & \ \ $^{+21}_{-19}$ \ \ & \ \ $^{+0.2}_{-0.3}$ \ \ & \ \ $^{+0.0}_{-0.0}$ \ \ & \ \ $^{+2.0}_{-2.0}$ \ \ & \ \ 6.1 \ \ & \ \ 0.1 \ \ & \ \ 1.2 \ \ &  & \\
\hline
\end{tabular}
}
\caption{
Measured double-differential dijet cross-sections for the range $1.5 \leq \ystar < 2.0$ and jet radius parameter $R=0.6$. 
Here, $\sigma$ is the measured cross-section. 
All uncertainties are given in \%. 
The variable $\delta_\mathrm{stat}^\mathrm{data}$ ($\delta_\mathrm{stat}^\mathrm{MC}$) is the statistical uncertainty from the data (MC simulation). 
The $\gamma$ components show the uncertainties due to those on the jet energy calibration from the in situ, pileup, close-by jet, and flavour components. 
The $u$ components show the uncertainty for the jet energy and angular resolution, the unfolding, the quality selection, and the integrated luminosity. 
While all columns are uncorrelated with each other, the in situ, pileup, and flavour uncertainties shown here are the sum in quadrature of multiple uncorrelated components. 
The full set of cross-section values and uncertainty components, each of which is fully correlated in dijet mass and \ystar but uncorrelated with the other components, can be found in HepData \cite{Buckley:2010jn}. 
}
\label{tab:sysunc_r06_Eta4}
\end{table*}

\renewcommand{\arraystretch}{1}

\renewcommand{\arraystretch}{1.8}

\begin{table*}[!ht]
\tiny
\centering
{\fontfamily{ptm}\selectfont
\begin{tabular}{|@{}c@{}@{}c@{}@{}c@{}@{}c@{}@{}r@{}@{}r@{}@{}r@{}@{}r@{}@{}r@{}@{}r@{}@{}r@{}@{}r@{}|@{}r@{}@{}r@{}|}
\hline
\ \ \mass--\ystar \ \ & \ \ \mass range \ \ & \ \ $\sigma$ \ \ & \ \ $\delta_{\mathrm{stat}}^{\mathrm{data}}$ \ \ & \ \ $\delta_{\mathrm{stat}}^{\mathrm{MC}}$ \ \ & \ \ $\gamma_{\mathrm{in~situ}}$ \ \ & \ \ $\gamma_{\mathrm{pileup}}$ \ \ & \ \ $\gamma_{\operatorname{close-by}}$ \ \ & \ \ $\gamma_{\mathrm{flavour}}$ \ \ & \ \ u$_{\mathrm{JER}}$ \ \ & \ \ u$_{\mathrm{JAR}}$ \ \ & \ \ u$_{\mathrm{unfold}}$ \ \ & \ \ u$_{\mathrm{qual}}$ \ \ & \ \ u$_{\mathrm{lumi}}$ \ \ \\ \relax
\ \ bin \# \ \ & \ \ [TeV] \ \ & \ \ [pb/TeV] \ \ &\ \  \% \ \ & \ \ \% \ \ & \ \  \% \ \ & \ \ \% \ \ & \ \ \% \ \ & \ \ \% \ \ & \ \ \% \ \ & \ \ \% \ \ & \ \ \% \ \ & \ \ \% \ \ & \ \ \% \ \ \\
\hline
\ \ $78$ \ \ & \ \ 1.31--1.45 \ \ & \ \ 3.39e+03 \ \ & \ \ 4.4 \ \ & \ \ 1.1\ \  & \ \ $^{+18}_{-17}$ \ \ & \ \ $^{+2.6}_{-2.2}$ \ \ & \ \ $^{+2.8}_{-2.8}$ \ \ & \ \ $^{+4.1}_{-4.1}$ \ \ & \ \ 5.1 \ \ & \ \ 0.6 \ \ & \ \ 0.1 \ \ & \multirow{8}{*}{\ \ 0.5 \ \ } & \multirow{8}{*}{\ \ 1.8 \ \ } \\
\ \ $79$ \ \ & \ \ 1.45--1.60 \ \ & \ \ 1.78e+03 \ \ & \ \ 5.1 \ \ & \ \ 1.1\ \  & \ \ $^{+19}_{-17}$ \ \ & \ \ $^{+2.5}_{-2.0}$ \ \ & \ \ $^{+2.6}_{-2.7}$ \ \ & \ \ $^{+3.9}_{-3.9}$ \ \ & \ \ 4.8 \ \ & \ \ 0.6 \ \ & \ \ 0.1 \ \ &  & \\
\ \ $80$ \ \ & \ \ 1.60--1.76 \ \ & \ \ 8.32e+02 \ \ & \ \ 6.2 \ \ & \ \ 1.3\ \  & \ \ $^{+19}_{-17}$ \ \ & \ \ $^{+2.4}_{-1.9}$ \ \ & \ \ $^{+2.5}_{-2.6}$ \ \ & \ \ $^{+3.8}_{-3.7}$ \ \ & \ \ 4.4 \ \ & \ \ 0.6 \ \ & \ \ 0.1 \ \ &  & \\
\ \ $81$ \ \ & \ \ 1.76--2.12 \ \ & \ \ 2.93e+02 \ \ & \ \ 6.4 \ \ & \ \ 0.81\ \  & \ \ $^{+20}_{-17}$ \ \ & \ \ $^{+2.3}_{-1.8}$ \ \ & \ \ $^{+2.5}_{-2.4}$ \ \ & \ \ $^{+3.7}_{-3.4}$ \ \ & \ \ 4.0 \ \ & \ \ 0.5 \ \ & \ \ 0.0 \ \ &  & \\
\ \ $82$ \ \ & \ \ 2.12--2.55 \ \ & \ \ 5.75e+01 \ \ & \ \ 8.8 \ \ & \ \ 0.72\ \  & \ \ $^{+22}_{-19}$ \ \ & \ \ $^{+2.5}_{-2.1}$ \ \ & \ \ $^{+2.5}_{-2.4}$ \ \ & \ \ $^{+3.5}_{-3.2}$ \ \ & \ \ 4.0 \ \ & \ \ 0.3 \ \ & \ \ 0.0 \ \ &  & \\
\ \ $83$ \ \ & \ \ 2.55--3.04 \ \ & \ \ 9.59e+00 \ \ & \ \ 12 \ \ & \ \ 0.48\ \  & \ \ $^{+24}_{-20}$ \ \ & \ \ $^{+2.8}_{-2.3}$ \ \ & \ \ $^{+2.5}_{-2.3}$ \ \ & \ \ $^{+3.3}_{-3.1}$ \ \ & \ \ 4.9 \ \ & \ \ 0.3 \ \ & \ \ 0.0 \ \ &  & \\
\ \ $84$ \ \ & \ \ 3.04--3.61 \ \ & \ \ 1.34e+00 \ \ & \ \ 12 \ \ & \ \ 0.62\ \  & \ \ $^{+26}_{-22}$ \ \ & \ \ $^{+3.0}_{-2.5}$ \ \ & \ \ $^{+2.2}_{-2.3}$ \ \ & \ \ $^{+3.2}_{-3.1}$ \ \ & \ \ 6.1 \ \ & \ \ 0.4 \ \ & \ \ 0.1 \ \ &  & \\
\ \ $85$ \ \ & \ \ 3.61--5.04 \ \ & \ \ 4.50e-02 \ \ & \ \ 29 \ \ & \ \ 1.2\ \  & \ \ $^{+34}_{-28}$ \ \ & \ \ $^{+4.3}_{-3.3}$ \ \ & \ \ $^{+0.6}_{-0.6}$ \ \ & \ \ $^{+3.2}_{-3.2}$ \ \ & \ \ 9.5 \ \ & \ \ 0.4 \ \ & \ \ 1.6 \ \ &  & \\
\hline
\end{tabular}
}
\caption{
Measured double-differential dijet cross-sections for the range $2.0 \leq \ystar < 2.5$ and jet radius parameter $R=0.6$. 
Here, $\sigma$ is the measured cross-section. 
All uncertainties are given in \%. 
The variable $\delta_\mathrm{stat}^\mathrm{data}$ ($\delta_\mathrm{stat}^\mathrm{MC}$) is the statistical uncertainty from the data (MC simulation). 
The $\gamma$ components show the uncertainties due to those on the jet energy calibration from the in situ, pileup, close-by jet, and flavour components. 
The $u$ components show the uncertainty for the jet energy and angular resolution, the unfolding, the quality selection, and the integrated luminosity. 
While all columns are uncorrelated with each other, the in situ, pileup, and flavour uncertainties shown here are the sum in quadrature of multiple uncorrelated components. 
The full set of cross-section values and uncertainty components, each of which is fully correlated in dijet mass and \ystar but uncorrelated with the other components, can be found in HepData \cite{Buckley:2010jn}. 
}
\label{tab:sysunc_r06_Eta5}
\end{table*}

\renewcommand{\arraystretch}{1}

\renewcommand{\arraystretch}{1.8}

\begin{table*}[!ht]
\tiny
\centering
{\fontfamily{ptm}\selectfont
\begin{tabular}{|@{}c@{}@{}c@{}@{}c@{}@{}c@{}@{}r@{}@{}r@{}@{}r@{}@{}r@{}@{}r@{}@{}r@{}@{}r@{}@{}r@{}|@{}r@{}@{}r@{}|}
\hline
\ \ \mass--\ystar \ \ & \ \ \mass range \ \ & \ \ $\sigma$ \ \ & \ \ $\delta_{\mathrm{stat}}^{\mathrm{data}}$ \ \ & \ \ $\delta_{\mathrm{stat}}^{\mathrm{MC}}$ \ \ & \ \ $\gamma_{\mathrm{in~situ}}$ \ \ & \ \ $\gamma_{\mathrm{pileup}}$ \ \ & \ \ $\gamma_{\operatorname{close-by}}$ \ \ & \ \ $\gamma_{\mathrm{flavour}}$ \ \ & \ \ u$_{\mathrm{JER}}$ \ \ & \ \ u$_{\mathrm{JAR}}$ \ \ & \ \ u$_{\mathrm{unfold}}$ \ \ & \ \ u$_{\mathrm{qual}}$ \ \ & \ \ u$_{\mathrm{lumi}}$ \ \ \\ \relax
\ \ bin \# \ \ & \ \ [TeV] \ \ & \ \ [pb/TeV] \ \ &\ \  \% \ \ & \ \ \% \ \ & \ \  \% \ \ & \ \ \% \ \ & \ \ \% \ \ & \ \ \% \ \ & \ \ \% \ \ & \ \ \% \ \ & \ \ \% \ \ & \ \ \% \ \ & \ \ \% \ \ \\
\hline
\ \ $86$ \ \ & \ \ 2.12--2.55 \ \ & \ \ 1.09e+02 \ \ & \ \ 18 \ \ & \ \ 1.1\ \  & \ \ $^{+27}_{-22}$ \ \ & \ \ $^{+3.1}_{-2.5}$ \ \ & \ \ $^{+2.5}_{-2.1}$ \ \ & \ \ $^{+2.9}_{-2.9}$ \ \ & \ \ 5.2 \ \ & \ \ 1.2 \ \ & \ \ 0.3 \ \ & \multirow{4}{*}{\ \ 0.5 \ \ } & \multirow{4}{*}{\ \ 1.8 \ \ } \\
\ \ $87$ \ \ & \ \ 2.55--3.04 \ \ & \ \ 9.75e+00 \ \ & \ \ 26 \ \ & \ \ 1.8\ \  & \ \ $^{+30}_{-24}$ \ \ & \ \ $^{+3.1}_{-2.2}$ \ \ & \ \ $^{+2.5}_{-2.0}$ \ \ & \ \ $^{+2.9}_{-2.7}$ \ \ & \ \ 4.8 \ \ & \ \ 1.0 \ \ & \ \ 0.1 \ \ &  & \\
\ \ $88$ \ \ & \ \ 3.04--3.93 \ \ & \ \ 2.42e+00 \ \ & \ \ 36 \ \ & \ \ 1.1\ \  & \ \ $^{+38}_{-29}$ \ \ & \ \ $^{+3.3}_{-2.3}$ \ \ & \ \ $^{+2.6}_{-2.6}$ \ \ & \ \ $^{+2.8}_{-2.6}$ \ \ & \ \ 5.9 \ \ & \ \ 1.0 \ \ & \ \ 0.1 \ \ &  & \\
\ \ $89$ \ \ & \ \ 3.93--5.04 \ \ & \ \ 2.10e-02 \ \ & \ \ 76 \ \ & \ \ 2.4\ \  & \ \ $^{+61}_{-43}$ \ \ & \ \ $^{+5.3}_{-4.5}$ \ \ & \ \ $^{+2.9}_{-3.1}$ \ \ & \ \ $^{+3.1}_{-3.2}$ \ \ & \ \ 4.2 \ \ & \ \ 2.4 \ \ & \ \ 0.4 \ \ &  & \\
\hline
\end{tabular}
}
\caption{
Measured double-differential dijet cross-sections for the range $2.5 \leq \ystar < 3.0$ and jet radius parameter $R=0.6$. 
Here, $\sigma$ is the measured cross-section. 
All uncertainties are given in \%. 
The variable $\delta_\mathrm{stat}^\mathrm{data}$ ($\delta_\mathrm{stat}^\mathrm{MC}$) is the statistical uncertainty from the data (MC simulation). 
The $\gamma$ components show the uncertainties due to those on the jet energy calibration from the in situ, pileup, close-by jet, and flavour components. 
The $u$ components show the uncertainty for the jet energy and angular resolution, the unfolding, the quality selection, and the integrated luminosity. 
While all columns are uncorrelated with each other, the in situ, pileup, and flavour uncertainties shown here are the sum in quadrature of multiple uncorrelated components. 
The full set of cross-section values and uncertainty components, each of which is fully correlated in dijet mass and \ystar but uncorrelated with the other components, can be found in HepData \cite{Buckley:2010jn}. 
}
\label{tab:sysunc_r06_Eta6}
\end{table*}

\renewcommand{\arraystretch}{1}

\clearpage

\ifdraft
\bibliographystyle{atlasnote}
\bibliography{paper}
\else
\providecommand{\href}[2]{#2}\begingroup\raggedright\endgroup

\fi

\ifdraft
\clearpage
\input{auxiliary}
\else
\onecolumn
\clearpage 
\begin{flushleft}
{\Large The ATLAS Collaboration}

\bigskip

G.~Aad$^{\rm 48}$,
T.~Abajyan$^{\rm 21}$,
B.~Abbott$^{\rm 112}$,
J.~Abdallah$^{\rm 12}$,
S.~Abdel~Khalek$^{\rm 116}$,
O.~Abdinov$^{\rm 11}$,
R.~Aben$^{\rm 106}$,
B.~Abi$^{\rm 113}$,
M.~Abolins$^{\rm 89}$,
O.S.~AbouZeid$^{\rm 159}$
% ...
\bigskip
\\
% ...
$^{11}$ Institute of Physics, Azerbaijan Academy of Sciences, Baku, Azerbaijan\\
$^{12}$ Institut de F{\'\i}sica d'Altes Energies and Departament de F{\'\i}sica de la Universitat Aut{\`o}noma de Barcelona, Barcelona, Spain\\
% ...
$^{21}$ Physikalisches Institut, University of Bonn, Bonn, Germany\\
% ...
$^{48}$ Fakult{\"a}t f{\"u}r Mathematik und Physik, Albert-Ludwigs-Universit{\"a}t, Freiburg, Germany\\
% ...
$^{89}$ Department of Physics and Astronomy, Michigan State University, East Lansing MI, United States of America\\
% ...
$^{106}$ Nikhef National Institute for Subatomic Physics and University of Amsterdam, Amsterdam, Netherlands\\
% ...
$^{112}$ Homer L. Dodge Department of Physics and Astronomy, University of Oklahoma, Norman OK, United States of America\\
$^{113}$ Department of Physics, Oklahoma State University, Stillwater OK, United States of America\\
% ...
$^{116}$ LAL, Universit{\'e} Paris-Sud and CNRS/IN2P3, Orsay, France\\
% ...
$^{159}$ Department of Physics, University of Toronto, Toronto ON, Canada\\
% ...
\end{flushleft}


\fi

\end{document}